\documentclass[twocolumn]{aastex63}

\newcommand{\sna}{SN~1987A}
\newcommand{\sleak}{Sanduleak~\nobreakdash-69$^\circ$~202}

\newcommand{\kmps}{~km~s$^{-1}$}
\newcommand{\ergps}{~erg~s$^{-1}$}
\newcommand{\ergpspcm}{~erg~s$^{-1}$~cm$^{-2}$}
\newcommand{\pcm}{~cm$^{-2}$}

\newcommand{\Msun}{M$_{\sun}$}

\newcommand{\Lsun}{L$_{\sun}$}

\newcommand{\hi}{H\textsc{\,i}}
\newcommand{\hii}{H\textsc{\,ii}}

\newcommand{\xray}{X\nobreakdash-ray}

\newcommand{\xmm}{\mbox{XMM-Newton}}
\newcommand{\chandra}{\mbox{Chandra}}
\newcommand{\nustar}{\mbox{NuSTAR}}

\newcommand{\fermi}{\mbox{Fermi}}

\newcommand{\ti}{$^{44}$Ti}
\newcommand{\tisc}{$^{44}$Sc}
\newcommand{\tiscca}{$^{44}$Ca}
\newcommand{\co}{$^{55}$Co}
\newcommand{\fe}{$^{55}$Fe}
\newcommand{\femn}{$^{55}$Mn}

\usepackage{amsmath}
\usepackage{textcomp}
\usepackage{pifont}

\newcommand{\cmark}{\ding{51}}%
\newcommand{\xmark}{\ding{55}}%

\submitjournal{ApJ}

\shorttitle{Thermal X-Rays, Radioactive Lines, No Pulsar}
\shortauthors{Alp et al.}

\begin{document}

\title{Thermal Emission and Radioactive Lines, but No Pulsar, in the
  Broadband X-Ray Spectrum of Supernova 1987A}

\correspondingauthor{Dennis Alp}
\email{dalp@kth.se}

\author[0000-0002-0427-5592]{Dennis Alp}
\affiliation{Department of Physics, KTH Royal Institute of Technology,
  and The Oskar Klein Centre, SE\nobreakdash-10691 Stockholm, Sweden}             

\author[0000-0003-0065-2933]{Josefin Larsson}
\affiliation{Department of Physics, KTH Royal Institute of Technology,
  and The Oskar Klein Centre, SE\nobreakdash-10691 Stockholm, Sweden}             

\author[0000-0001-8532-3594]{Claes Fransson} \affiliation{The Oskar
  Klein Centre, Department of Astronomy, Stockholm University,
  AlbaNova, SE\nobreakdash-10691 Stockholm, Sweden}

\begin{abstract}
  Supernova 1987A offers a unique opportunity to study an evolving
  supernova in unprecedented detail over several decades. The \xray{}
  emission is dominated by interactions between the ejecta and the
  circumstellar medium, primarily the equatorial ring (ER). We analyze
  3.3~Ms of \nustar{} data obtained between 2012 and 2020, and two
  decades of \xmm{} data. Since ${\sim}$2013, the flux below 2~keV has
  declined, the 3--8~keV flux has increased, but has started to
  flatten, and the emission above 10~keV has remained nearly
  constant. The spectra are well described by a model with three
  thermal shock components. Two components at 0.3 and 0.9~keV are
  associated with dense clumps in the ER, and a 4~keV component may be
  a combination of emission from diffuse gas in the ER and the
  surrounding low-density \hii{} region. We disfavor models that
  involve non-thermal \xray{} emission and place constraints on
  non-thermal components, but cannot firmly exclude an underlying
  power law. Radioactive lines show a \ti{} redshift of
  $670^{+520}_{-380}$\kmps{}, \ti{} mass of
  $1.73_{-0.29}^{+0.27}\times{}10^{-4}$~\Msun{}, and \fe{} mass of
  $<4.2\times{}10^{-4}$~\Msun{}. The 35--65~keV luminosity limit on
  the compact object is $2\times{}10^{34}$\ergps{}, and $<15$\,\% of
  the 10--20~keV flux is pulsed. Considering previous limits, we
  conclude that there are currently no indications of a compact
  object, aside from a possible hint of dust heated by a neutron star
  in recent ALMA images.
\end{abstract}


\section{Introduction}\label{sec:intro}
Supernova (SN) 1987A is the closest observed SN in more than four
centuries \citep{arnett89, mccray93, mccray16}. It is a core-collapse
SN located in the Large Magellanic Cloud (LMC) and was first detected
on 1987 February 23. The progenitor star,
\sleak{}~\citep{sanduleak70}, was identified as a B3~Ia blue
supergiant~\citep{west87}. The progenitor shaped the circumstellar
medium (CSM) around \sna{} and produced a triple-ring structure
\citep{wampler90, burrows95c}. This structure consists of two larger
outer rings and a smaller equatorial ring (ER). Different explanations
for this based on interacting winds \citep{blondin93, martin95},
binary mergers \citep{morris07, menon17, utrobin21}, or rapid rotation
\citep{chita08} have been proposed. The three rings are nearly
coaxial, forming an hourglass-like shape \citep{chevalier95,
  larsson19b}. They are nearly circular, but appear elliptical because
they are viewed at an inclination of ${\sim}$38--45\degr{}
\citep{crotts00, tziamtzis11}. The \xray{} emission that we study in
this paper is primarily due to the interaction with this complex CSM.

At early times, the ejecta from \sna{} propagated through a
low-density \hii{} region inside of the ER, which gave rise to faint
radio and \xray{} emission \citep{staveley-smith92, chevalier95,
  hasinger96}. After approximately 3000~d, the ejecta started
interacting with the denser ER. This interaction produced a
significant brightening in all wavebands \citep{bouchet06, park06,
  groningsson08b, ng08, dwek10}. Notably, high-density clumps inside
the ER produced bright optical ``hotspots'' \citep{sonneborn98,
  lawrence00}. More recent observations show that the optical emission
is declining and that new hotspots have appeared outside of the ER
\citep{fransson15, larsson19b}. Furthermore, the infrared (IR)
emission has been declining since ${\sim}$9000~d \citep{arendt16,
  arendt20}, \xray{}s below 2~keV started decreasing at 10,000~d
\citep{frank16, ravi19c}, and the radio blast wave is now
reaccelerating \citep{cendes18}. These observations all show that most
of the blast wave is now propagating past the ER, while the ER is
being dissolved.

In addition to the long monitoring of \sna{} in radio, optical, and
\xray{}s, a potential GeV detection has recently been reported using
Fermi data \citep{malyshev19, petruk20}. A major uncertainty is if the
emission is truly originating from \sna{} because the angular
resolution of Fermi is insufficient to resolve \sna{} from other
nearby sources. If associated with \sna{}, the data suggest an
increase in the GeV flux by a factor of ${\sim}$2 between 2010 and
2018, which could indicate efficient cosmic ray acceleration in the
shocks \citep{dwarkadas13, berezhko15}.

The \xray{} emission produced by the interactions with the ER has been
extensively studied in the literature \citep{burrows00, michael02,
  park02, park04, park05, park06, park11, zhekov05, zhekov06,
  zhekov09, zhekov10, haberl06, dewey08, dewey12, heng08, racusin09,
  sturm09, sturm10, maggi12, helder13, orlando15, orlando19, frank16,
  miceli19, bray20, greco21, sun21}. For brevity, we only summarize
the most relevant findings from the previous studies. The ER is
resolved by Chandra, which allows for studies of the morphology and
expansion velocity. Comparisons with other wavelengths show that
\xray{}s below 2~keV are generally more strongly correlated with
optical, whereas harder \xray{}s better follow the radio evolution.
The \xray{} emission has been interpreted as predominantly or
completely of thermal origin. The detection of the Fe~K line complex
clearly shows that thermal emission is important at energies up to
${\sim}$10~keV.

Thermal models for the X-ray spectra require multiple components of
different temperatures. This is not unexpected given the complex
structure of the CSM around \sna{}, but it is difficult to uniquely
associate each spectral component with one of the many possible
emission regions. The most likely origins for the \xray{} emission
are: the dense ER clumps, the diffuse ER, the \hii{} region, the
reverse shock propagating into the ejecta, and reflected shocks off
the ER clumps. However, it remains uncertain if there is a
contribution from a non-thermal component. Possible origins for an
additional component are shock-accelerated electrons, high-temperature
free-free emission, or a pulsar wind nebula (PWN).

In addition to the thermal emission, there are a number of lines that
result from radioactive decay \citep{diehl18}. Explosive
nucleosynthesis in SN explosions produces unstable proton-rich
isotopes that decay by electron capture. This gives rise to different
types of radioactive lines, one of which is nuclear de-excitation
lines. The most prominent example at current epochs is the pair of
lines at 67.87 and 78.32~keV, which are produced by the
$\text{\ti{}}\rightarrow{}\text{\tisc{}}\rightarrow\text{\tiscca{}}$
chain \citep{grebenev12, boggs15}. Technically, \ti{} decays into an
excited nuclear state of \tisc{}, and these lines are emitted when the
\tisc{} nucleus de-excite to its ground state
\citep{cameron99}. However, these lines are conventionally referred to
as the \ti{} lines.

The \ti{} is produced in the SN explosion and serves as an important
probe of the explosion physics (e.g., \citealt{janka17}). From its
high-energy nuclear decay lines, it is possible to infer the \ti{}
redshift, which is a measure of the explosion asymmetries, as well as
the initial \ti{} mass. Important advantages of these nuclear
radioactive lines are that the emission escapes the ejecta practically
unattenuated at current epochs \citep{alp18c} and that the emission is
proportional to the mass. The proportionality constant is known
through the 85~yr lifetime \citep{ahmad06} and is not dependent on any
shock dynamics or radiation reprocessing.

Another type of radioactive line is electron capture
K\nobreakdash-shell emission lines \citep{leising01}. This is a result
of the electron capture process leaving a vacancy in the
K\nobreakdash-shell. When the vacancy is filled by electron
de-excitation, a characteristic \xray{} below 10~keV is emitted. The
$\text{\ti{}}\rightarrow{}\text{\tisc{}}$ decay also produces \tisc{}
K$\alpha$ emission. Another example is the
$\text{\co{}}\rightarrow{}\text{\fe{}}\rightarrow{}\text{\femn{}}$
chain \citep{junde08}, which emits \femn{} K$\alpha$ lines.

Both \co{} and \fe{} are produced in SNe, but \co{} has a lifetime of
only 25~h. Therefore, we will henceforth refer to this as the \fe{}
chain and treat \co{} as \fe{}. \fe{} has a lifetime of 3.9~yr and its
daughter isotope \femn{} produces K$\alpha$ lines around
5.9~keV. Furthermore, the daughter isotope of \ti{}, \tisc{}, emits
lines around 4.1~keV. Detections or strong limits on these lines would
constrain nucleosynthesis yields and absorption properties of the
ejecta, which in turn are related to the explosion mechanism and
progenitor properties.

The line energies and, in particular, the electron capture decay rates
are often treated as constants, but depend on the level of ionization
\citep{mochizuki99,laming01,motizuki04}. This is primarily significant
for ionization to H\nobreakdash-like, He\nobreakdash-like, or
completely ionized states. The inner ejecta where the bulk of the
radioactive elements reside are not expected to be highly ionized
\citep{jerkstrand11}. Therefore, the standard values measured for
neutral atoms can be applied to the radioactive line energies and
decay rates relevant for \sna{}.

Neutrinos were detected from the core collapse of \sna{}
\citep{hirata87}, which signaled the formation of a neutron star. No
further unequivocal evidence for the neutron star has been presented,
and it remains uncertain if it collapsed further into a black
hole. Numerous searches have been performed \citep{graves05,
  manchester07, alp18c, alp18b, esposito18, cigan19, page20, greco21},
but only tentative indications have been found. \citet{cigan19}
reported a possible sign of localized energy input by a neutron star
in a 679~GHz ALMA image. They suggested heating by thermal surface
neutron star emission \citep{alp18b, page20} or the emergence of a
PWN, which was also suggested by \citet{zanardo14}. Recently,
\citet{greco21} interpreted the \xray{} emission above 10~keV as
indications of a PWN.

In this paper, we study new \nustar{} data from 2020, together with
archival \nustar{} data from 2012 to 2014 \citep{boggs15, greco21}. To
extend the energy range and time span, we also include archival \xmm{}
observations spanning nearly two decades. These \xmm{} observations
have previously been analyzed in a number of separate studies focused
on different aspects of the data \citep{haberl06, heng08, sturm10,
  maggi12, sun21}

The main focus is the interpretation of the 0.45--24~keV spectrum and
a possible contribution from a non-thermal component, which could be
connected to the GeV emission or a PWN. We analyze the radioactive
\ti{} lines using \nustar{}, and place limits on the radioactive
\femn{} and \tisc{} lines using \xmm{}. \nustar{} is also used to
constrain the pulsed fraction and put a deep upper limit on the
contribution from a compact object. Finally, we put the compact object
limit into context by comparing it with previous multi-wavelength
observations.

This paper is organized as follows. We present the observations
constituting the main data set in Section~\ref{sec:obs} and the
methods used for the analysis in Section~\ref{sec:met}. The results
are contained in Section~\ref{sec:res}, and discussed in
Section~\ref{sec:dis}. We provide a summary and highlight the
conclusions in Section~\ref{sec:sum}. Two independent parts of the
analysis are separated into appendices. In Appendix~\ref{app:ps}, we
search for pulsed emission in the 2012--2014 \nustar{} data. We
analyze the radioactive K\nobreakdash-shell lines from \femn{} and
\tisc{} using 2000--2003 \xmm{} data in
Appendix~\ref{app:rka}. Finally, we explore the calibration
uncertainties of \nustar{} and \xmm{} in Appendices~\ref{app:sys}
and~\ref{app:pn}.

\section{Observations and Data Reduction}\label{sec:obs}
We focus on the \nustar{} \citep{harrison13} observations of \sna{},
but include \xmm{} \citep{jansen01} data to extend the lower energy
bound to 0.45~keV and resolve the \xray{} line emission. Accurate
modeling of the line emission is important for the interpretation of
the entire \xray{} spectrum. Including \xmm{} data also extends the
temporal baseline for flux measurements back to 2003. We use \xmm{}
instead of \chandra{} primarily because we do not need to spatially
resolve the ER. Instead, we prioritize the better photon statistics
provided by the larger effective area of \xmm{}. For the data
reduction and analysis, we use \mbox{HEAsoft} 6.27.2.

\subsection{\nustar{}}\label{sec:nus}
\begin{deluxetable*}{ccccccccccccccc}
  \tablecaption{\nustar{} Observations\label{tab:obs_nus}}
  \tablewidth{0pt}
  \tablehead{\colhead{Obs.\ ID} & \colhead{Start Date} & \colhead{Epoch} & \colhead{Exp.} & $C_\text{A, 3--24~keV}$ & $C_\text{B, 3--24~keV}$ & $C_\text{A, 10--24~keV}$ & $C_\text{B, 10--24~keV}$ & \colhead{Group} \\
             \colhead{}       & \colhead{(YYYY-mm-dd)} & \colhead{(d)}   & \colhead{(ks)} & \colhead{}              & \colhead{}              & \colhead{}               & \colhead{}               & \colhead{}      }
           \startdata
    40001014002 & 2012-09-07 & \hphantom{1,}9328 & \hphantom{0}69 &  898 &  840 &  42 &  45 & 1 \\
    40001014003 & 2012-09-08 & \hphantom{1,}9329 &            137 & 1898 & 1846 & 110 & 109 & 1 \\
    40001014004 & 2012-09-11 & \hphantom{1,}9332 &            199 & 2751 & 2817 & 176 & 165 & 1 \\
    40001014006 & 2012-10-20 & \hphantom{1,}9371 & \hphantom{0}54 &  796 &  737 &  35 &  18 & 2 \\
    40001014007 & 2012-10-21 & \hphantom{1,}9372 &            200 & 3021 & 3095 & 186 & 176 & 2 \\
    40001014009 & 2012-12-12 & \hphantom{1,}9424 & \hphantom{0}28 &  293 &  342 &   0 &  18 & 3 \\
    40001014010 & 2012-12-12 & \hphantom{1,}9424 &            186 & 2828 & 2686 & 171 & 182 & 3 \\
    40001014012 & 2013-06-28 & \hphantom{1,}9622 & \hphantom{0}19 &  274 &  280 &   0 &  18 & 4 \\
    40001014013 & 2013-06-29 & \hphantom{1,}9623 &            473 & 7099 & 6803 & 459 & 415 & 4 \\
    40001014015 & 2014-04-21 & \hphantom{1,}9919 & \hphantom{0}97 & 1404 & 1465 &  58 &  51 & 5 \\
    40001014016 & 2014-04-22 & \hphantom{1,}9920 &            432 & 6996 & 6924 & 432 & 391 & 5 \\
    40001014018 & 2014-06-15 & \hphantom{1,}9974 &            200 & 3030 & 2863 & 168 & 145 & 6 \\
    40001014020 & 2014-06-19 & \hphantom{1,}9978 &            275 & 4442 & 4134 & 313 & 253 & 6 \\
    40001014022 & 2014-08-01 &            10,021 & \hphantom{0}48 &  738 &  683 &  19 &  39 & 7 \\
    40001014023 & 2014-08-01 &            10,021 &            428 & 7046 & 6260 & 445 & 361 & 7 \\
    40501004002 & 2020-05-13 &            12,133 &            180 & 3828 & 3868 & 167 & 199 & 8 \\
    40501004004 & 2020-05-27 &            12,147 &            230 & 3544 & 3477 & 195 & 169 & 8 \\
    \enddata

    \tablecomments{The net source counts in each of the two \nustar{}
      modules are given by $C_\mathrm{A}$ and $C_\mathrm{B}$. Counts
      are given for two different energy ranges as indicated by the
      subscripts. The observations with 0 counts are results of no
      bins (after binning to a minimum of 25 counts per bin) passing
      the energy range requirement as implemented by the
      \texttt{ignore} task in XSPEC due to short exposures.}

\end{deluxetable*}
\begin{figure*}
  \centering
  \includegraphics[width=0.49\textwidth]{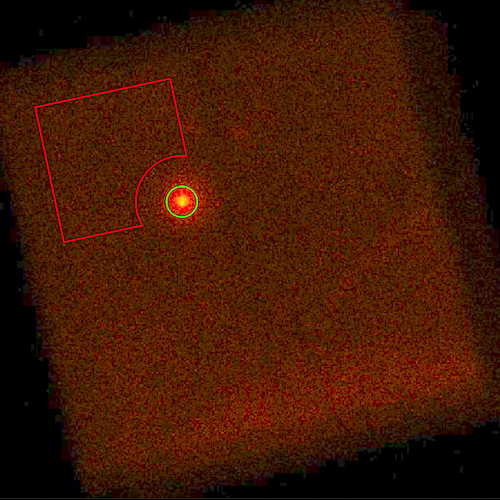}
  \includegraphics[width=0.49\textwidth]{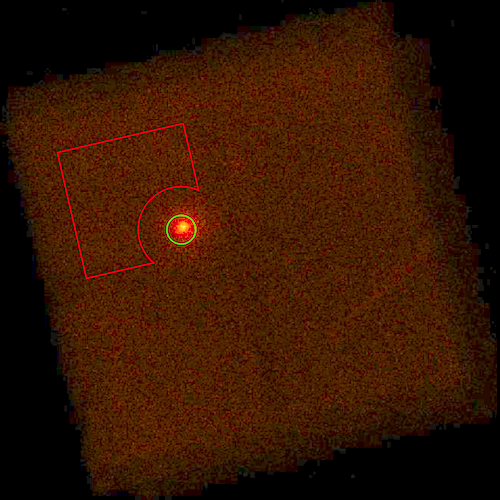}

  \caption{\nustar{} images showing the source extraction regions
    (green) and background regions (red) for FPMA (left) and FPMB
    (right). We only show the observation from 2014 April 22 because
    other observations are similar. The intensity scales are
    approximately linear and have been slightly offset to better show
    the background. The pixel size is $2\farcs{}46\times2\farcs{}46$
    and the FoV is approximately $13\arcmin{}\times{}13\arcmin$. North
    is up and east is left.\label{fig:nus}}
  
\end{figure*}
The \nustar{} telescope has an energy range of 3--78.4~keV and an
angular resolution of 18\arcsec{} (full width at half maximum;
FWHM). The spatial resolution is insufficient to resolve the ER
(${\sim}$1\arcsec{}) but clearly separates \sna{} from any other
sources. The energy resolution has a FWHM response of 0.4~keV at
10~keV and 0.9~keV at 60~keV. This spectral resolution allows for
separation of isolated lines and line complexes from the continuum,
but does not resolve finer structures or crowded spectral
regions. \nustar{} has two mirror assemblies and two corresponding
focal plane CCD modules (FPMs), referred to as FPMA and FPMB. The two
parallel setups are practically identical but all data are analyzed
separately and only combined for presentation purposes.

We use all on-axis \nustar{} observations of \sna{}
(Table~\ref{tab:obs_nus}), which constitute a total exposure time of
3.3~Ms. The majority of the data are from 2012--2014. These were
previously used to study the radioactive \ti{} lines \citep{boggs15}
and the continuum \citep{greco21}. To complement these data, we
obtained new observations in May 2020, which allow us to study the
recent temporal evolution. There are 17 individual observations in
total, but these constitute eight groups of observations that are
quasi-simultaneous (within two weeks, see
Table~\ref{tab:obs_nus}). Henceforth, the observations within each
group are treated as simultaneous, but the data are not combined
except for presentation purposes.

The \nustar{} data reduction largely follows standard procedures. To
reduce the data, we use NuSTARDAS 1.9.2 and \nustar{} CALDB version
20200726. First, calibrations and filters are applied using the
\texttt{nupipeline} task. The exposures are filtered for passages
through the South Atlantic Anomaly. This is a trade-off between
exposure time and background. We check the SAA filter reports and use
the options \texttt{saacalc=1}, \texttt{saamode=strict}, and
\texttt{tentacle=yes}.

Spectra are then generated using the standard task
\texttt{nuproducts}. Source counts are extracted from a circular
region centered on \sna{}, shown in Figure~\ref{fig:nus}. We choose a
radius of 30\arcsec{}, which is recommended for weaker sources by the
\nustar{} Observatory
Guide\footnote{\url{https://heasarc.gsfc.nasa.gov/docs/nustar/NuSTAR_observatory_guide-v1.0.pdf}}. This
results in an encircled energy fraction of approximately 50\,\%.
\sna{} is relatively weak above 8~keV, which is the energy range that
is most important for the analysis. Background regions are selected
from the same CCD chip as the source. The vicinity of \sna{} is free
of strong NuSTAR sources, but bright \xray{} sources outside the field
of view (FoV) introduce stray light into the FoV \citep{madsen17c}. In
general, we aim to select background regions that are squares with
sides of $4.5$\arcmin, while excluding a region around \sna{} and
stray light. The excluded region around \sna{} is a circle with a
radius of 90\arcsec{} (see Figure~\ref{fig:nus}).

We limit the energy range used for the continuum analysis to
3--24~keV. Both the flux and effective area quickly decrease toward
higher energies, resulting in a signal-to-background ratio (S/B) lower
than $0.3$. The low S/B above 24~keV results in a very high
sensitivity to background systematics.
Finally, the spectra are grouped to at least 25 counts per bin. 

\subsection{\xmm{}}\label{sec:xmm}
\begin{deluxetable*}{ccccccccccccccc}
  \tablecaption{\xmm{} Observations\label{tab:obs_xmm}}
  \tablewidth{0pt}
  \tablehead{\colhead{Obs.\ ID} & \colhead{Start Date} & \colhead{Epoch} & \colhead{RGS Exp.\tablenotemark{a}} & \colhead{pn Exp.\tablenotemark{a}} & $C_\text{RGS1}$ & $C_\text{RGS2}$ & $C_\text{pn}$ & \colhead{\nustar{} Groups\tablenotemark{b}} \\
             \colhead{}       & \colhead{(YYYY-mm-dd)} & \colhead{(d)}   & \colhead{(ks)}                      & \colhead{(ks)}                     & \colhead{}      & \colhead{}      & \colhead{}    & \colhead{}}
  \startdata
    0144530101\hphantom{\tablenotemark{c}} & 2003-05-10 & \hphantom{1,}5920 &            112(113)            & 59(107)            &   2996 &   3698 &  21,122 & \hphantom{.}\nodata{} \\
    0406840301\hphantom{\tablenotemark{c}} & 2007-01-17 & \hphantom{1,}7268 & \hphantom{0}82(111)            & 64(107)            &   5682 &   9030 &  91,386 & \hphantom{.}\nodata{} \\
    0506220101\hphantom{\tablenotemark{c}} & 2008-01-11 & \hphantom{1,}7627 &            114(115)            & 70(110)            & 10,169 & 15,965 & 136,480 & \hphantom{.}\nodata{} \\
    0556350101\hphantom{\tablenotemark{c}} & 2009-01-30 & \hphantom{1,}8012 & \hphantom{0}90(102)            & 68(100)            & 10,258 & 15,503 & 164,945 & \hphantom{.}\nodata{} \\
    0601200101\hphantom{\tablenotemark{c}} & 2009-12-11 & \hphantom{1,}8327 & \hphantom{0}92(92)\hphantom{0} & 77(90)\hphantom{0} & 11,689 & 17,587 & 212,221 & \hphantom{.}\nodata{} \\ 
    0650420101\hphantom{\tablenotemark{c}} & 2010-12-12 & \hphantom{1,}8693 & \hphantom{0}66(66)\hphantom{0} & 51(64)\hphantom{0} &   9162 & 14,248 & 157,450 & \hphantom{.}\nodata{} \\ 
    0671080101\hphantom{\tablenotemark{c}} & 2011-12-02 & \hphantom{1,}9048 & \hphantom{0}75(82)\hphantom{0} & 62(81)\hphantom{0} & 11,156 & 17,346 & 213,257 & \hphantom{.}\nodata{} \\ 
    0690510101\tablenotemark{c}            & 2012-12-11 & \hphantom{1,}9423 & \hphantom{0}70(70)\hphantom{0} & 60(68)\hphantom{0} & 10,663 & 16,665 & 212,129 & 1--4 \\
    0743790101\tablenotemark{c}            & 2014-11-29 &            10,141 & \hphantom{0}73(80)\hphantom{0} & 57(78)\hphantom{0} & 10,827 & 17,174 & 210,278 & 5--7 \\
    0763620101\hphantom{\tablenotemark{c}} & 2015-11-15 &            10,492 & \hphantom{0}66(66)\hphantom{0} & 54(64)\hphantom{0} &   9558 & 15,084 & 195,380 & \hphantom{.}\nodata{} \\ 
    0783250201\hphantom{\tablenotemark{c}} & 2016-11-02 &            10,845 & \hphantom{0}74(74)\hphantom{0} & 48(72)\hphantom{0} & 10,416 & 16,159 & 176,047 & \hphantom{.}\nodata{} \\ 
    0804980201\tablenotemark{c}            & 2017-10-15 &            11,192 & \hphantom{0}72(79)\hphantom{0} & 34(78)\hphantom{0} &   9458 & 14,983 & 120,854 & \hphantom{.}\nodata{} \\    
    0831810101\hphantom{\tablenotemark{c}} & 2019-11-27 &            11,965 & \hphantom{0}35(35)\hphantom{0} & 19(32)\hphantom{0} &   3925 &   6303 &  63,847 & 8 \\
  \enddata

  \tablecomments{The net source counts in the different instruments
    are given by $C_\mathrm{RGS1}$, $C_\mathrm{RGS2}$, and
    $C_\mathrm{pn}$. All counts are for the standard energy ranges we
    use for the analysis (see text). The RGS counts are for 1st order
    only. The differences between RGS1 and RGS2 are primarily driven
    by the different non-operational CCDs. The total counts in the 2nd
    order spectra are 45\,\% of the 1st order counts for RGS1 and
    26\,\% for RGS2.}

  \tablenotetext{a}{Exposure times after(before) excluding
    high-background intervals.}
  
  \tablenotetext{b}{Indicates which \xmm{} data sets are matched with
    which \nustar{} observation groups, cf.\ Table~\ref{tab:obs_nus}.}

  \tablenotetext{c}{Observations used for determining the fixed
    parameters (Section~\ref{sec:abu}).}

\end{deluxetable*}

\begin{figure}
  \centering
  \includegraphics[width=0.49\textwidth]{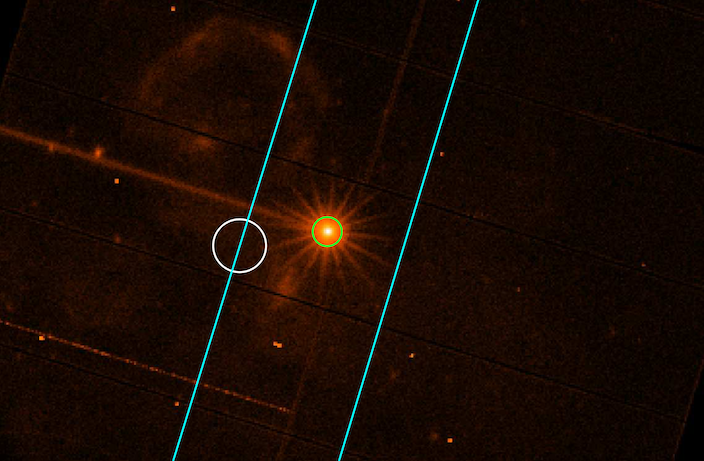}
  \caption{\xmm{}/pn image showing the source region (green),
    background region (white), and RGS FoV (cyan). The cyan lines are
    parallel with the dispersion direction. We only show the
    observation from 2014 November 29 because other observations are
    similar. The intensity scale is logarithmic and white corresponds
    to ${\sim}$24,000 counts per pixel. The pixel size is
    $4\arcsec\times{}4\arcsec{}$ and the FoV of the image is
    approximately $27\arcmin{}\times{}18\arcmin$. North is up and east
    is left.\label{fig:xmm}}
\end{figure}
\begin{figure}
  \centering
  
  \includegraphics[width=0.49\textwidth]{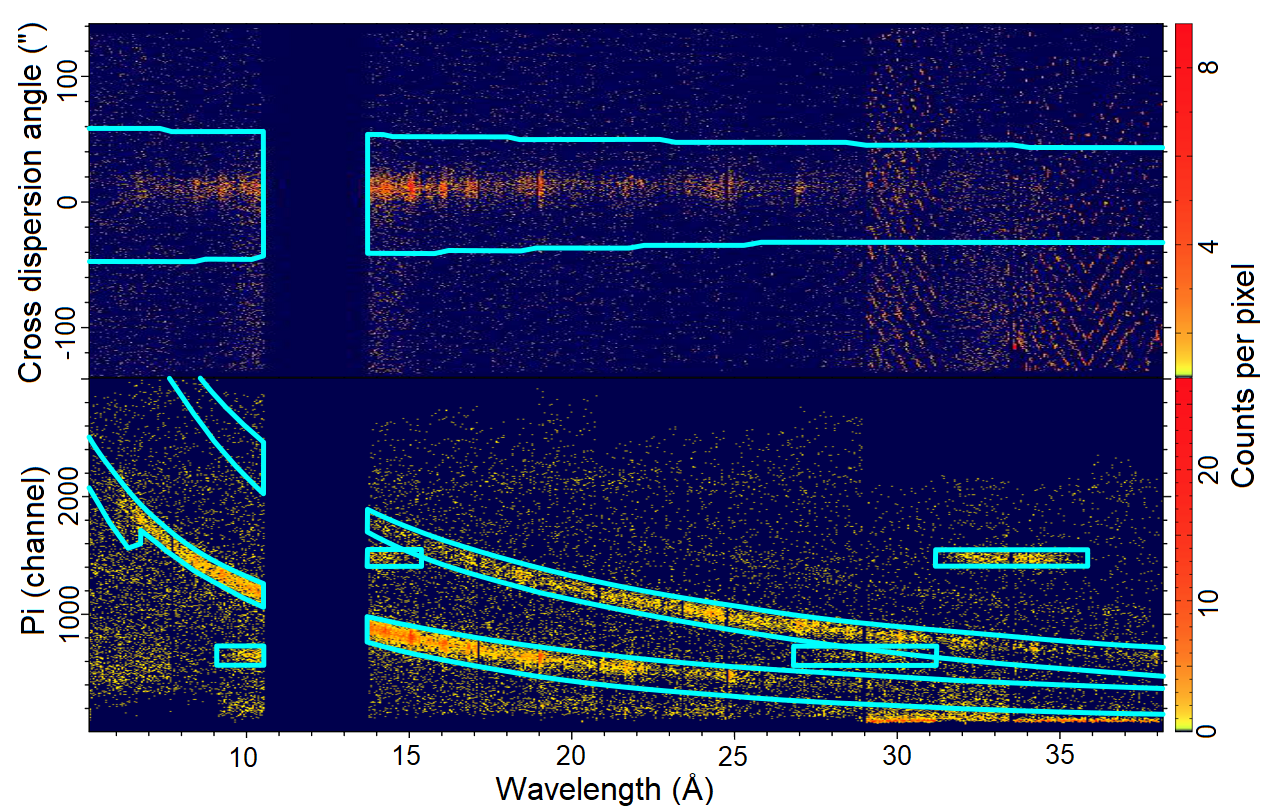}
  \caption{\xmm{}/RGS1 images showing a 1\nobreakdash-dimensional
    dispersed image in the upper panel and a PI-dispersion angle plot
    (banana plot) in the lower panel. The cyan regions in the upper
    panel are the source regions. The curved regions in the lower
    panel are source regions (1st and 2nd order) while the four
    rectangular regions show excluded areas (containing calibration
    sources). The spatial background region covers the entire exposed
    surface of the focal plane, except for a buffer zone around the
    source. We only show the observation from 2014 November 29 because
    other observations are similar.\label{fig:rgs}}
\end{figure}

\xmm{} operates six instruments in parallel: the pn CCD
\citep{struder01}; MOS1 and MOS2 \citep{turner01}; two grating
spectrometers \citep{den_herder01}; and the optical monitor
\citep{mason01}. Here, we focus on the pn CCD and the reflection
grating spectrometers (RGSs). The pn camera covers 0.3--10~keV with a
spectral resolution of $E/\Delta E \sim$ 20--50, which is comparable
to the \nustar{} energy resolution. The RGSs cover 0.35--2.5~keV and
provide $E/\Delta E \sim$ 200--600, which is sufficient to resolve the
large number of lines in this energy range. The angular resolution is
approximately 6\arcsec{} (FWHM) for both pn and the cross-dispersion
direction of the RGSs, which does not resolve the ER. The RGS
apertures extend across the entire 30\arcmin{} FoV along the
dispersion direction, but the effective area decreases quickly as a
function of off-axis angle. We verified that no other sources are
confused with \sna{}. Contamination along the dispersion direction of
the RGSs can be identified by comparing the dispersion angle to the
energy estimate from the RGS CCD readouts.

We primarily focus on the most recent \xmm{} observations for the
joint NuSTAR-XMM analysis, but include 13 publicly available \xmm{}
observations in total for a longer temporal baseline
(Table~\ref{tab:obs_xmm}). Due to modest levels of pile-up (detailed
below), we only use CCD data from the pn CCD, and not the MOS CCDs
\citep{turner01}. Pile-up is not present in the RGS data.

For the data reduction, we use XMM SAS 18.0.0 \citep{gabriel04} with
XMM CCF updated on 2020 August 3. Most steps of the data reduction
follow default standards. We reprocess the data with the latest
calibrations and filter the data for high-background intervals
(Table~\ref{tab:obs_xmm}).

For the pn data, we extract source spectra from a circular region
centered on \sna{} with a radius of 30\arcsec{}, shown in
Figure~\ref{fig:xmm}. This is a typical source extraction radius and
results in an encircled energy of 80--90\,\% depending on energy
(lower for higher energies). A larger source radius would only
increase the signal marginally while lowering the S/B. For reference,
a source radius of 90\arcsec{} results in an encircled energy of
90--95\,\%, while increasing the background from ${\sim}$1\,\% to
almost 10\,\%. Backgrounds are extracted from circular regions from
the same CCD chip (\#4) with radii of approximately 60\arcsec{}
(Figure~\ref{fig:xmm}). The very high S/B implies that the choice of
background has a negligible impact on the analysis. The RGS spectra
are constructed using the default source and background regions. A
heliocentric correction to the photon energies is also applied to the
RGS data by default.

We pay special attention to pile-up in the CCD spectra. Pile-up occurs
when multiple photons hit the same or adjacent pixels within the same
readout frame. The results are that some photons are rejected as
invalid events, and an artificial hardening of the spectrum as the
energies of multiple photons are combined. Pile-up affects
approximately 5\,\% of the events when the photon flux peaks around
the year 2014. This is estimated using the task \texttt{epatplot},
which compares the observed event patterns with expected pattern
distributions. Even though the magnitude of the effect is low, it
could significantly affect the fits. The bins with the most photons
have statistical uncertainties as low as 1\,\%. We use the task
\texttt{rmfgen} to produce response matrix files (RMFs) that account
for pile-up, which is enabled by the option
\texttt{correctforpileup=yes}. The ancillary response files (ARFs) are
then generated using the standard task \texttt{arfgen}.

The CCD spectra are grouped to a minimum of 25 counts per bin using
\texttt{specgroup}. Additionally, we impose a minimum bin size
corresponding to $1/3$ of the energy resolution FWHM using the
parameter \texttt{oversample=3}. The RGS spectra are also binned to a
minimum of 25 counts per bin, but using the general task
\texttt{grppha}. The large number of photons allows for this binning
without significantly downsampling the energy resolutions of the RGSs.

We limit the energy range of the pn data to 0.8--10~keV. The lower
bound of 0.8~keV is higher than the lower calibration limit of
0.3~keV. We ignore data below 0.8~keV because of inconsistencies
between the pn and RGS spectra of up to 5\,\% close to strong lines,
which are likely due to calibration uncertainties.  The pn spectral
bins have individual statistical uncertainties as low as $\pm1$\,\% at
low energies, which implies that slight calibration issues could
significantly reduce the goodness of fit. The calibration accuracy is
known to decrease toward lower energies (\citealt{plucinsky08,
  plucinsky17}; see also Figures~\ref{fig:cal2}
and~\ref{fig:dos_cal}), possibly due to the lowest-energy internal
calibration line of pn being at 1.5~keV \citep{struder01}. We choose
the 0.8~keV limit as a trade-off between using as much data as
possible and introducing calibration tensions into the fits.

We use RGS data from both RGS1 and RGS2, and 1st and 2nd order. An
RGS1 image showing the dispersed data is provided in
Figure~\ref{fig:rgs}. The two 1st order spectra cover the energy range
from 0.45 to 1.95~keV, whereas the 2nd order spectra range from 0.70
to 1.95~keV. The low energy cutoff of 0.45~keV implies that we include
the strong N\textsc{\,vii} Ly$\alpha$ line at 0.50~keV, below which
very few photons are detected. Additionally, the parts of the RGS
spectra that coincide with non-operational CCDs are excluded.

\section{Analysis}\label{sec:met}
We use XSPEC 12.11.0 for spectral fitting and use the photoabsorption
cross sections of \citet{verner95}. We follow the XSPEC convention of
the photon index $\Gamma$ being given by $N \propto E^{-\Gamma}$,
where $N$ is the photon flux density and $E$ the photon
energy. Furthermore, the spectral binning of at least 25 counts per
bin allows for the use of $\chi^2$ statistics.

Confidence intervals are computed using the \texttt{error} command in
XSPEC. All intervals are 90\,\% unless otherwise stated, whereas
one-sided limits are 3$\sigma{}$. The \texttt{error} command varies
the parameters until the $\chi^2$ value reaches a given threshold. We
adopt a critical value of 2.706, which is commonly used in \xray{}
analyses. However, we stress that this technically only represents the
90\,\% interval for one parameter of interest \citep{avni76,
  lampton76, cash76}. The same argument applies to the 3$\sigma{}$
limits, for which we use a corresponding threshold of 7.740.

We use different combinations of data sets for different purposes. For
the flux estimates, we prioritize good fits to individual
epochs. Therefore, we only use \nustar{} data for \nustar{} fluxes,
and pn and RGS data for \xmm{} fluxes. For the spectral analysis, we
perform joint fits to \nustar{} and RGS data. We omit the \xmm{}/pn
data in these fits due to calibration uncertainties, which could
introduce systematic effects. The significance of the calibration
uncertainties is exacerbated by the large number of photons in the pn
data. We discuss the magnitude and characteristics of the instrumental
uncertainties further in Appendix~\ref{app:sys}. Furthermore,
Appendix~\ref{app:pn} investigates the effects of including pn in the
data analysis, as well as explores a number of other alternatives for
how to manage the systematic uncertainties. The primary conclusion is
that different choices affect the best-fit values quantitatively but
the qualitative scientific conclusions remain unchanged.

For all fits, we leave a free cross-normalization constant between
different instruments to accommodate calibration uncertainties. The
cross-normalizations also allow for some freedom between the \nustar{}
and \xmm{} observations, which are separated in time. These constants
are generally fitted to values within 0.9--1.1, with most values being
close to unity. For some fits, the constants deviate by more than
10\,\% form unity, but this is predominantly due to the time
difference between some of the \nustar{} and \xmm{} observations. This
shows that the cross-normalizations primarily capture calibration
uncertainties and to some extent also the temporal evolution between
non-simultaneous observations. Importantly, these constants do not
drift to unreasonable values, which would indicate problems with the
fits.

\subsection{Spectral Modeling}\label{sec:mod}
For spectral modeling, we consistently use the same base model unless
otherwise stated. It consists of an absorption component, a Gaussian
smoothing kernel, and three shocked plasma components.\footnote{In
  XSPEC terms, this is
  \texttt{constant(TBabs(gsmooth(}\\\texttt{vpshock+vpshock+vpshock)))}.}
We use the Tuebingen-Boulder interstellar medium (ISM) absorption
model (\texttt{TBabs}; \citealt{wilms00}), \texttt{gsmooth} for
smoothing, and \texttt{vpshock} for the plasma emission. We use three
duplicates of the shock component to properly capture the range of
temperatures that contribute to the spectrum in the 0.45--24~keV range
(Section~\ref{sec:sa} and Appendix~\ref{app:pn}). Below, we describe
the components in more detail.

The absorption component has only one free parameter, which is the H
column density ($N_\mathrm{H}$). We stress that this is the total H
density, including both atomic H and molecular H$_2$
\citep{willingale13}. Solely for the absorption, we adopt the Milky
Way ISM abundances of \citet{wilms00}. Implications of this on the
estimated $N_\mathrm{H}$ are discussed in Section~\ref{sec:abu}.

The Gaussian smoothing represents the broadening of the spectral lines
due to the bulk and turbulent motion of the emitting plasma (Appendix
of \citealt{dewey12}). The amount of Doppler blurring in units of
energy is proportional to the energy. This implies that the exponent
($\alpha$) for the \texttt{gsmooth} energy dependence is always frozen
to 1. Thus, the only free parameter of the smoothing component is the
magnitude of the blurring measured by $\sigma{}$ of the Gaussian
kernel. Freezing $\alpha$ to 1 implicitly assumes that the emission is
produced by gas with the same bulk velocity. Neither assumption is
likely strictly fulfilled. However, we fix it at 1 since we find
negligible improvements when leaving it as a free parameter.

The \texttt{vpshock} component \citep{borkowski01} is based on
calculations of \xray{} spectra using a SN blast-wave model. The main
parameters are the shock temperature ($T$), ionization age ($\tau$),
and emission measure (EM). The EM is implemented as the XSPEC
normalization but can be converted to an EM using a distance of
51.2~kpc to \sna{} \citep{panagia91, mitchell02}. The \texttt{vpshock}
model also allows for variable abundances (Section~\ref{sec:abu}). The
model assumes an adiabatic, one-dimensional plasma shock with constant
temperature propagating into a uniform CSM. This is clearly
oversimplified compared to the complex ER structure. At \xray{}
energies below ${\sim}$2~keV, the shocks producing the emission in
\sna{} are also likely radiative and not adiabatic \citep{pun02,
  groningsson08b}. A radiative shock produces softer emission compared
to an adiabatic shock at the same temperature, as well as different
line ratios \citep{nymark06}. These effects add additional systematic
uncertainties into the absorption and abundance estimates discussed
below. The \texttt{vpshock} model also assumes a Maxwellian electron
velocity distribution and no cosmic ray modifications. Accelerated
cosmic rays could diffuse upstream and form a shock precursor that
decelerates the gas ahead of the shock \citep{borkowski01}. This would
lead to a lower temperature than in the standard, non-modified shock
of the same velocity.

\subsection{Fixed Parameters}\label{sec:abu}
\begin{deluxetable}{lcccccccccccccc}
  \tablecaption{Fixed Parameters\label{tab:fix}}
  \tablewidth{0pt}
  \tablehead{Parameter & Value}
  \startdata
  $N_\mathrm{H}$                                   & $2.60_{-0.05}^{+0.05}\times{}10^{21}$\pcm{} \\
  $\sigma(1\mathrm{~keV})$\hphantom{0000000000000} & $1.1\pm0.1$~eV \\
  $z$                                              & $(9\pm{}5)\times{}10^{-4}$ \\
  $V_\mathrm{r}(z)$                                & $260\pm{}120$\kmps{} \\
  \enddata

  \tablecomments{Parameters determined from simultaneous fits to three
    XMM-Newton observations. They are kept fixed in all other
    fits. The recession velocity ($V_\mathrm{r}$) is inferred from the
    fitted redshift.}
  
\end{deluxetable}
There are a number of spectral parameters that we keep constant
throughout the time range spanned by our observations. These
parameters are the ISM absorption column density, line broadening,
elemental abundances, and redshift (Tables~\ref{tab:fix}
and~\ref{tab:abu}). We initially perform a fit to determine these
parameters. This is done to avoid having to fit all spectra
simultaneously with an excessive number of free parameters. All these
constant parameters are primarily constrained by the RGS spectra, and
to some extent the pn spectra. Thus, for this particular fit, we
simultaneously fit the model to three \xmm{} observations: 9423;
10,141; and 11,192~d (Table~\ref{tab:obs_xmm}). The choice of these
observations offer a trade-off between exposure time, covering the
NuSTAR epochs, and a short enough time range during which spectral
variations are moderate.

For this fit, we tie all parameters that are expected to be
constant. Each observation epoch consists of one pn spectrum and four
RGS spectra. Among these five spectra, the plasma temperature,
ionization age, and EM are tied across the instruments. The setup is
the same for each of the three plasma components. These plasma
parameters are not tied between observations since they evolve
significantly with time. The cross-normalization constant is frozen to
1.0 for the 1st order RGS1 spectra and left free for the other spectra
from the same observation. The global fit statistic is $\chi^2=5459$
for 4617 degrees of freedom (DoF), with relatively similar goodnesses
of fit for different spectra.

From the fit, we obtain a best-fit $N_\mathrm{H}$ of
$2.60_{-0.05}^{+0.05}\times{}10^{21}$\pcm{}. This is low compared to
estimates of the Galactic $N_\mathrm{H}$ from \hi{} surveys. For
example, \citet{willingale13} report $4.09\times{}10^{21}$\pcm{} and
\citet{hi4pi16} $3.86\times{}10^{21}$\pcm{} (assuming 20\,\% of H$_2$
for the latter), which are too high to result in statistically
acceptable fits to the \xmm{} (mainly RGS) data. Additionally, optical
extinction estimates show that the LMC contribution is greater than
the Milky Way contribution \citep{fitzpatrick90, france11}. This
implies that the \xray{} absorption to \sna{} should be even higher
than the Galactic \hi{} estimates.

There are two likely contributing factors to our apparent
underestimation of $N_\mathrm{H}$. First, the employed adiabatic model
(Section~\ref{sec:mod}) produces harder spectra than radiative shocks
\citep{nymark06}. This could artificially suppress $N_\mathrm{H}$ in
order to produce a softer absorbed spectrum. Second, we use the
Galactic abundances of \citet{wilms00} for the absorption component to
reduce complexity. In reality, the gas along the line of sight could
be of lower metallicity, especially the LMC absorption component. This
would also have the effect of lowering the fitted $N_\mathrm{H}$ to
compensate for an assumed metallicity that is too high. However, we
note that our \xray{} estimate of $N_\mathrm{H}$ is comparable to
those obtained by other \xray{} analyses. For example,
\citet{zhekov09} report $1.44_{-0.12}^{+0.16}\times{}10^{21}$\pcm{}
and \citep{park06} report
$2.35_{-0.08}^{+0.09}\times{}10^{21}$\pcm{}. These differences are
small in light of the different modeling techniques and instruments
used. In summary, we conclude that our best-fit $N_\mathrm{H}$ of
$2.60\times{}10^{21}$\pcm{} may be underestimated due to systematic
uncertainties, but we use it in all subsequent fits since it provides
the best fit quality and is comparable to previous \xray{} estimates.

The best-fit Doppler blur, measured as $\sigma$ of the Gaussian
smoothing kernel, is $1.1\pm0.1$~eV at 1~keV. For reference,
\citet{dewey12} find a value of 1.0~eV in their analysis of RGS data.

The best-fit redshift with statistical uncertainties obtained in XSPEC
is $8.81_{-0.01}^{+0.02}\times{}10^{-4}$. The total uncertainty is
dominated by the RGS absolute calibration of $\pm{}5$~m\AA{}
($4\times{}10^{-4}$~keV at 1~keV). Therefore, we adopt
$(9\pm{}5)\times{}10^{-4}$ as the redshift estimate from our \xray{}
data, which corresponds to a recession velocity of
$260\pm{}120$\kmps{}. This is consistent with an optical estimate of
286.74\kmps{} \citep{groningsson08b}. However, we note that these
values represent redshifts integrated over velocities along the line
of sight and the entire ER, which has spatial differences in the
relative brightness in \xray{} and optical. Therefore, it is likely
that the \xray{} and optical redshifts are truly slightly
different. For our purpose, we prioritize a good fit to the data and
choose to use the value we obtain from the \xray{} fit.

\begin{deluxetable}{lcccccccccccccc}
  \tablecaption{ER Abundances\label{tab:abu}}
  \tablewidth{0pt}
  \tablehead{\colhead{Element} & \colhead{Abundance\tablenotemark{a}} & \colhead{$A/A_\mathrm{ISM}$\tablenotemark{b}} & \colhead{$A/A_\mathrm{LMC}$\tablenotemark{c}} & \colhead{Ref.} \\
             \colhead{}        & \colhead{(dex)}                      & \colhead{}                                                            & \colhead{}                                                            & \colhead{}}
  \startdata
  H  &           $\equiv{}12$ &            $\equiv{}1$ &            $\equiv{}1$ & \nodata{} \\
  He &                $11.40$ &                 $2.56$ &                 $2.87$ & 1 \\
  C  &                 $7.51$ &                 $0.13$ &                 $0.29$ & 1 \\
  N  & $8.30_{-0.02}^{+0.03}$ & $2.65_{-0.12}^{+0.20}$ & $14.58_{-0.68}^{+1.08}$ & \nodata{} \\
  O  & $8.26_{-0.02}^{+0.01}$ & $0.37_{-0.01}^{+0.01}$ & $0.82_{-0.03}^{+0.03}$ & \nodata{} \\
  Ne & $7.82_{-0.01}^{+0.01}$ & $0.76_{-0.02}^{+0.02}$ & $1.63_{-0.03}^{+0.03}$ & \nodata{} \\
  Mg & $7.28_{-0.01}^{+0.01}$ & $0.76_{-0.02}^{+0.02}$ & $0.65_{-0.02}^{+0.01}$ & \nodata{} \\
  Si & $7.28_{-0.01}^{+0.01}$ & $1.03_{-0.02}^{+0.02}$ & $0.30_{-0.01}^{+0.01}$ & \nodata{} \\
  S  & $7.02_{-0.02}^{+0.02}$ & $0.85_{-0.04}^{+0.04}$ & $2.08_{-0.11}^{+0.10}$ & \nodata{} \\
  Ar &                 $6.23$ &                 $0.66$ &                 $0.87$ & 2 \\
  Ca &                 $5.89$ &                 $0.49$ &                 $1.00$ & 3 \\
  Fe & $7.02_{-0.004}^{+0.01}$ & $0.39_{-0.004}^{+0.01}$ & $0.62_{-0.01}^{+0.02}$ & \nodata{} \\
  Ni &                 $6.04$ &                 $0.98$ &                 $1.00$ & 3 \\
  \enddata
  \tablerefs{(1) Sect.~3.1 of \citet{lundqvist96}, (2) Table~7 of
    \citet{mattila10}; (3) Table~1 of \citet{russell92}.}

  \tablecomments{Values with error bars were determined from our fits,
    whereas the others are taken from the provided references.}

  \tablenotetext{a}{Expressed in terms of the astronomical log scale
    $12 + \log_{10}[A(\mathrm{X})/A(\mathrm{H})]$, where
    $A(\mathrm{X})$ is abundance of element X.}

  \tablenotetext{b}{$A_\mathrm{ISM}$ is the Galactic ISM abundances
    of \citet{wilms00}.}
  
  \tablenotetext{c}{$A_\mathrm{LMC}$ is the LMC abundances of
    \citet{russell92}.}

\end{deluxetable}
All fitted abundances are provided in Table~\ref{tab:abu}. It also
includes abundances of He, C, Ar, Ca, and Ni, which are taken from the
literature. These abundances cannot be fitted for because they are
very weakly constrained by the \xray{} data. Overall, the fitted
abundances are largely within the ranges of values reported by
previous studies \citep{lundqvist96, mattila10, sturm10, dewey12,
  bray20}. For example, our estimated Fe abundance of
$0.62_{-0.01}^{+0.02}$ relative to the LMC abundance can be compared
to other \xray{} estimates (all expressed relative to the LMC
abundance): $0.55_{-0.03}^{+0.03}$ \citep{zhekov09},
$0.36_{-0.01}^{+0.01}$ \citep{sturm10}, $0.62_{-0.01}^{+0.02}$
\citep{dewey12}, and $0.44_{-0.03}^{+0.03}$ \citep{bray20}. However,
we caution that all \xray{} estimates of the abundances likely suffer
from similar significant systematic uncertainties due to the modeling
and absorption uncertainties described above.

We leave the abundances of all trace elements to their default values,
namely 1.0 relative to our adopted ISM abundances. All fits are
insensitive to these trace elements. We test this by setting the
abundances of all trace elements to 0.0 using the command
\texttt{NEI\_TRACE\_ABUND}. The $\chi^{2}$ value fluctuates by a few
and all fitted parameters are practically unchanged.

\subsection{Flux Estimation}\label{sec:flx}
To estimate fluxes, we use the model and parameters described
above. The model is fitted simultaneously to all spectra within the
groups of \nustar{} observations and within each \xmm{}
observation. The free parameters are the temperatures, ionization
ages, and EMs for each of the shock components, and a free
cross-normalization between the instruments. The \nustar{} data are
unable to robustly constrain the coolest plasma component. Therefore,
we freeze its parameters to the values from the fits to the \xmm{}
data that are closest in time. This component is almost completely
below the lower energy limit of \nustar{} and does not affect the
fluxes significantly. The parameters of each plasma component that are
fitted for are tied across instruments since the spectra are
(quasi-)simultaneous. The average reduced fit statistic is
$\chi^2_\mathrm{red}=0.93$ with an average of 347 DoF for the
\nustar{} epochs. The corresponding numbers for the \xmm{}
observations are $\chi^2_\mathrm{red}=1.2$ and 1555 DoF.

We measure fluxes using the XSPEC component \texttt{cflux}, which has
a parameter that corresponds to the flux. After fitting the model, the
cross-calibration constant is replaced by \texttt{cflux} and the model
is refitted. This results in a flux estimate from each spectrum, which
also captures the calibration differences between the instruments. We
report the weighted average as the best-estimate flux, but the
individual flux measurements provide a handle on the level of
systematics (Appendix~\ref{app:sys}). All reported fluxes are observed
(not correcting for ISM absorption) due to the uncertain amount of
absorption and to allow for comparisons with previous studies.

\subsection{Continuous Temperature Model Setup}\label{sec:dos}
In addition to the standard model, we use a more complex model for the
joint NuSTAR and RGS spectral analysis. It is the same as the base
model except that the three shock components are replaced by a
distribution of shocks over a continuous temperature interval
\citep{zhekov06, zhekov09}, analogous to the XSPEC model
\texttt{c6pvmkl} \citep{lemen89, singh96}. The distribution of shock
EMs is given by
\begin{equation}
  \label{eq:phi}
  \phi{} = \delta_1 \exp\!\left(\sum_{i=0}^{6} a_i P_i\!\left(T^\prime\right)\right),
\end{equation}
where $\delta_1$ is a normalization, $a_i$ are the fitting
coefficients, and $P_i$ is the Chebyshev polynomial of the first kind
of order $i$. The polynomial argument $T^\prime{}$ is a rescaling of
the plasma temperature $T$ such the temperature interval $0.125$ to
$10$~keV is mapped logarithmically to the domain $-1$ to $1$. The
seemingly arbitrary parametrization of $\phi$ reduces the number of
free parameters and makes the fitting better conditioned. Numerically,
the continuous temperature is implemented as 28 \texttt{vpshock}
components with $T$ logarithmically spaced between 0.125 and
10~keV. To further reduce the number of free parameters, we
parametrize the ionization age of each shock component as
\begin{equation}
  \label{eq:tau}
  \tau = \delta_2 T_\mathrm{keV}^p,
\end{equation}
where $T_\mathrm{keV}$ is the temperature in keV, and both the
normalization $\delta_2$ and power-law (PL) index $p$ are fitted for. To
summarize, this means that we fit a shock with a continuous
distribution of $T$ using ten free parameters: two normalizations,
seven polynomial coefficients, and the ionization PL index.

The primary advantage of this continuous model is the ability to
capture the complex underlying physics. It has only one more free
parameter than the three-shock standard model. The tradeoff is
increased freedom in temperature at the cost of a more constrained
ionization age. Importantly, the continuous model allows for the
possibility to separate different components by analyzing the EM
distribution. Statistically, typical improvements compared to the
standard three-shock model is $\Delta \chi^2=-20$ for an average
number of DoF of
${\sim}$1800. However, the number of shock components in ordinary
models is arbitrary and harder to interpret physically. A priori, it
is not clear how each discrete model component translates to physical
components.

\section{Results}\label{sec:res}
\subsection{Light Curves}\label{sec:lc}
\begin{deluxetable*}{ccccccccccccccc}
  \tablecaption{\nustar{} Fluxes\label{tab:nus_flx}}
  \tablewidth{0pt}
  \tablehead{\colhead{Group} & \colhead{Epoch} &                     \colhead{$F_\text{3--8}$} & \colhead{$F_\text{10--24}$}                   & \colhead{$F_\text{3--24}$} \\
             \colhead{}      & \colhead{(d)}   & \colhead{($10^{-13}$~erg~s$^{-1}$~cm$^{-2}$)} & \colhead{($10^{-13}$~erg~s$^{-1}$~cm$^{-2}$)} & \colhead{($10^{-13}$~erg~s$^{-1}$~cm$^{-2}$)}}
  \startdata
  1 &   9331 & $10.5_{-0.2}^{+0.2}\pm0.8$ & $1.78_{-0.07}^{+0.07}\pm0.14$ & $13.3_{-0.2}^{+0.2}\pm1.1$ \\
  2 &   9372 & $11.3_{-0.2}^{+0.3}\pm0.9$ & $1.67_{-0.10}^{+0.09}\pm0.13$ & $14.0_{-0.3}^{+0.3}\pm1.1$ \\
  3 &   9424 & $11.0_{-0.3}^{+0.3}\pm0.9$ & $1.67_{-0.11}^{+0.12}\pm0.13$ & $13.7_{-0.3}^{+0.3}\pm1.1$ \\
  4 &   9623 & $11.2_{-0.2}^{+0.2}\pm0.9$ & $1.73_{-0.08}^{+0.08}\pm0.14$ & $14.0_{-0.2}^{+0.2}\pm1.1$ \\
  5 &   9920 & $12.6_{-0.2}^{+0.2}\pm1.0$ & $1.82_{-0.09}^{+0.07}\pm0.15$ & $15.6_{-0.2}^{+0.2}\pm1.2$ \\
  6 &   9976 & $11.7_{-0.2}^{+0.2}\pm0.9$ & $1.90_{-0.08}^{+0.09}\pm0.15$ & $14.7_{-0.3}^{+0.3}\pm1.2$ \\
  7 & 10,021 & $12.1_{-0.2}^{+0.2}\pm1.0$ & $1.69_{-0.08}^{+0.08}\pm0.13$ & $15.0_{-0.2}^{+0.2}\pm1.2$ \\
  8 & 12,140 & $16.2_{-0.3}^{+0.3}\pm1.3$ & $1.92_{-0.11}^{+0.08}\pm0.15$ & $19.6_{-0.3}^{+0.3}\pm1.6$ \\
  \enddata

  \tablecomments{The flux $F_{a\text{--}b}$ denotes the flux from $a$
    to $b$~keV. Asymmetric error bars are statistical and the
    symmetric uncertainties are systematic (Appendix~\ref{app:sys}).}
\end{deluxetable*}

We report observed fluxes in a number of different energy bands. For
\nustar{}, we study the 3--8, 10--24, and 3--24~keV ranges. The
3--8~keV is the ``hard'' band in \xmm{} and \chandra{}
contexts. \xmm{} fluxes are provided in the 0.5--2, 3--8, and
0.5--8~keV bands, in line with previous studies (e.g.\
\citealt{frank16}). Additionally, we also compute the 0.45--0.7~keV
flux, which is dominated by N\textsc{\,vii} and O\textsc{\,viii}
Ly$\alpha$. All \nustar{} fluxes are provided in
Table~\ref{tab:nus_flx}, whereas the \xmm{} fluxes are provided in
Appendix~\ref{app:xmm_flx}.

\begin{figure*}
  \centering
  \includegraphics[width=\textwidth]{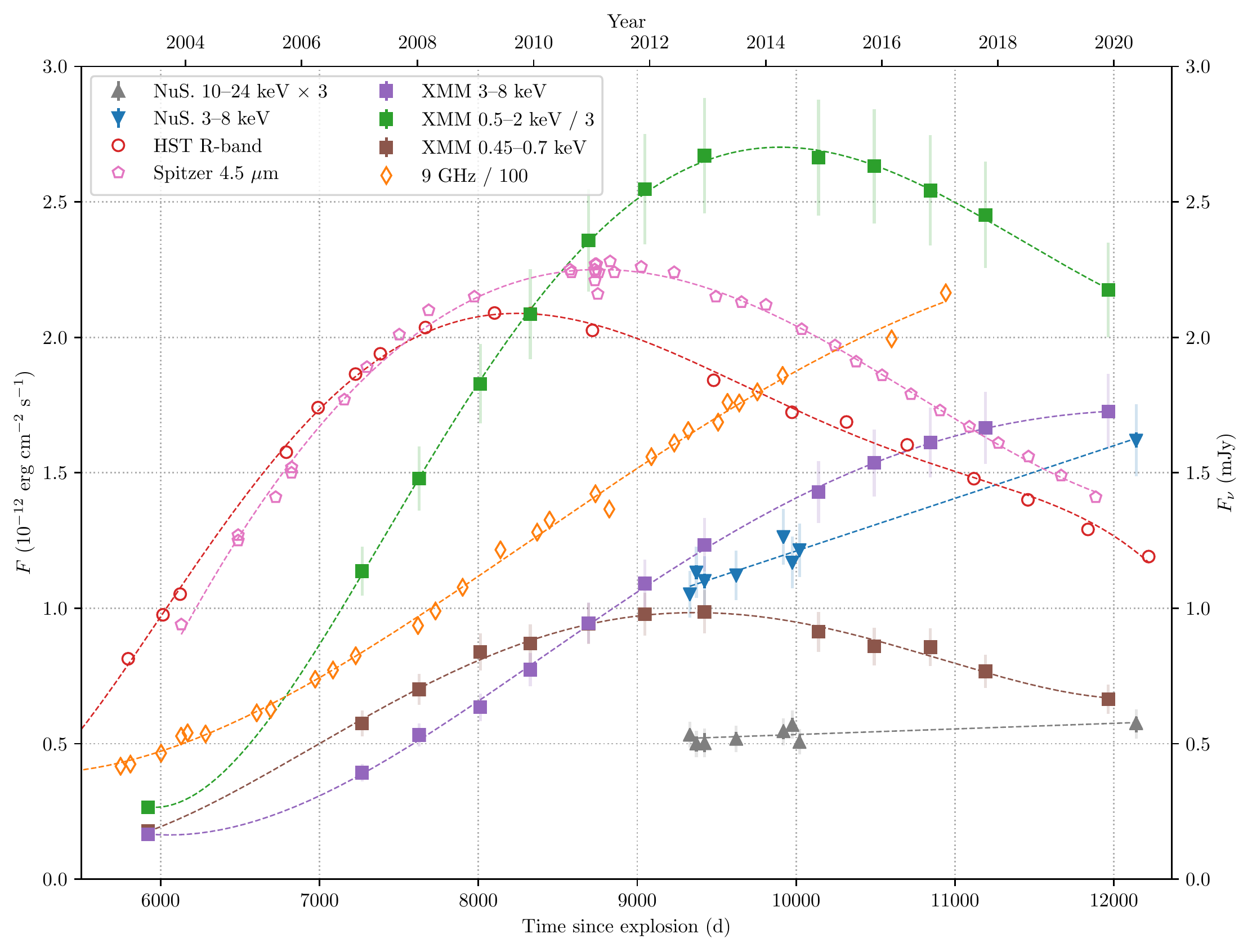}

  \caption{\nustar{} (triangles), \xmm{} (squares), HST
    R\nobreakdash-band (circles; \citealt{larsson19b}), Spitzer
    $4.5$~\textmu{}m (pentagons; \citealt{arendt20}), and 9~GHz radio
    light curves (diamonds; \citealt{cendes18}). Solid symbols
    represent \xray{} data and hollow symbols show other wavelengths.
    The two $y$-axes are not equivalent. The right axis
    applies to the Spitzer and radio data, whereas the left applies to the
    other data. We note that the 0.5--2~keV, 10--24~keV, and radio
    data have been rescaled for clarity as shown in the legend. The
    dashed lines are polynomial fits to guide the eye. The fits to
    NuSTAR data are of first order, while the others are fourth or
    fifth order polynomials. The solid \xray{} error bars are
    statistical (often smaller than the markers) and the
    semi-transparent error bars approximately represent the systematic
    uncertainties (Appendix~\ref{app:sys}).\label{fig:lc}}
\end{figure*}

\begin{figure*}
  \centering
  \includegraphics[width=\textwidth]{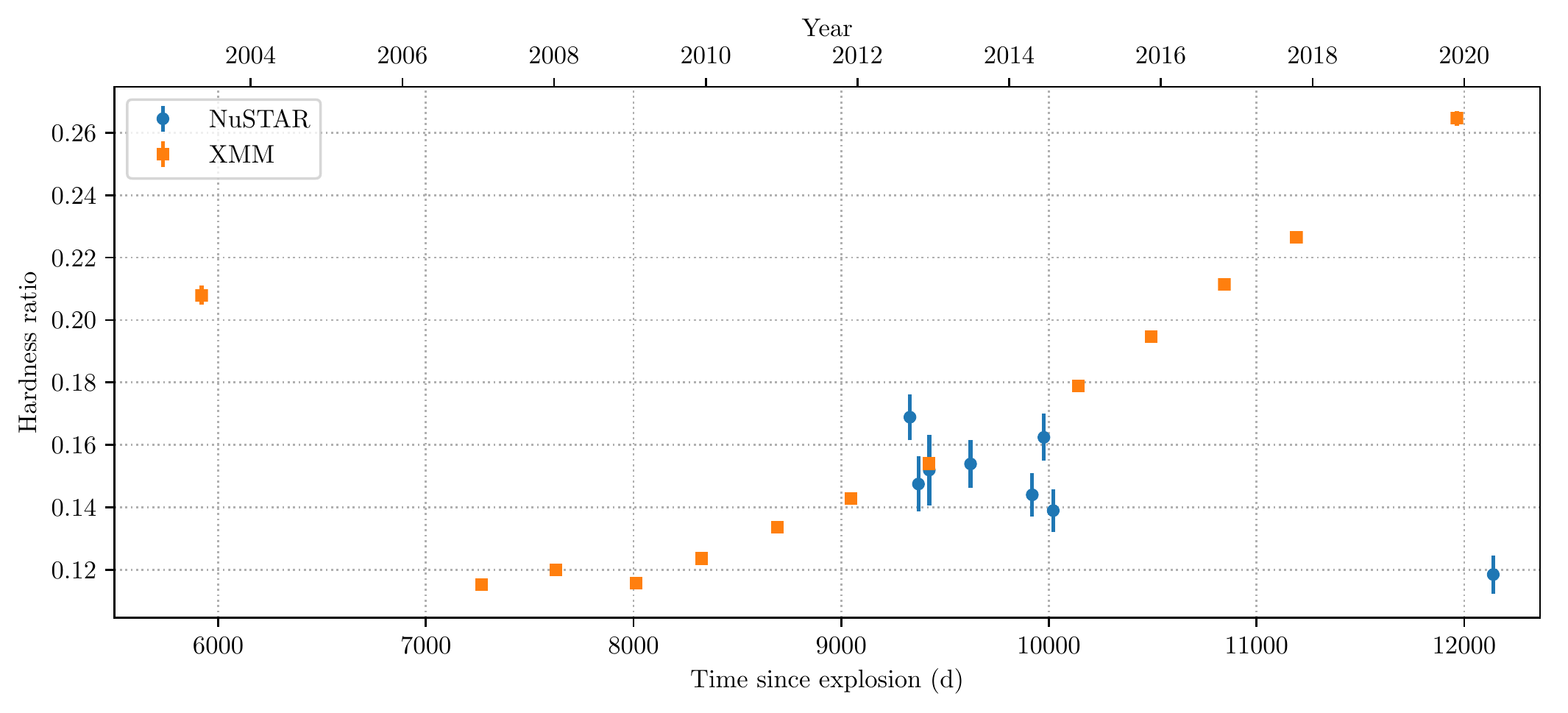}

  \caption{Hardness ratios, defined as the ratio in a hard over a soft
    band. For \nustar{}, the hard band is 10--24~keV and soft
    3--8~keV. The soft \xmm{} range is 0.5--2~keV, whereas the hard is
    3--8~keV. These error bars only include the statistical
    uncertainty since ratios are computed using homogeneous
    instrumental setups in each band.\label{fig:hr}}
\end{figure*}

The light curves are shown in Figure~\ref{fig:lc}. In addition to
\nustar{} and \xmm{}, we show HST R\nobreakdash-band (WFPC2/F675W,
ACS/F625W, WFC3/F625W; \citealt{larsson19b}), $4.5$~\textmu{}m Spitzer
\citep{arendt20}, and 9~GHz radio \citep{cendes18} light curves. The
two most recent HST data points are previously unpublished, but are
computed using the same methods as the other data points and include
only the ER emission. The observations were obtained on 2019 July 22
(11,837~d) and 2020 August 6 (12,218~d).

It is apparent that the 3--8~keV light curves and the 9~GHz radio show
similar temporal evolutions. In contrast, the 10--24~keV component
increases at a lower relative rate. We note that the 0.5--2~keV flux
clearly starts decreasing after around 10,000~d, and that the rise of
the 3--8~keV flux slows slightly after 11,000~d. This is also reported
by \citet{sun21} using \xmm{} data, as well as in a preliminary study
of recent Chandra observations \citep{ravi19c}. These spectral changes
are captured by the hardness ratios shown in Figure~\ref{fig:hr}. This
illustrates how the 3--8~keV flux becomes increasingly dominant, with
the spectra hardening in the \xmm{} energy range and softening in the
\nustar{} range. Finally, we note that the 0.45--0.7~keV evolution is
more strongly correlated with the decaying optical flux than the
0.5--2~keV light curve.

The offset of ${\sim}$10--15\,\% between the \nustar{} and \xmm{}
3--8~keV light curves reveal a systematic difference
(Appendix~\ref{app:sys}). However, for our conclusions, we focus on
the evolution as observed by the same instrument, which should be free
from large systematics.

There are indications of variability of approximately 5\,\% between
the \nustar{} epochs on timescales of a few weeks (among groups 1--3
and 5--7). We caution that these variations might be due to
observational uncertainties, instrumental origin, or the data
reduction process.

\subsection{Standard Shock Model}\label{sec:sa}
\begin{figure*}
  \centering
  \includegraphics[width=\textwidth]{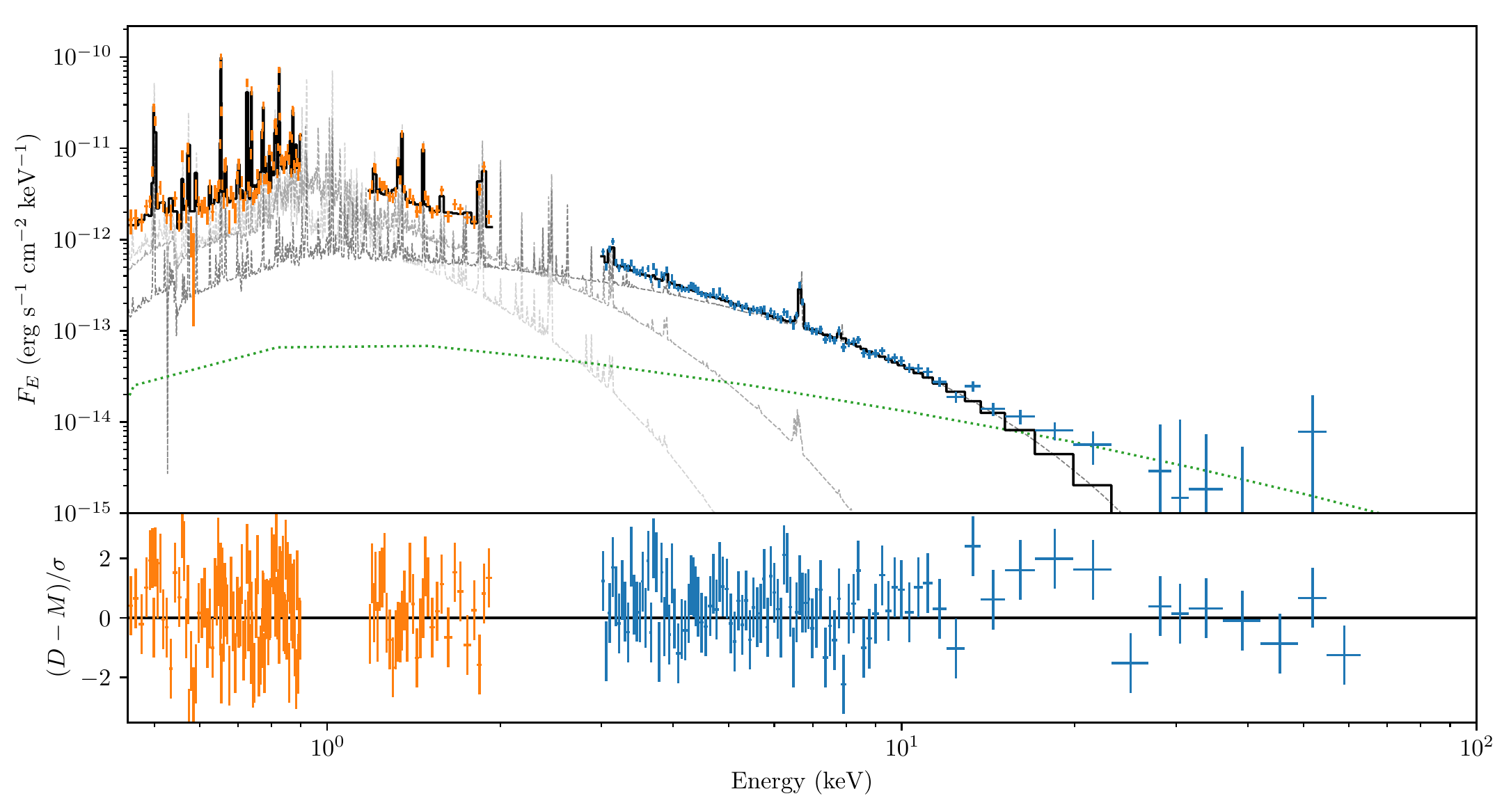}
  
  \caption{Upper panel: Unfolded observed spectra (absorbed but
    corrected for the instrument responses) from \nustar{} (blue), RGS
    (orange), and the model (black) in flux space. The \nustar{} data
    are plotted up to 65~keV (just below the \ti{} lines), but the
    fits are performed using only data up to 24~keV. For visual
    clarity, we show the sum of the six individual \nustar{} exposures
    from the first epoch at 9331~d and only the 1st order RGS1 data
    from 9423~d. Both the \nustar{} and RGS data are also rebinned
    (represented by the horizontal error bars). The vertical error
    bars are 1$\sigma{}$ intervals. The thin dashed lines are the low-
    (light gray), mid- (gray), and high-temperature (dark gray) shock
    components that constitute the model. The dotted green line is the
    non-thermal synchrotron prediction from \citet{berezhko15}. It is
    not part of the fitted model, but is overplotted for
    reference. The synchrotron component is shown for 9280~d, close to
    the predicted time of the synchrotron maximum. Although the green
    synchrotron prediction appears to match two bins around 20~keV, an
    additional component is not statistically motivated
    (Section~\ref{sec:lim}) and the synchrotron prediction is only
    marginally consistent with the data above 24~keV
    (Section~\ref{sec:cr}). Lower panel: Normalized residuals between
    data and model. This difference is computed in count space,
    implying that the model is forward-folded and then compared to the
    data. The non-thermal synchrotron component (dotted green line) is
    not part of the model used for computing the
    residuals.\label{fig:con}}
  
\end{figure*}

\begin{figure}
  \centering
  \includegraphics[width=\columnwidth]{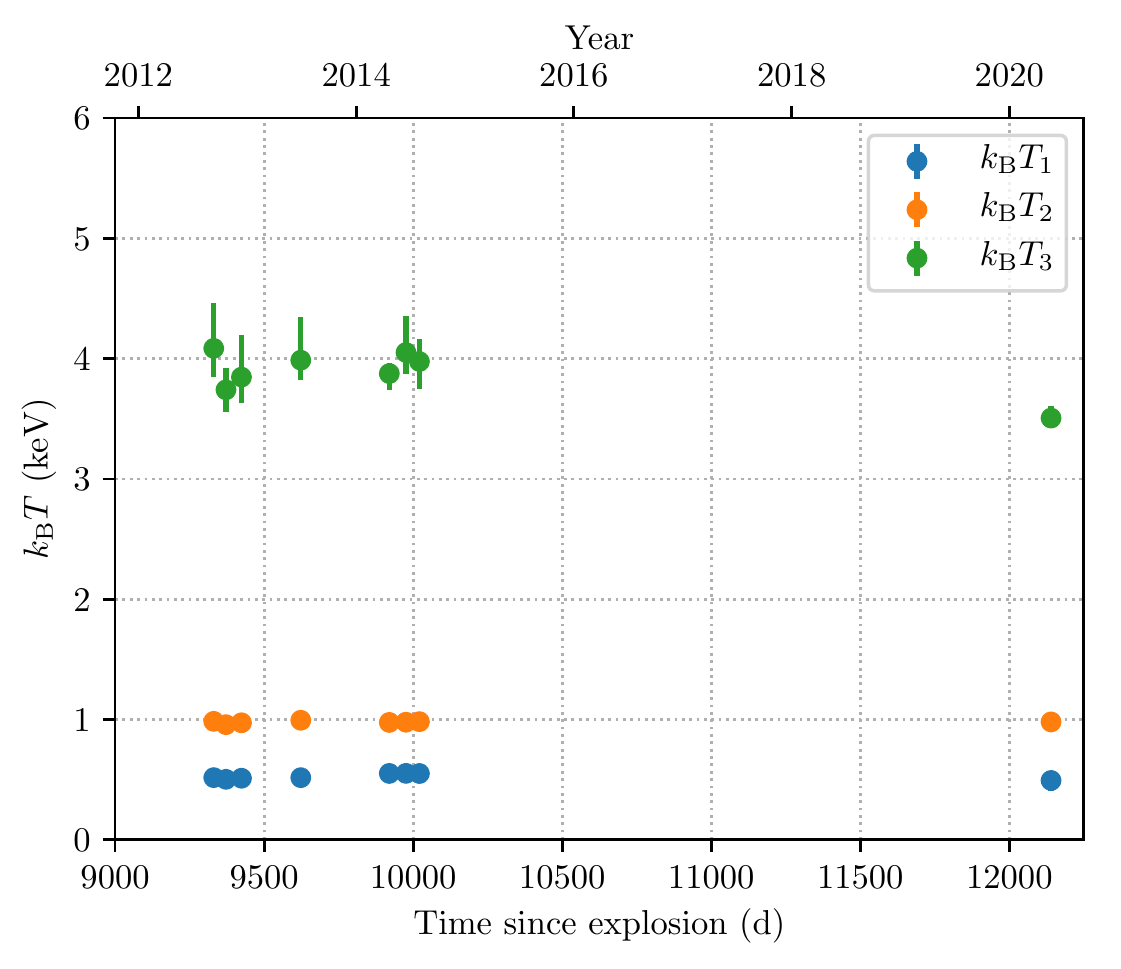}
  
  \caption{Evolution of the best-fit temperatures of the standard
    three-shock model. The subscripts denote the different shock
    components: low (1), mid (2), and high (3) temperature. The error
    bars for the low- and mid-temperature components are small
    compared to the markers (Tables~\ref{tab:par}
    and~\ref{tab:com}).\label{fig:com}}

\end{figure}

We fit the spectral model with three shock components to all \nustar{}
and RGS data sets. Figure~\ref{fig:con} shows the best-fit model, its
three components, and data from the first \nustar{} epoch
(9331~d). The standard model provides an adequate fit and no
additional component is statistically necessary
(Section~\ref{sec:lim}).

We verify that the 0.45--24~keV spectrum cannot be adequately modeled
by a two-shock component model. This worsens the average goodness of
fit to the combined \nustar{} and RGS data by
$\Delta\chi^2\approx{}320$ for ${\sim}$1800 DoF. We provide a much
more extensive investigation of the effects of different set-ups for
the spectral analysis in Appendix~\ref{app:pn}.

Evolutions of the temperatures of the shock components are shown in
Figure~\ref{fig:com}. Typical temperatures for the three components
are 0.3, 0.9, and 4~keV. The only variation in best-fit temperature is
a slight decrease of the hottest component at the last \nustar{} epoch
(12,140~d).

For completeness, we provide all best-fit parameters, fluxes of shock
components, and fit statistics in Appendix~\ref{app:fit}. We caution
that the ionization ages for the three components may be
unreliable. This is most evident for the mid-temperature
  component, which has a pegged ionization age at all epochs
  (Table~\ref{tab:par}). In particular, there appears to be too much
freedom in the fits when the ionization ages of the three components
are completely free. It is also probable that the fitted ionization
ages are sensitive to underlying model uncertainties
(Section~\ref{sec:mod}). The fitted parametrization (Eq.~\ref{eq:tau})
of the ionization age used in the continuous shock model is likely
more robust (Section~\ref{sec:ssm}).

\subsection{The Fe K$\alpha$ Line}
\begin{figure}
  \centering
  \includegraphics[width=\columnwidth]{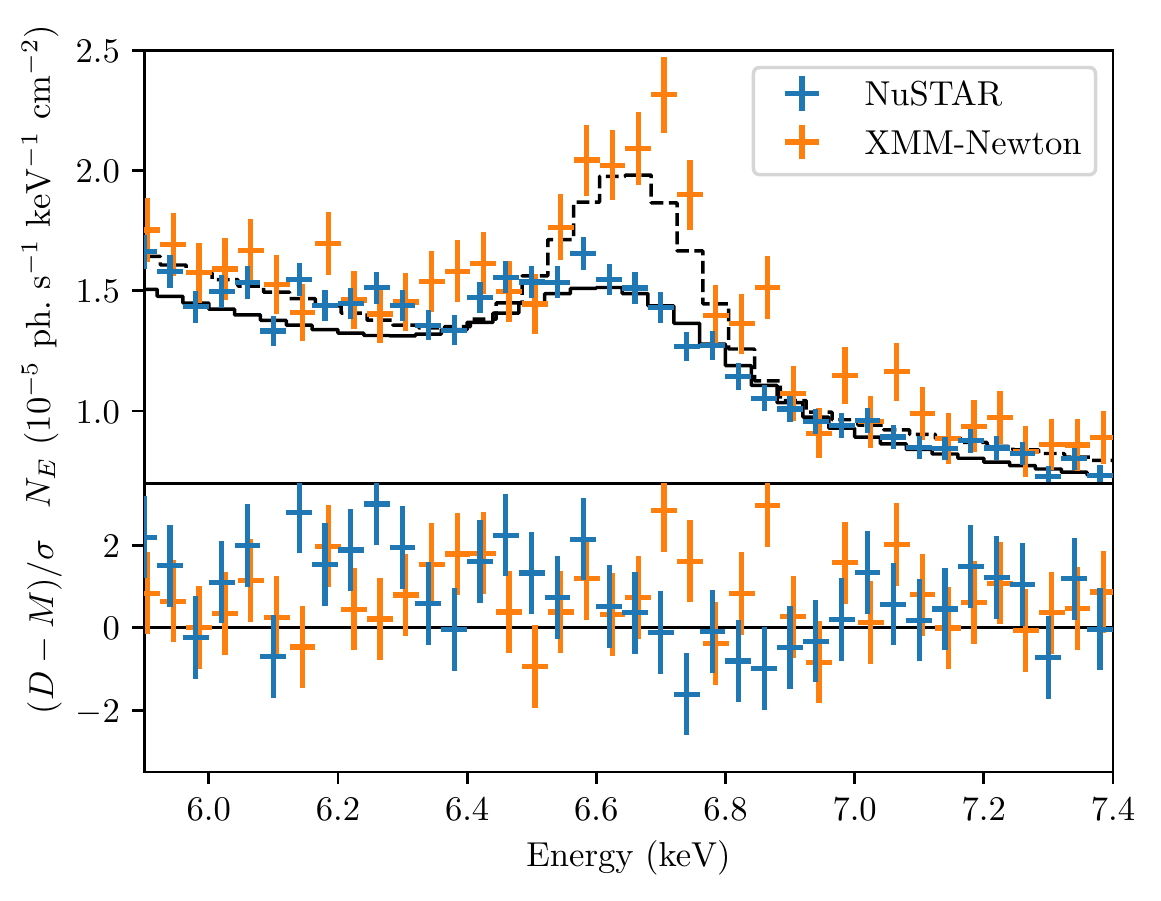}
  
  \caption{Spectral fits around the Fe~K complex in count space. The
    panels are structured similarly to Figure~\ref{fig:con}. For
    visual clarity, the data shown are the sum of all \nustar{} data
    (blue crosses) and the sum of the eight \xmm{}/pn observations
    after 8500~d. The summed data are not suited for detailed spectral
    fitting. Instead, we overplot the best-fit to the \nustar{} data
    at 9920~d, which is close to the average best-fit. The solid
    (\nustar{}) and dashed (\xmm{}) black lines are the same
    underlying model, but produce different fluxes because of the
    different instrumental responses. We stress that the large amount
    of data (${\sim}$500 photons per bin in \nustar{}) makes even
    small flux and energy differences appear highly significant. The
    bin size is 0.04~keV for both \nustar{} and \xmm{}. For reference,
    the absolute energy calibration around the Fe~K complex is
    0.04~keV for \nustar{} \citep{madsen15} and 0.01~keV for
    \xmm{}/pn.\label{fig:fek}}

\end{figure}

Due to the importance of the Fe~K lines, we show the integrated Fe~K
complex in Figure~\ref{fig:fek}. The Fe~K blend is clearly detected
and implies that at least a significant fraction of the emission at
high energies is thermal (see further Section~\ref{sec:nt}). The
observed line centroid in the \nustar{} data is ${\sim}$6.59~keV and
${\sim}$6.66~keV in the \xmm{} data. We note that \citet{sun21}
performed a more detailed, time-resolved analysis of the Fe line. They
find that the lines centroid increases from around 6.60~keV at 8500~d
to 6.67~keV at 12,000~d.

A line energy of 6.66~keV approximately corresponds to
Fe\textsc{\,xxiii}, which is consistent with a thermal origin of
temperature $>1$~keV \citep{makishima86, kallman04}. The
identification as Fe\textsc{\,xxiii} is only indicative and the
spectrum clearly has contributions from a range of ionization
levels. For reference, \citet{maggi12} reported an emission-line
centroid of $6.60\pm0.01$~keV and a width of ${\sim}$100~eV. From
this, they infer the presence of ionization stages from
Fe\textsc{\,xvii} to Fe\textsc{\,xxiv}, which is in agreement with our
result.

Our model spectrum is completely dominated by the 4~keV shock
component at these energies (Figure~\ref{fig:con}) and it captures the
line profile well. Previous, more careful \xmm{} analyses of the Fe
line have reported that there is a significant contribution of lower
ionization levels around 6.5~keV, which is not captured by a thermal
shock model \citep{sturm10, maggi12}. \citet{maggi12} suggested that
this could originate from fluorescence from near-neutral Fe, including
Fe in the unshocked ejecta. We do not investigate this in detail, but
Figure~\ref{fig:fek} shows that any low-ionization contribution is
clearly much weaker than the dominant component from the thermal shock
model.

\subsection{Continuous Shock Model}\label{sec:ssm}
\begin{figure}
  \centering
  \includegraphics[width=\columnwidth]{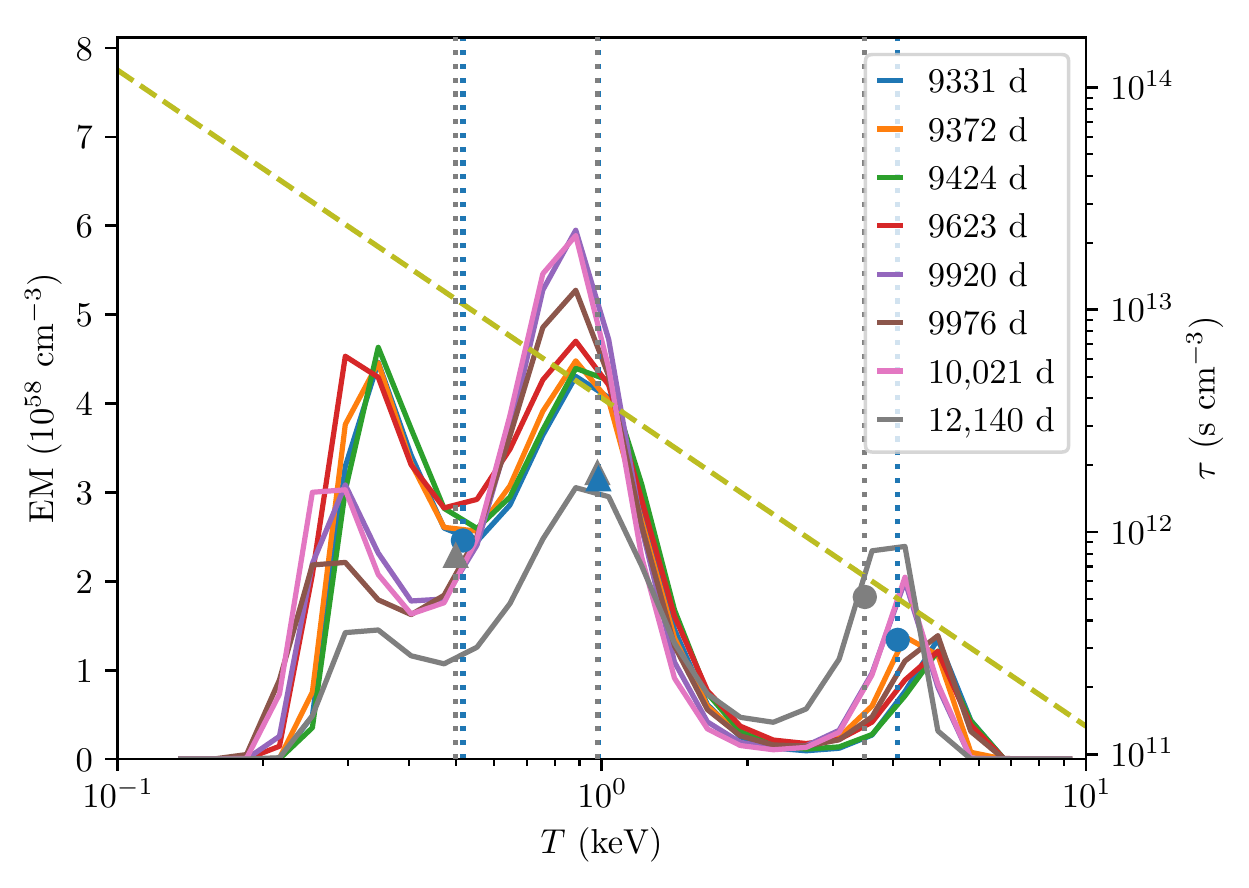}

  \caption{Results of the spectral analysis. Left axis: Solid lines
    are the fitted EM distributions (Eq.~\ref{eq:phi}) for the
    continuous shock temperature model. Right axis: The dashed olive
    line is the parametrized ionization age (Eq.~\ref{eq:tau}) from
    the continuous temperature fits. It is shown for the average
    $\delta_2$ and $p$ for visual clarity, but the $\tau$
    parametrizations for the different \nustar{} epochs are
    similar. The vertical dotted lines are the temperatures for the
    first (blue; September 2012) and last (gray; May 2020) \nustar{}
    epochs that are obtained by fitting the standard model with three
    shock components. We note that the mid-temperature lines close to
    1~keV are almost overlapping. The points are the fitted $\tau$ for
    the respective plasma components and the triangles are lower
    limits in cases where $\tau$ is unconstrained. All parameter
    values from the fits using three shock components are provided in
    Table~\ref{tab:par}.\label{fig:dos}}
  
\end{figure}
In this section, we report the results of the continuous temperature
shock model (Section~\ref{sec:dos}). The EM distributions are shown in
Figure~\ref{fig:dos}. The distributions at all epochs are
qualitatively similar, with three peaks at shock temperatures around
0.3, 0.9, and 4~keV. The isolated peak around 4~keV appears to be
robustly separated from the lower-temperature components. The reduced
$\chi^2$ for all fits are in the range 1.05--1.15, with an average
number of DoF of 1685.

We caution that the apparent bimodality separating the 0.3 and 0.9~keV
peaks is only marginally significant. We draw this conclusion because
some fits using different initial conditions find solutions with a
single, broad hump ranging from 0.3--1~keV. These solutions are only
marginally statistically worse. Regardless of the detailed structure
of the distribution below $2$~keV, it is still clear that there is a
contribution from a broad range of temperatures in the range
0.3--0.9~keV. We reiterate that the shock model relies on a number of
assumptions, which introduce additional uncertainties
(Section~\ref{sec:mod}). It is not computationally feasible to perform
a formal error analysis for the EM distribution. Consequently, we note
that some of the EM distribution variability between epochs could be
insignificant. For analyzing the temporal evolution, the light curves
(Section~\ref{sec:lc}) and three-shock model fits
(Section~\ref{sec:sa}) are more robust.

The parametrized ionization age (Eq.~\ref{eq:tau}) is a decreasing
function of temperature at all epochs (Figure~\ref{fig:dos}). The
best-fit parameters are
$\delta_2 = 4_{-2}^{+16}\times{}10^{12}$~s~cm$^{-3}$ and
$p=(-1.5)_{-1.3}^{+0.7}$. The ionization age is formally defined as
the product of the remnant age and the density. For a remnant age of
10,000~d, the parameters above yield a density of
$9[0.5$--$50]\times{}10^{4}$~cm$^{-3}$ at 0.3~keV,
$6[2$--$20]\times{}10^{3}$~cm$^{-3}$ at 0.9~keV, and
$4[2$--$6]\times{}10^{2}$~cm$^{-3}$ at 4~keV. The values quoted above
are averages across the eight \nustar{} epochs and the intervals in
brackets show the minimum and maximum values. As noted in
Section~\ref{sec:sa}, the free ionization ages for the standard
three-shock model are poorly constrained by the fits. The fit of the
parametrized ionization age for the continuous model discussed here is
likely more robust, but we caution that systematics certainly are
present at some level. In particular for shock temperatures as low as
0.3~keV, since these shocks are likely not adiabatic
(Section~\ref{sec:mod}).

Figure~\ref{fig:dos} also shows the best-fit temperatures for fits
using the standard three-shock-component model. Only temperatures for
the first and last epochs are shown, but similar temperatures are
obtained for all data sets (Figure~\ref{fig:com}). This indicates that
the simpler three-component model captures the same three peaks in the
EM distribution reasonably well. The slight shift of the coolest
component to ${\sim}$0.5~keV, in contrast to ${\sim}$0.3~keV for the
continuous model, could possibly be explained by a bias due to the
absorption. The absorption drastically reduces the number of photons
below ${\sim}$0.7~keV, which implies that it could be statistically
favorable to fit the coolest component to slightly higher temperatures
when the model is restricted to only three shock components.

\subsection{Constraints on Non-thermal Emission}\label{sec:lim}
\begin{figure*}
  \centering
  \includegraphics[width=\textwidth]{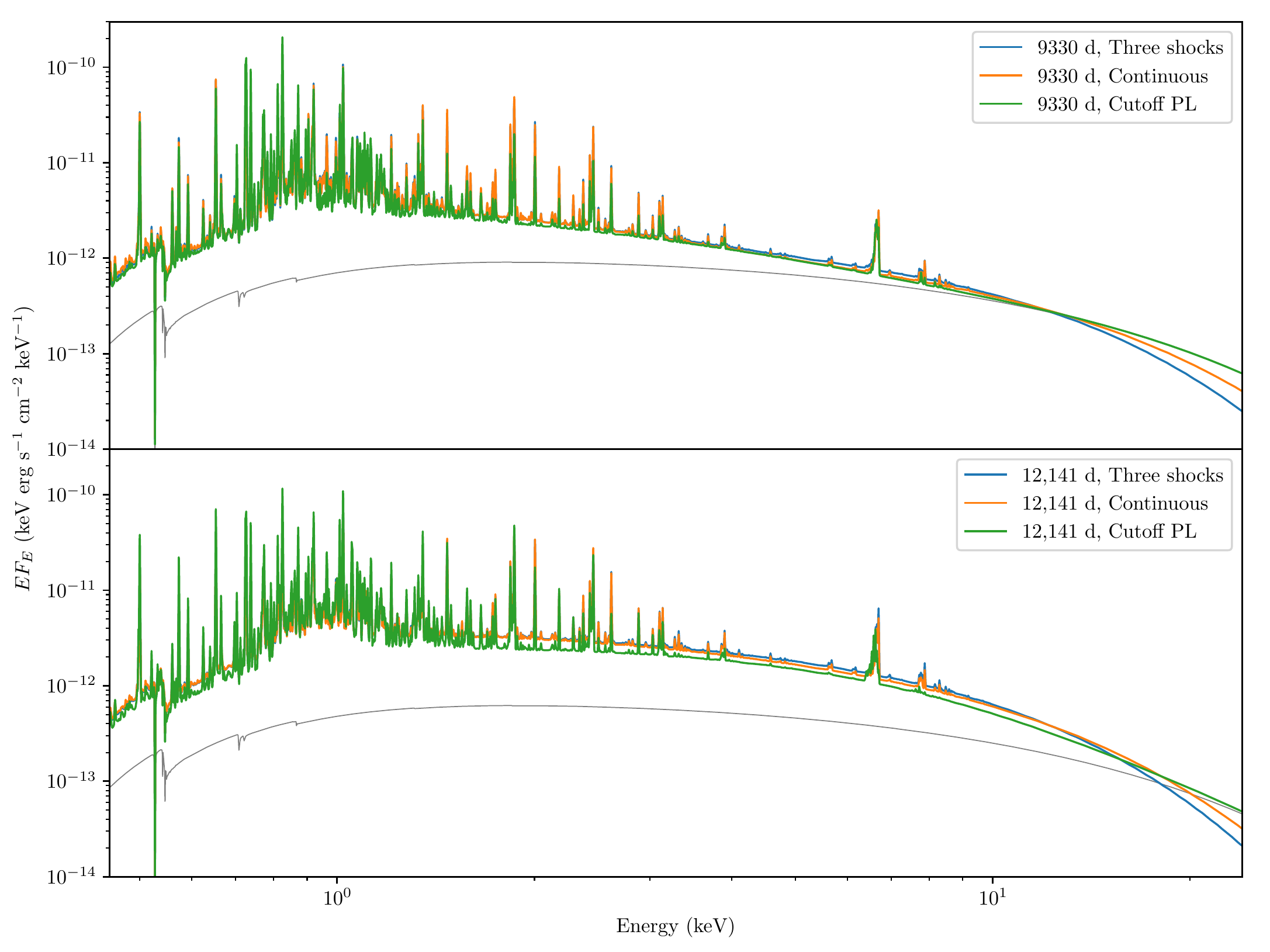}
  
  \caption{Three different models fitted to the first (upper panel)
    and last (lower panel) \nustar{} epochs. The models are the
    standard three thermal shocks model (blue), the continuous
    temperature model (orange), and two thermal shocks with an
    additional cutoff PL (green). We also show the cutoff PL component
    separately (thin gray lines). The blue line in the upper panel
    (largely covered by the other lines) shows the same model as in
    Figure~\ref{fig:con}, but we note that the $y$-axes are different
    to better follow conventions.\label{fig:mod}}

\end{figure*}

We find no indications of a non-thermal contribution to the \xray{}
spectrum. The thermal models describe the data well and neither
replacing the hottest plasma component with a cutoff PL, nor adding an
additional PL component, results in a better model. Below, we
motivate this conclusion in more detail.

First, we use the standard three-shock model, but replace the hottest
shock component with a cutoff PL. This phenomenological cutoff PL can
be interpreted as a hot 10--20~keV free-free component, synchrotron
emission, or potentially represent a PWN. The photon index of the
cutoff PL is frozen to $\Gamma=2$ to prevent degeneracies with the
shock components. This value for $\Gamma$ is similar to that of the
Crab Nebula \citep{buhler14}. We verify that choosing $\Gamma=1.5$ or
$2.5$ does not significantly affect the conclusions.

Freezing the photon index effectively forces the cutoff PL to dominate
at higher energies. The cutoff energies and normalizations are left
free and fitted to all \nustar{} epochs. This fit focuses on higher
energies and, therefore, we omit the RGS data and instead freeze the
coolest shock component. All non-thermal components are harder than
the average shape of the thermal spectra. This implies that any
non-thermal contribution is increasingly subdominant toward lower
energies (e.g.\ Figures~\ref{fig:con} and~\ref{fig:mod}). Thus, the
RGS data will not affect the hard non-thermal components.

Importantly, the Fe abundance needs to be free in these fits, but
still constant across different epochs. This is necessary because the
Fe abundance was fitted using the three-shock model
(Section~\ref{sec:abu}), where the hottest component captured the Fe~K
line complex. With the hottest component replaced, the Fe abundance
will naturally need to increase. This is because the cutoff PL
dominates at those energies and the line strength is proportional to
the abundance.

We show a comparison of the three-shock model, continuous temperature
model, and two-shock plus cutoff PL model in
Figure~\ref{fig:mod}. From the fits of the cutoff PL model, we obtain
an average fitted cutoff energy of 8~keV and an Fe abundance of
$2.0_{-0.5}^{+1.0}$ relative to the LMC Fe abundance. For reference,
the corresponding Fe abundances are $1.8_{-0.6}^{+1.1}$ for
$\Gamma=1.5$ and $2.5_{-0.8}^{+2.0}$ for $\Gamma=2.5$. Statistically,
the goodness with the $\Gamma=2$ cutoff PL is $\chi^2=2594$ for 2778
DoF, compared to $\chi^2=2597$ for 2771 DoF for the three-shock
model. However, we consider the Fe abundance to be unreasonably high
(further discussed in Section~\ref{sec:nt}) and a strong indication
that the thermal contribution around 6.5~keV is underestimated using
this model. Furthermore, the fits to different epochs result in cutoff
energies ranging from 2.4--13~keV and normalizations spanning a factor
of 3. This is compensated by changes of the best-fit temperature of
the warmer of the two shocks from 1.5--5.4~keV. These factors all
indicate that it is possible to include a non-thermal component, but
it is not statistically nor physically motivated, and adds unnecessary
complexity to the model. We note that we consider adding a cutoff PL
as introducing additional complexity since it is composed of a
qualitatively new component even though, formally, the number of DoF
is reduced from 2778 to 2771\footnote{Each shock component has three
  free parameters whereas the cutoff PL has two since $\Gamma$ is
  frozen, and the Fe abundance is free when fitting the cutoff PL
  model.}. Therefore, we reject the model with the cutoff PL.

Another possibility is to simply add a non-thermal component to the
three-shock model. In this case, a cutoff energy cannot be
constrained, so we add a PL with $\Gamma=2$. We note that the Fe
abundance remains frozen to our adopted standard value of 0.62 of LMC
since the hottest shock component still dominates around
6.5~keV. Fitting yields an improvement of $\Delta\chi^2=-2.71$ for 1
additional DoF (2770 DoF in total) compared to the standard
three-shock model. This means that the PL results in a negligible
improvement and it is not motivated to include it.

To put a limit on the non-thermal emission, we start from the fit
above with the additional PL component with $\Gamma=2$. By fitting
simultaneously to all observations, we obtain one average upper limit
instead of an upper limit at each epoch. The upper limit is computed
by finding the PL normalization that results in an increase of
$\Delta\chi^2=7.740$ relative to the best fit (using the XSPEC task
\texttt{error}). We take the flux of the PL component with this new
normalization as a 3$\sigma{}$ upper limit. The resulting limits are
$7\times{}10^{-14}$~\ergpspcm{} between 3--8~keV and coincidentally
also $7\times{}10^{-14}$~\ergpspcm{} between 10--24~keV.

These limits are clearly model-dependent, as further discussed in
Section~\ref{sec:nt}. A very conservative, model-independent limit is
to consider the total fluxes as upper limits
(Table~\ref{tab:nus_flx}), with the 10--24~keV range likely being the
most relevant.

\subsection{35--65~keV Limit on the Compact Object}\label{sec:lim_co}
We obtain a 3$\sigma{}$ upper limit on any flux from \sna{} in the
35--65~keV range of $8\times{}10^{-14}$\ergpspcm{}. This was computed
using a $\Gamma=2$ PL component without an underlying thermal
component. This limit does not require model assumptions because
\sna{} is not detected in this energy range, which is above the
continuum emission from the ER and below the radioactive $^{44}$Ti
line emission. This limit is primarily relevant for the compact object
but, naturally, also applies to any other emission component.

\subsection{Radioactive $^{44}Ti$ Line Emission}\label{sec:tle}
The \ti{} decay chain produces high-energy \xray{} lines, which have
been detected by \nustar{} (see \citealt{boggs15} for details). The
radioactive K$\alpha$ lines in the \xmm{} energy range are studied
separately in Appendix~\ref{app:rka}. We repeat the analysis of
\citet{boggs15}, but with the inclusion of our new data.

Following the baseline case of \citet{boggs15}, we tie the fluxes,
energies, and widths of the two lines. This can be done because the
relative fluxes ($96.4/93.0$) and energies
($78.32\text{~keV}/67.87$~keV) are given by the relative yields and
transition energies, respectively. The widths are simply proportional
to the energy. This implies that both lines are fitted simultaneously,
using the free parameters of only one line. We choose to report the
fitted parameters for the 67.87~keV line below.

\begin{figure}
  \centering
  \includegraphics[width=\columnwidth]{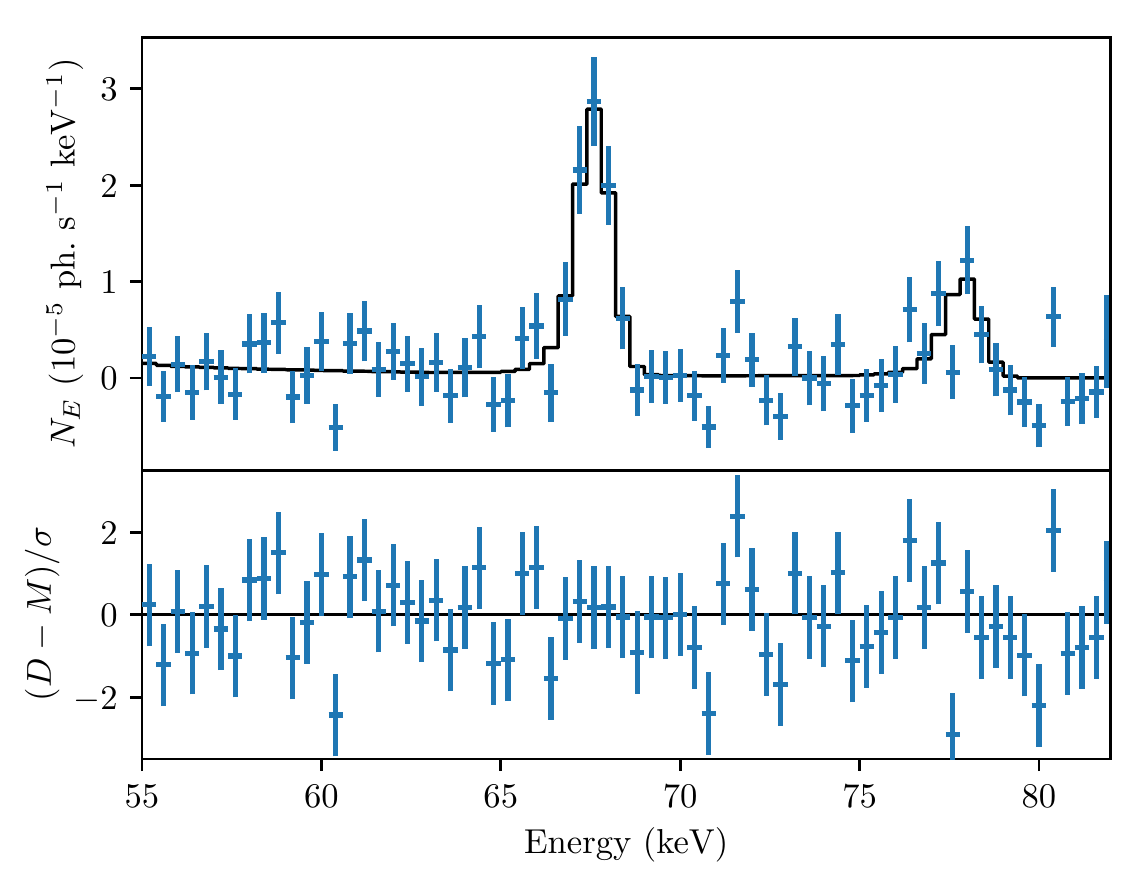}
  
  \caption{Fits of the \ti{} lines. The panels are structured as in
    Figure~\ref{fig:fek}. The large difference between the two lines
    are due to the highly varying instrumental response within this
    energy range, which is close to the upper energy limit of
    \nustar{}. For visual clarity, the data shown are the rebinned sum
    of all \nustar{} observations.\label{fig:til}}

\end{figure}
We show the fitted \ti{} lines in Figure~\ref{fig:til}. The best-fit
energy of the 67.87~keV line is $67.64^{+0.12}_{-0.09}$~keV, compared
to $67.64^{+0.09}_{-0.09}$~keV of \citet{boggs15}. We note that
separate fits to the individual modules yield
$67.70^{+0.17}_{-0.13}$~keV for FPMA and $67.61^{+0.13}_{-0.14}$~keV
for FPMB. In terms of velocities, our redshifts (after correcting for
recession velocity and the look-back effect; \citealt{boggs15}) are
$670^{+520}_{-380}$\kmps{} for the simultaneous fit to both modules,
$420^{+740}_{-560}$\kmps{} for FPMA, and $820^{+580}_{-640}$\kmps{}
for FPMB.

The difference between the modules indicates that systematic
uncertainties could be comparable to the statistical
uncertainties.\footnote{We reiterate that these confidence intervals
  are at a 90\,\% level, as in \citet{boggs15}.}  In this context,
systematic uncertainties include all factors related to the data
reduction and analysis, in addition to the instrumental calibration
uncertainty (see also Appendix~\ref{app:sys}). For reference, the
reported energy calibration uncertainty at 67.87~keV is 0.06~keV
\citep{madsen15}, corresponding to 270\kmps{}.

Analogous to the redshift, we provide \ti{} mass estimates based on
both the joint fit and fits to the individual modules. The combined
analysis yields a 67.87~keV line flux of
$3.7_{-0.6}^{+0.6}\times{}10^{-6}$~photons~s$^{-1}$~cm$^{-2}$, which
corresponds to an initial \ti{} mass of
$1.73_{-0.29}^{+0.27}\times{}10^{-4}$~\Msun{}. When converting \ti{}
fluxes to masses, we use a distance of 51.2~kpc and correct the
observed fluxes by 3\,\% to account for Compton scattering by the
ejecta (\citealt{alp18c} and Section~\ref{sec:co} below).

Our mass estimate is comparable to the value
$1.5_{-0.3}^{+0.3}\times{}10^{-4}$~\Msun{} reported by
\citet{boggs15}, and $1.5_{-0.5}^{+0.5}\times{}10^{-4}$~\Msun{}
obtained from modeling of an optical spectrum by
\citet{jerkstrand11}. Furthermore, a slight difference is apparent
between FPMA and FPMB also for the inferred masses. The FPMA estimate
is $1.86_{-0.40}^{+0.43}\times{}10^{-4}$~\Msun{}, compared with
$1.57_{-0.38}^{+0.43}\times{}10^{-4}$~\Msun{} obtained from FPMB. This
further highlights the subtle systematic uncertainties, which are
important to consider.

\section{Discussion}\label{sec:dis}
\subsection{Physical Interpretation of the Thermal Emission}\label{sec:ec}
Thermal shock models adequately describe the observed \xray{}
spectra. We find no indications of a non-thermal component in the
spectrum. Instead, we favor an interpretation where the low- and
mid-temperature components originate from the ER clumps, while the
high-temperature component is produced in low-density regions. The ER
clumps could give rise to a range of temperatures as a result of
varying incident shock angles, hydrodynamic disruption of the clumps,
and a mix of radiative and adiabatic shocks of varying
temperatures. The low-density regions refer to both the diffuse gas
between the ER clumps and the high-latitude \hii{} region. This
interpretation is motivated by a number of independent simulations and
observations across the electromagnetic spectrum. We present these
points separately for the ER clumps in Section~\ref{sec:er} and
low-density regions in Section~\ref{sec:ldr}, before tying things
together in Section~\ref{sec:cp}.

\subsubsection{Emission from the ER Clumps}\label{sec:er}
Previous studies of the ER clumps have shown the following.
\begin{itemize}
\item Optical and ultraviolet (UV) observations resolve the hotspots
  in the ER and show slow, radiative shocks with velocities up to
  ${\sim}700$~\kmps{} into clumps with pre-shock densities of up to
  $6\times{}10^{4}$~cm$^{-3}$ \citep{pun02, groningsson08b,
    larsson19b}.
  
\item There is a clear east-west asymmetry in the \xray{} emission
  \citep{frank16, cendes18}. The \xray{} emission below 2~keV is
  stronger to the west and follows the optical ER emission. This
  implies that the ER clumps dominate the $<2$~keV \xray{} flux.

\item The inferred ER radii in optical and \xray{}s also indicate that
  the $<2$~keV \xray{}s originate from the same region as the optical
  ER \citep{frank16, larsson19b}. Furthermore, the $<2$~keV radius
  appears to separate into two components after 6000~d when divided at
  0.8~keV, with the $<0.8$~keV component expanding the slowest.

\item Simulations of shock interactions with the ER clumps reveal a
  complex picture \citep{borkowski97, pun02, orlando15,
    orlando19}. The interactions clearly disrupt the clumps, disperse
  the clump material, and produce both radiative and adiabatic shocks
  of varying temperatures. The temperatures are expected to range from
  ${\sim}$0.2 to 1~keV depending on the shock velocities, densities,
  and incidence angle.
\end{itemize}

In this paper, we have presented new light curves of the soft \xray{}
emission (Figure~\ref{fig:lc}). These are in line with previous
conclusions and further corroborate the hypothesis that the soft
\xray{} emission is primarily produced in the ER clumps.
Figure~\ref{fig:lc} shows that \xray{}s below 2~keV have a similar
evolution as optical and there are indications that the 0.45--0.7~keV
light curve is more strongly correlated with the optical emission than
the 0.5--2~keV emission after 10,000~d. This may be explained as a
result of the cooling time ($t_\mathrm{cool}$) depending on the shock
velocity ($V_\mathrm{s}$) and density ($n$) as
\begin{equation}
  \label{eq:cool}
  t_\mathrm{cool} = 47 \left(\frac{V_\mathrm{s}}{500\text{\kmps}}\right)^{3.4} \left(\frac{n}{10^{4}\text{~cm}^{-3}}\right)^{-1}\text{~yr}
\end{equation}
\citep{groningsson08b}. Densities much higher than $10^{4}$~cm$^{-3}$
are therefore required to produce optical emission for shocks faster
than $500$\kmps{}. Consequently, a decreasing fraction of the \xray{}s
above ${\sim}$2~keV is expected to follow the optical emission.

\subsubsection{Emission from the Low-density Regions}\label{sec:ldr}
In addition to studies indicating that the \xray{} emission below
2~keV originates from the ER clumps, the \xray{}s above 2~keV likely
originate from a region of wider latitudinal extent and lower
density. This is primarily motivated by the following arguments.
\begin{itemize}
\item Already in 1997, HST observations revealed Ly$\alpha$ and
  H$\alpha$ emission extending to $\pm$30\degr{} above and below the
  ER \citep{michael98}. Later, radio and \xray{} observations showed
  similar latitudinal extents \citep{ng09, ng13, cendes18}. This is
  also consistent with VLT observations, which show H$\alpha$ emission
  extending to velocities higher than ${\sim}$13,000~\kmps{} in 2012
  \citep{fransson13}. Later VLT observations continue to show similar
  velocities \citep{larsson19b}.
  
\item In contrast to the $<2$~keV \xray{}s, the asymmetries of the
  $>2$~keV \xray{}s follows the radio torus and are brighter to the
  east. The new optical emission beyond the ER is also stronger to the
  east, indicating stronger interaction at high latitudes in this
  direction \citep{larsson19b}.

\item The inferred ER radii in radio and \xray{}s indicate that the
  $>2$~keV \xray{} and radio emission is produced by shocks with
  velocities of ${\sim}$3000~\kmps{} \citep{frank16, cendes18}. We
  note that small discrepancies between the \xray{} and radio radii
  could be due to a combination of limited spatial resolution,
  projection effects, and overlap between different components
  (energies) in \xray{}s.

\item \xray{} emission from the fastest shocks moving in the
  low-density regions are expected to produce a faint, broad component
  in the line profiles. This could potentially reveal the relative
  contributions of different shock velocities to the \xray{} emission
  below 2~keV. A 20\,\% contribution from a broad component
  corresponding to a velocity of ${\sim}$9000~\kmps{} has been
  reported by \citet{dewey12}, but the limited energy resolution and
  line blending complicate analyses. The \xray{} line profiles have
  also been successfully fitted without a broad component
  \citep{zhekov09}. We do not investigate the line profiles in detail,
  but obtain good fits using models without an additional broad
  component. However, optical and UV lines show directly that a
  ${\sim}$13,000\kmps{} component is present \citep{michael98,
    fransson13}. For these very fast shocks, the slow equipartition
  between the electrons and ions likely leads to a considerably lower
  electron than ion temperature, which we discuss below.

\end{itemize}

We present the 3--8 and 10--24~keV light curves to 2020 in
Figure~\ref{fig:lc}. The 3--8~keV \xray{} light curve follows the
radio light curve relatively well. The 10--24~keV light curve has a
flatter evolution, but this is likely the effect of a slight decrease
in temperature (Figure~\ref{fig:com}), rather than an indication of a
different physical origin (Section~\ref{sec:nt}). This shows that the
\xray{} light curves agree with the picture painted by the previous,
independent observations.

\subsubsection{The Combined Picture}\label{sec:cp}
The points above, along with further pieces of information, form a
combined picture. Assuming equipartition between electrons and protons
behind the shocks, the temperature is related to the shock velocities
by
\begin{equation}
  \label{eq:tvs}
  k_\mathrm{B}T = 1.4\left(\frac{V_\mathrm{s}}{1000\text{\kmps{}}}\right)^{2}\text{~keV}
\end{equation}
for our adopted abundances. Using observed velocities in the optical
and \xray{}s one can therefore relate these to the observed
temperatures we find.

The optical ER expands with a velocity of $680\pm{}50$\kmps{}
\citep{larsson19b} since ${\sim}$7700~d, and line profiles of
the optical hotspots show velocities up to 700--1000\kmps{}, depending
on the geometry, with a FWHM of ${\sim}300$\kmps{}
\citep{fransson15}. This corresponds to a shock temperature of
$k_\mathrm{B}T=0.1$--1~keV. Therefore, these shocks may be at least
partially responsible for the two lower temperature components. These
shocks will be radiative for densities up to
${\sim}6\times 10^{4}$~cm$^{-3}$.

From the Chandra \xray{} observations later than 6000~d,
\citet{frank16} find an expansion velocity of $1851 \pm 105$~\kmps{}
for the 0.5--2~keV band and $3071 \pm 299$\kmps{} for the 2--10~keV
band. Assuming equipartition, the former corresponds to
$k_\mathrm{B}T=4.8$~keV, while the latter corresponds to 13.2~keV. To
reconcile the high velocity in the 0.5--2~keV band with the
temperature one needs to assume that the emission in the 0.5--2~keV
band has contributions from both slower radiative shocks, also
emitting in the optical, and faster, adiabatic shocks emitting mainly
in \xray{}s. Deviations from equipartition could also be
contributing. This is primarily relevant for the hottest component and
is described in more detail below. We note that \citet{frank16} find
an expansion velocity of $-110\pm{}313$\kmps{} in the 0.3--0.8~keV
range after 6000~d. This shows that the inferred velocities within
\xray{} energy ranges are combinations from shocks of varying
expansion velocities.

Finally, to connect the higher velocities seen by \citet{frank16} in
the 2--10~keV band to the 4~keV component that we find, one has to
assume only partial equipartition of the ion and electron temperatures
behind these faster shocks. Slow equilibration behind fast shocks has
been inferred from supernova remnants, where a ratio of electron to
ion temperature of
$T_\mathrm{e}/T_\mathrm{p} \propto 1/V_\mathrm{s}^2$ has been proposed
for Balmer dominated shocks into neutral media \citep{ghavamian13}.
In the context of \sna{}, \citet{france11} find
$T_\mathrm{e}/T_\mathrm{p} \approx 0.14$--0.35 from observations of
the reverse shock with a velocity of ${\sim}$10,000\kmps{}. \xray{}
emission from the reverse shock, with an electron temperature of
20--49~keV (assuming $T_\mathrm{e}/T_\mathrm{p}$ in the range above),
is presumably too faint to be seen, because of the low ejecta and CSM
densities.

In addition to the temperatures, the densities inferred from the
fitted parametrized ionization age (Section~\ref{sec:ssm}) are
comparable to previous estimates. Based on optical data, the densities
of the unshocked ER clumps are estimated to reach
$6\times{}10^{4}$~cm$^{-3}$ \citep{pun02,
  groningsson08b}. Furthermore, modeling based on radio and optical
data find densities of $10^{3}$~cm$^{-3}$ in the diffuse ER
\citep{mattila10} and a density of $10^{2}$~cm$^{-3}$ in the
surrounding \hii{} region \citep{chevalier95, lundqvist99}. Our
estimates are uncertain, but capture the trend of a decreasing
ionization age as a function of increasing energy. This is an
indication that cooler, slower shocks originate from regions of higher
density.

There are a number of observables that are not completely explained by
our model. We are unable to provide a natural reason for why the
temperature distribution produced by the clumps would be bimodal. It
could simply be that the apparent bimodality is a smoother broad
distribution in reality. Furthermore, we assume that the hottest
component arises from regions spanning a wide range of
densities. Thus, it is somewhat surprising that the temperature
distribution shows such a narrow peak around 4~keV. A possibility is
that the relatively homogeneous \hii{} region dominates this
component.

However, given the complexity of the system, we believe that these are
minor issues. There are also significant modeling uncertainties
underlying the shock model used for the analysis, especially at
temperatures below ${\sim}$1~keV where radiative losses become
important (Section~\ref{sec:mod}). In addition, other emission
components, such as the reverse shock propagating back into the
ejecta, are predicted to contribute at some level at current epochs
\citep{orlando15, orlando19}. This is likely an underlying component,
which is absorbed into our three-component interpretation.

The spectra from different epochs are relatively similar except for a
decrease in temperature of the hottest component in the last epoch
(Figure~\ref{fig:com}). The geometry is too complex for detailed
predictions of the temporal evolution. We simply note that a slight
decrease in temperature is physically reasonable based on a general
deceleration of the ejecta and the blast wave. Therefore, a completely
thermal interpretation of the spectrum is consistent with the observed
temporal evolution.

To summarize, the combination of all available information implies
that the \xray{} emission likely is of purely thermal origin. This
emission would be produced by shocks with typical temperatures of
0.3--0.9~keV in the ER clumps, and 4~keV in the surrounding diffuse ER
and \hii{} region. However, we conclude by cautioning not to
overinterpret the data. This is simply our favored simplification of a
clearly more complex reality.
\vspace*{10pt}

\subsection{Previous Thermal Modeling}\label{sec:ptm}
\begin{deluxetable*}{ccccccccccccccc}
  \tablecaption{Comparison with Previously Fitted Thermal Model Temperatures\label{tab:lit}}
  \tablewidth{0pt}
  \tablehead{\colhead{Date}         & \colhead{Epoch} & \colhead{Model} & \colhead{Energy range} & \colhead{$k_\mathrm{B}T_1$\tablenotemark{a}} & \colhead{$k_\mathrm{B}T_2$\tablenotemark{a}} & \colhead{$k_\mathrm{B}T_3$\tablenotemark{a}} & \colhead{Reference} \\
             \colhead{(YYYY-mm-dd)} &   \colhead{(d)} & \colhead{}      & \colhead{(keV)}        & \colhead{(keV)}             & \colhead{(keV)}             & \colhead{(keV)}             & \colhead{}       }
  \startdata
    2019-11-27 & 11,965         & Two shocks       & 0.3--10   & 0.6 & \nodata{} & 2.5 & \citet{sun21}    \\
    2018-09-15 & 11,527         & Two shocks       & 0.5--3    & 0.5 & \nodata{} & 1.5 & \citet{bray20}   \\
    2017-10-15 & 11,192         & Two shocks       & 0.3--10   & 0.6 & \nodata{} & 2.3 & \citet{sun21}    \\
    2016-11-02 & 10,845         & Two shocks       & 0.3--10   & 0.7 & \nodata{} & 2.5 & \citet{sun21}    \\
    2015-09-17 & 10,433         & Two shocks       & 0.5--8    & 0.3 & \nodata{} & 2.1 & \citet{frank16}  \\
    2014-09-18 & 10,069         & Two shocks       & 0.5--3    & 0.5 & \nodata{} & 1.4 & \citet{bray20}   \\
    2014-07-11 & ${\sim}$10,000 & Three shocks     & 0.45--24  & 0.5 &         1 & 4   & This paper       \\
    2014-07-11 & ${\sim}$10,000 & Continuous       & 0.45--24  & 0.3 &       0.9 & 4.2 & This paper       \\
    2011-03-03 & 8774           & Three shocks     & 0.6--5    & 0.5 &       1.2 & 2.7 & \citet{dewey12}  \\
    2009-01-30 & 8012           & Two shocks       & 0.4--8    & 0.5 & \nodata{} & 2.4 & \citet{sturm10}  \\
    2007-09-10 & 7504           & Two shocks       & 0.4--7    & 0.5 & \nodata{} & 1.9 & \citet{zhekov09} \\
    2007-09-10 & 7504           & Continuous       & 0.4--7    & 0.5 & \nodata{} & 2.2 & \citet{zhekov09} \\
    2007-03-25 & 7335           & Three shocks     & 0.6--5    & 0.5 &       1.2 & 4.3 & \citet{dewey12}  \\
    2007-01-20 & 7271           & Three shocks     & 0.3--8    & 0.3 &       1.7 & 2.7 & \citet{zhekov10} \\
    2007-01-17 & 7268           & Two shocks       & 0.2--10   & 0.4 & \nodata{} & 3.0 & \citet{heng08}   \\
    2005-07-14 & 6716           & Two shocks       & 0.3--8    & 0.3 & \nodata{} & 2.3 & \citet{park06}   \\
    2004-08-30 & 6398           & Two shocks       & 0.4--7    & 0.5 & \nodata{} & 2.7 & \citet{zhekov06} \\
    2004-08-30 & 6398           & Continuous       & 0.4--7    & 0.5 & \nodata{} & 3   & \citet{zhekov06} \\
    2003-05-10 & 5920           & Two shocks       & 0.2--9    & 0.3 & \nodata{} & 3.1 & \citet{haberl06} \\ 
    2000-12-07 & 5036           & Two shocks       & 0.3--8    & 0.2 & \nodata{} & 3.2 & \citet{park06}   \\
    2000-12-07 & 5036           & Three shocks     & 0.3--8    & 0.2 &       2.9 & 5.1 & \citet{zhekov10} \\
    2000-12-07 & 5036           & Three shocks     & 0.4--8.1  & 0.7 &       1.2 & 4.2 & \citet{dewey12}  \\
    2000-01-20 & 4714           & Two shocks       & 0.3--9    & 0.2 & \nodata{} & 2.0 & \citet{haberl06} \\
  \enddata

  \tablenotetext{a}{The subscripts denote individual shock component
    temperatures or temperature distribution peaks. Models with only
    two characteristic temperatures lack the mid-temperature
    component.}

\end{deluxetable*}

Our spectral analysis indicates the presence of three components.
Only a few previous analyses have used three shock components
\citep{zhekov10, dewey12, maggi12}. In contrast, most earlier studies
found bimodal temperature distributions \citep{zhekov06, zhekov09,
  dewey08} or used two shock components \citep{haberl06, heng08,
  sturm10, park11, helder13, frank16, bray20}.

We provide a comparison of our fitted temperatures with a selection of
literature values in Table~\ref{tab:lit}. Our fits find higher maximum
temperatures than most other analyses, especially at comparable
epochs. Notably, even the previous analyses with continuous
temperature models only found two peaks. The differences are primarily
due to the wider energy range provided by \nustar{}, which we show in
Appendix~\ref{app:pn} (Table~\ref{tab:pn}).

Some previous fits found comparable maximum temperatures at epochs
earlier than ${\sim}7000$~d, before the ER interaction peak. This is
likely because the continuum from the surrounding low-density region
was dominant even below ${\sim}$5~keV, which is the range where other
analyses are primarily sensitive. In addition to the values in
Table~\ref{tab:lit}, \citet{maggi12} mention that the hottest
temperature is $>3.5$~keV in \xmm{} observations from 7000--9000~d,
but they did not present a complete spectral analysis. There appears
to be a trend that \xmm{} data result in higher temperatures than
Chandra data, likely due to the higher high-energy sensitivity of
\xmm{}. This strengthens the argument that \nustar{} data is required
to more robustly fit shock components with temperatures above
${\sim}$3~keV (see Appendix~\ref{app:pn} for further details).

\subsection{Non-Thermal Emission}\label{sec:nt}
Even though we favor a thermal interpretation of the \xray{} emission,
it remains possible that there is a non-thermal contribution. In
Section~\ref{sec:lim}, we showed that a model where the hottest shock
component is replaced by a cutoff PL provides a fit of comparable
quality as the purely thermal model. However, an important feature for
the interpretation of the \xray{} spectrum is the Fe~K line complex
around 6.4--6.9~keV (Figure~\ref{fig:fek}).

The presence of a clear Fe~K line in the observed \xray{} spectrum
implies that at least part of the hard \xray{} emission is of thermal
origin. The strength of the line predicted by the thermal model is
proportional to the fitted Fe abundance. Therefore, in principle, it
is possible to accommodate an additional non-thermal component by
increasing the Fe abundances. This results in a higher modeled thermal
line-to-continuum ratio, which would then need an additional
non-thermal continuum component to properly match the observed
line-to-continuum ratio. Consequently, the Fe abundance would be
artificially suppressed if we attempt to fit a purely thermal model to
a spectrum that has a significant fraction of non-thermal emission.

We find that our best-fit Fe abundance of 0.62 relative to the LMC
(Table~\ref{tab:abu}) provides adequate fits to the Fe~K line
(Figure~\ref{fig:fek}; see also \citealt{sturm10, maggi12}). We
consider an Fe abundance of $0.62^{+0.02}_{-0.01}$ relative to the LMC
to be reasonable and, hence, not an indication of an underlying
non-thermal component. This is further strengthened by the independent
Fe estimate of $0.56^{+0.31}_{-0.31}$ relative to the LMC based on
optical spectroscopy \citep{mattila10}. If there is a significant
non-thermal contribution, we expect a substantially lower fitted Fe
abundance.

The correlation between the 3--8~keV and radio light curves
(Figure~\ref{fig:lc}) could be interpreted as a suggestion that (part
of) the hard \xray{}s are non-thermal. However, the 10--24~keV light
curve rises much more slowly than the 3--8~keV light curve. This is
opposite to what would naively be expected if the hard \xray{}s are of
non-thermal origin. The 10--24~keV flux would be expected to increase
faster because thermal \xray{}s would contribute more to the 3--8~keV
than the 10--24~keV range in relative terms. In this scenario, it is
implicitly assumed that the thermal flux would not increase as fast as
the radio flux, based on comparisons between radio and soft
\xray{}s. Since the 10--24~keV flux increases more slowly than
3--8~keV, a non-thermal model would need to have a contrived spectral
evolution, which we disfavor against a simple, slight decrease in
maximum temperature in a completely thermal model.

Following the above arguments, we conclude that the evidence for a
non-thermal component is weak. Instead, we use the data to compute
upper limits (Sections~\ref{sec:lim} and \ref{sec:lim_co}) as
constraints on a non-thermal component. These limits constrain cosmic
ray acceleration and the properties of the compact object, as
discussed below.

\subsection{Cosmic Ray Acceleration}\label{sec:cr}
The shocks from the CSM interactions around \sna{} are expected to
accelerate relativistic particles \citep{berezhko11, berezhko15,
  dwarkadas13}. The relativistic electrons from this acceleration
process are thought to produce the observed radio emission and it is
possible that the same non-thermal synchrotron component extends to
\xray{} energies. From a theoretical perspective, \citet{berezhko15}
present a prediction for the non-thermal \xray{} contribution
associated with cosmic ray acceleration. Using their model for cosmic
ray acceleration \citep{berezhko00, berezhko06}, they find a peak
35--65~keV flux of $5\times{}10^{-14}$\ergpspcm{} at 9280~d
(Figure~\ref{fig:con}), which is marginally consistent with our upper
limit of $8\times{}10^{-14}$~\ergpspcm{}. However, the predicted
fluxes at 3--8 ($1.4\times{}10^{-13}$\ergpspcm{}) and 10--24~keV
($1.2\times{}10^{-13}$\ergpspcm{}) exceed the corresponding
model-dependent limits by a factor of ${\sim}$2
(Section~\ref{sec:lim}).

Recently, it has been reported that a brightening GeV source has been
detected by \fermi{}/LAT, possibly associated with \sna{}
\citep{malyshev19, petruk20}. The GeV emission could be of either
hadronic or leptonic origin. In the hadronic scenario, accelerated
protons interact with protons in the ISM and produce neutral pions,
which decay into gamma rays. On the other hand, a leptonic origin
implies relativistic bremsstrahlung and inverse Compton scattering by
accelerated electrons. The scenario that most directly affects the
observed \xray{} emission is if the GeV emission is leptonic and a
significant part of the \xray{} emission is non-thermal. In this case,
a steep rise in the GeV emission would be correlated with a more rapid
increase in the (hard) \xray{} light curve due to an increasing
non-thermal component.

The observed GeV light curve increases by at least a factor of 2
between 2012 and 2018 \citep{malyshev19, petruk20}. This time range
roughly coincides with the first and last \nustar{} observation.  Both
the \xray{} and radio light curves continue to smoothly increase at
steady rates by factors of ${\sim}$1.5 over this time interval
(Figure~\ref{fig:lc}). Notably, the high-energy \xray{} component
above 10~keV is the flattest and barely increases during this
time. Therefore, we conclude that there is no obvious correlation
between any part of the \xray{} or radio light curve with the GeV
increase. This indicates that the GeV emission is of hadronic origin
or unrelated to \sna{}.

\subsection{Limits on the Compact Object}\label{sec:co}
The upper limit on a non-thermal component in the \nustar{} data
places new constraints on the compact object at hard \xray{} energies
and complements the multi-wavelength limits presented in
\citet{alp18b}. Currently, the only tentative indication of a neutron
star is the bright spot observed in a 679~GHz ALMA image
\citep{cigan19}. This could potentially be the result of local dust
heating by thermal emission from a neutron star, as suggested in
\citet{alp18b}. A PWN can also match the observation, but this model
is practically unconstrained and can fit almost any data. Therefore,
we find no reason to favor the PWN model over passive thermal neutron
star heating. However, we stress that the conclusions from the ALMA
image are uncertain due to a low S/B, possible structure in the
underlying dust emission, and poorly constrained spectrum and
luminosity of the possible source. More data are needed to clarify the
nature of the ALMA hotspot.

Recently, \citet{greco21} interpreted the 2012--2014 \nustar{} data as
indications of a PWN and rejected a purely thermal interpretation of
the data. This is in stark contrast to our results. As we show in
Section~\ref{sec:lim}, we are unable to motivate a model with a
non-thermal component.

We also search for pulsed emission between 10 and 20~keV
(Appendix~\ref{app:ps}), which could be present if a PWN is
contributing. No pulsations are detected and the pulsed fraction is
constrained to less than $15$\,\% of the total source flux between
10--20~keV.

Adding a PWN component to a thermal model will, of course, still fit
the data, but the additional complexity cannot be statistically nor
physically motivated. Furthermore, even if the adopted model contains
a non-thermal component, we believe that more likely explanations are
synchrotron emission from shock accelerated electrons
\citep{berezhko11, berezhko15, dwarkadas13} or the high-energy tail of
a free-free component. As outlined in Section~\ref{sec:cp}, the
reverse shock is expected to produce a faint component with
$k_\mathrm{B}T=20$--49~keV. Both shock synchrotron and free-free
components are expected, and are likely to be present at low levels
compared to the dominant thermal emission.

Finally, any proposed model for the compact object in \sna{} must form
a coherent physical picture with observations at other wavelengths and
the total energy budget of the system. Due to the excellent
observational coverage, the allowed bolometric luminosity of the
neutron star is constrained to be lower than 10--100~\Lsun{}
\citep{alp18b}. The quoted luminosity interval accounts for the
uncertainty in the amount of newly formed dust in the ejecta along the
line of sight.

Our 35--65~keV flux limit provides the deepest constraints on any
\xray{} emission from a compact object in \sna{}. \xray{} emission
from the compact object is affected by absorption since it resides
close to the center of the ejecta \citep{alp18c}. The high metallicity
of the SN ejecta results in photoabsorption dominating up to
${\sim}$30~keV. This implies that the cross section above 30~keV is
dominated by Compton scattering, and that the opacity at these
energies is many order of magnitudes lower than in the 0.3--10~keV
band. We estimate the 35--65~keV absorption using a three-dimensional
neutrino-driven SN explosion model \citep{alp18c}, specifically the
B15 model from \citet{wongwathanarat15}. This is the same absorption
model as was used for \sna{} by \citet{alp18b}. We find that
approximately 5\,\% of the 35--65~keV flux is expected to be absorbed
at 10,000~d, which we henceforth correct for. For reference, the
optical depth to center of the ejecta at current epochs is ${\sim}100$
at 1~keV, ${\sim}$3 at 5~keV, and ${\sim}$0.03 at 50~keV
\citep{alp18c}.

The absorption-corrected 35--65~keV limit is
$8\times{}10^{-14}$\ergpspcm{} (Section~\ref{sec:lim_co}) and
corresponds to approximately $2\times{}10^{34}$\ergps{} (7~\Lsun{})
for a distance of 51.2~kpc. This is equivalent to a fraction of
${\sim}$0.05 of the Crab Pulsar (excluding the Nebula;
\citealt{buhler14}) and is the most constraining direct limit on a
Crab Pulsar-like spectrum. This is deeper than previous hard \xray{}
limits, but does not qualitatively affect any conclusions in
\citet{alp18b}.

\section{Summary and Conclusions}\label{sec:sum}
We have studied new NuSTAR observations of \sna{} obtained in 2020,
and performed a combined analysis with archival \nustar{} data from
2012 to 2014 and \xmm{} data from 2000 to 2019. This combination
covers a wide \xray{} interval ranging from 0.45 to 78~keV. In
conjunction with the long temporal baseline, it provides a unique
\xray{} view of the complex \sna{} system.

The \xray{} light curves in different energy bands have evolved
significantly over the past few epochs. The 10--24~keV flux remained
practically constant between 9300 and 12,000~d (2012--2020). In
contrast, the 3--8~keV flux was rising, but has tentatively started to
flatten after 11,000~d (2017). Previously published radio light curves
only extend to 10,942~d (2017; \citealt{cendes18}), but followed the
rising phase of the 3--8~keV emission up to that point. The 0.5--2~keV
emission has clearly decreased since 10,000~d (2015). In particular,
we find that the correlation between optical and \xray{}s is
increasingly strong for decreasing \xray{} energies.

We favor a purely thermal description of the \xray{}
spectra. Acceptable fits require three shocked plasma components at
temperatures of approximately 0.3, 0.9, and 4~keV. We obtain very
similar results using a model with a continuous temperature
distribution. Previous studies have mostly used two components with
temperatures of approximately 0.5 and 2.5~keV. The difference is most
likely due to the extended energy range offered by \nustar{}, which
has not been used for fits of purely thermal spectra in previous
studies.

Our favored interpretation associates the 0.3 and 0.9~keV components
with the dense ER clumps, whereas the 4~keV emission originates from
the diffuse ER gas and the high-latitude \hii{} region. There is no
obvious explanation for why the ER clumps would produce emission at
two distinct temperatures. The bimodality between the 0.3 and 0.9~keV
components is, however, uncertain and could be the manifestation of a
smoother broad distribution.

We find no evidence for a non-thermal contribution to the \xray{}
data, neither from the temporal nor spectral analysis. The 10--24~keV
flux does not follow the synchrotron radio light curve, which would be
the most natural expectation if the \xray{} emission is increasingly
dominated by synchrotron emission from shock-accelerated electrons at
higher energies. The spectral fits do not improve when a PL is added,
and the clear Fe~K line complex is also well captured by a purely
thermal model. Replacing the hottest thermal component with a
non-thermal component results in a statistically acceptable fit, but
requires an unreasonably high Fe abundance. Furthermore, the thermal
interpretation forms a consistent picture with temporal, spectral, and
spatial multi-wavelength information from the literature. However, we
note that this does not firmly exclude a, possibly weak, non-thermal
contribution.

A brightening of \sna{} at GeV energies has recently been reported
\citep{malyshev19, petruk20}. We find no clear correlation between the
GeV and \xray{} emission. This indicates that the GeV emission is
dominated by hadronic processes or unrelated to \sna{}. However, in
the leptonic scenario, predictions for the level of \xray{} emission
and the high-energy synchrotron cooling cutoff are both
uncertain. This leaves the leptonic channel possible, but requires it
to be consistent with the observational upper limits
($7\times{}10^{-14}$\ergpspcm{} from 10 to 24~keV and
$8\times{}10^{-14}$\ergpspcm{} from 35 to 65~keV).

We also analyze the radioactive \ti{} lines using all the \nustar{}
data to repeat the analysis first performed by \citet{boggs15}. We
find a redshift of $670^{+520}_{-380}$\kmps{} and an initial mass of
$1.73_{-0.29}^{+0.27}\times{}10^{-4}$~\Msun{}. However, there appear
to be potentially significant systematic uncertainties, indicated by
differences in the results from different analysis methods. These
uncertainties are of approximately the same magnitude as the
statistical uncertainties in both the redshift and mass, which should
be considered in subsequent analyses and interpretations.

\xmm{} observations constrain the radioactive K$\alpha{}$ lines from
$\text{\fe{}}\rightarrow{}\text{\femn{}}$ and
$\text{\ti{}}\rightarrow{}\text{\tisc{}}$ (Appendix~\ref{app:rka}). No
lines are firmly detected, but the initial \fe{} mass is inferred to
be $<4.2\times{}10^{-4}$~\Msun{}, whereas the
$1.7\times{}10^{-4}$~\Msun{} \ti{} mass from \nustar{} is consistent
with a non-detection at 4.1~keV in \xmm{}. Nucleosynthesis simulations
predict \fe{} masses in the range 1--20$\times{}10^{-4}$~\Msun{}
\citep{leising06, popov14, sieverding20}. Our results favor a \fe{}
mass toward the lower end of this range. However, this relies on the
10HMM model. The masses from simulations are also based on a diversity
of SN progenitors and the progenitors are exploded using different
methods.

Since we favor a purely thermal model for the full \xray{} spectrum,
we argue that there are no indications of a pulsar. Even if a faint
non-thermal component is present, it is more likely to be shock
accelerated synchrotron emission or the high-energy tail of a
free-free component. We also constrain the pulsed fraction of the
10--20~keV flux to be less than $15$\,\% (Appendix~\ref{app:ps}).

The \nustar{} data constrain the 35--65~keV luminosity of the compact
object in \sna{} to be lower than $2\times{}10^{34}$\ergps{}
(7~\Lsun{}; $8\times{}10^{-14}$\ergpspcm{}). We put this into context
by comparing the limit with previous multi-wavelength
observations. These show that the bolometric luminosity of the compact
object must be lower than 10--100~\Lsun{} (range due to uncertain
properties of the dust). Currently, an ALMA image at 679~GHz is the
only tentative hint of a neutron star in \sna{}, which needs to be
verified with higher quality data.

Our analysis shows that \nustar{} observations are required for a
complete \xray{} view of \sna{}. \nustar{} is the only instrument
sensitive enough to detect the \xray{} continuum above 10~keV, which
affects the interpretation of the entire \xray{} spectrum. The optical
depth to the compact object is currently ${\sim}$1 at 10~keV, which
implies that practically no \xray{}s of lower energy escape, unless
clumping is stronger than simulations indicate. This means that
\nustar{} could be the most sensitive telescope to search for \xray{}
emission from a potential neutron star. If the GeV emission keeps
increasing, it is also possible that a non-thermal component will
emerge in the \nustar{} range. Future \nustar{} observations of \sna{}
are therefore critical to complement the regular monitoring at other
wavelengths.

\acknowledgments{We thank the anonymous referee for the comments. This
  work was supported by the Knut and Alice Wallenberg Foundation, the
  Swedish Research Council, and the Swedish National Space Board. This
  research has made use of data obtained through the High Energy
  Astrophysics Science Archive Research Center Online Service,
  provided by the NASA/Goddard Space Flight Center. This research has
  made use of the NuSTAR Data Analysis Software (NuSTARDAS) jointly
  developed by the ASI Science Data Center (ASDC, Italy) and the
  California Institute of Technology (Caltech, USA). Based on
  observations obtained with XMM-Newton, an ESA science mission with
  instruments and contributions directly funded by ESA Member States
  and NASA. The timing analysis was enabled by resources provided by
  the Swedish National Infrastructure for Computing (SNIC) at PDC
  Center for High Performance Computing, KTH Royal Institute of
  Technology, partially funded by the Swedish Research Council through
  grant agreement no.\ 2018-05973 (Dnr: PDC-2018-56 and
  PDC-2018-101). This research has made use of NASA's Astrophysics
  Data System.}

\vspace{5mm}
\facilities{NuSTAR, XMM (EPIC, RGS).}

\software{
  ADS \citep{kurtz00},
  astropy (3.0.4; \citealt{astropy13, astropy18}),
  \mbox{CALDB} (20200726),
  \mbox{CCF} (2020 August 3),
  \mbox{FTOOLS} \citep{blackburn95},
  \mbox{HEAsoft} (6.27.2; \citealt{heasarc14}),
  \mbox{matplotlib} (2.0.2; \citealt{hunter07}),
  \mbox{numpy} (1.13.1; \citealt{van_der_walt11}),
  \mbox{NuSTARDAS} (1.9.2; \citealt{van_der_walt11}),
  \mbox{SAOImage} DS9 (8.1; \citealt{joye03}),
  \mbox{SAS} (18.0.0; \citealt{gabriel04}),
  \mbox{scipy} (1.1.0; \citealt{virtanen20}),
  \mbox{XSPEC} (12.11.0; \citealt{arnaud96}).
}

\clearpage
\appendix
\vspace*{-13pt}
\section{Search for Pulsed Emission}\label{app:ps}
We search for pulsed \xray{} emission from \sna{} since this would be
the smoking gun of a pulsar. For this timing analysis, we use the
observations from 2012 to 2014 (all \nustar{} epochs except for the
last). Restricting the temporal baseline to two years improves the
sensitivity. A number of factors contribute to this. First, a shorter
temporal baseline requires a lower resolution in rotational period
since period errors propagate proportionally to the baseline. Second,
a lower resolution in pulsar spin-down rate is required since this
also propagates into the phase solution. Furthermore, limiting the
time range to two years reduces the risk of including more complex
phenomena not included in the modeled phase solution, such as
non-linear spin evolution or discrete glitches. Finally, including the
2020 data only increases the total number of photons by $<10$\,\%.

\begin{figure}
  \centering
  \figurenum{A.1}
  \includegraphics[width=\textwidth]{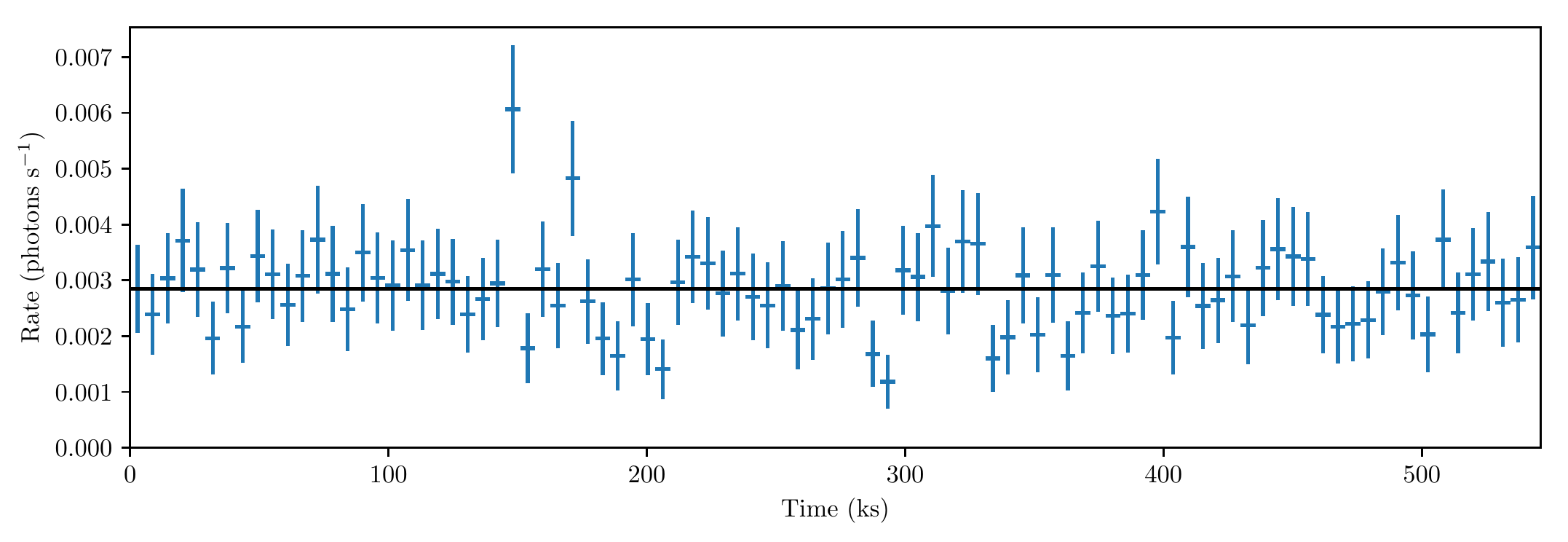}
  \caption{Combined FPMA and FPMB 10--20~keV source light curve of the
    2013 June 29 \nustar{} observation. The black line shows the
    average count rate, which has a total $\chi^2$ value of 88 for 93
    DoF (the error bars are 1$\sigma$). The bin size is 5.8~ks
    (horizontal error bars), matching the orbital period of
    \nustar{}. The count rate has been corrected for Earth occultation
    periods and time filtering criteria (Section~\ref{sec:nus}), which
    result in varying livetimes within each bin.\label{fig:nus_lc}}
\end{figure}
Before performing a detailed search for pulsed emission, we visually
inspect the light curves. The light curve from the 2013 June 29
observation is shown in Figure~\ref{fig:nus_lc} as an example (the
other light curves are similar). There are no indications of
variability within individual observations. A priori, a detection is
also unlikely given our favored model for \sna{} and limited data
quality. However, a rigorous search for pulsed emission is motivated
by the importance of a potential detection and the fact that the
expected spin period cannot be probed visually.

\subsection{Data Reduction}
The reduction of this data was performed separately and used the
latest software at an earlier time of analysis, specifically NuSTARDAS
1.8.0 and CALDB version 20180126. We transform the timestamps of the
events to Barycentric Dynamical Time (TDB) at the solar system
barycenter using the task \texttt{barycorr} (for an overview of
measures of time, see \citealt{eastman10}), with JPL Planetary
Ephemeris DE\nobreakdash-405.

We combine the data from the FPMA and FPMB modules, and restrict the
energy range to 10--20~keV. Below 10~keV, Chandra images clearly show
that the ER dominates. The optical depth through the ejecta is also
very high below 10~keV (Section~\ref{sec:co}). The upper 20~keV limit
is imposed to optimize the S/B.

We extract source photons from a circular region with a radius of
75\arcsec{} pixels, which optimizes the signal-to-noise ratio. No
background sample is necessary for the timing analysis. The source
region contains a total of 22,381 events and the number of source
events is estimated to be ${\sim}$10,000, assuming a spatially uniform
background.

\subsection{Timing Analysis}\label{sec:met_tim}
Overall, our timing method is similar to that of the Fermi pulsar
search Einstein@Home \citep{clark17}. Here, we provide an overview
of the method and motivate the choice of parameters.

The source extraction radius of 75\arcsec{} is relatively large
compared to the 18\arcsec{} FWHM of the telescope, which implies that
a large number of background events are included. This is not an issue
because we assign a weight to each event based on the likelihood that
it is produced by \sna{} \citep{bickel08, kerr11}. The likelihood is
computed by considering the position of the event, the background
level, and the known point spread function of the telescope.

To test the significance of pulsed emission, we use the $H$-test
introduced by \citet{de_jager89}. The test is named after
\citet{hart85} and belongs to the class of tests for uniformity on the
circle \citep{rayleigh19, beran69}. The $H$\nobreakdash-test is
designed to be powerful for typical, general pulsar pulse profiles,
which is useful when the pulse profile is unknown. The null cumulative
distribution function for the $H$\nobreakdash-statistic is accurately
known from simulations \citep{de_jager10}. The
$H$\nobreakdash-statistic from different observations can be summed
incoherently and its approximate null distribution is also known
\citep{de_jager10}.

To reduce computational complexity, we divide the total exposure of
2.8~Ms into 32 time intervals with an average exposure of 90~ks. A
search is performed on each interval separately and the test
statistics are then combined incoherently. The incoherent summing over
long timescales is possible because the timing accuracy of \nustar{}
is stable to within ${\sim}2$~ms (\citealt{madsen15}; see
\citealt{bachetti21} for a recent significant improvement to
$<100$~\textmu{}s).

The neutron star spin is characterized by the period ($P$) and
spin-down rate ($\dot{P}$). We search all periods from 30~ms to
1~ks. For reference, the temporal resolution of \nustar{} is 3~ms
\citep{madsen15} and the deadtime per event is 2.5~ms
\citep{bachetti15, bachetti18}.

For each $P$, we search $\dot{P}$ from 0 to a maximum limit on
$\dot{P}$. This limit is either set by the spin-down luminosity ($L$)
or the spin-down age ($\tau_\mathrm{sd}$), depending on which is more
constraining for a given $P$. The spin-down is related to the
luminosity by
\begin{equation}
  \label{eq:Pdot}
  \dot{P} = \frac{5LP^{3}}{8\pi{}^{2}MR^{2}},
\end{equation}
where $M$ is neutron star mass and $R$ the neutron star radius. We
adopt parameter values of $L=$10,000~\Lsun{}, $M=1.4$~\Msun{}, and
$R=12$~km \citep{ozel16}. This spin-down luminosity 
is much higher than the bolometric limit on the order of 100~\Lsun{}
\citep{alp18b}, but the spin-down luminosity could potentially be much
higher than the bolometric luminosity \citep{abdo13, caraveo14}.

The spin-down age is given by
\begin{equation}
  \label{eq:tsd}
  \tau_\mathrm{sd} = \frac{P}{2\dot{P}}
\end{equation}
and we choose the limit $\tau_\mathrm{sd}>10^8$~s (${\sim}3$ years).
With the aforementioned values, the limit on $\dot{P}$ from
Eq.~\eqref{eq:Pdot} is $<6\times10^{-10}P^3$~s$^{-3}$ and the limit
from Eq.~\eqref{eq:tsd} is $<5\times{}10^{-9}P$~s$^{-1}$. The
spin-down must fulfill both criteria, implying a sufficiently low $L$
to be consistent with multi-wavelength observations and high enough
$\tau_\mathrm{sd}$ to be reasonably consistent with the age of
\sna{}. For completeness, we also search $-\dot{P}$ for each $\dot{P}$
to allow for potential spin-up.

\subsection{Results}
We detect no pulsed emission from \sna{} and verify that the observed
$H_{T}$ distribution is consistent with the null distribution.
Instead, we put a constraint on the pulsed fraction of the emission in
the 10--20~keV range. We compute the limit by simulating a pulsed
signal. The signal is modeled by a single Gaussian with a duty cycle
of 20\,\% (FWHM).

Assuming the 10--20~keV emission to be from a PWN, the 3$\sigma{}$
upper limit on the source pulsed fraction is 15\,\% (background
subtracted). Importantly, however, the 10--20~keV range certainly
includes photons from ER interactions. This implies that the upper
limit on the pulsed fraction of the compact object emission is less
constraining than 15\,\%. For example, if the thermal-to-PWN flux
ratio is $>7$, the pulsed fraction is completely unconstrained since a
100\,\% pulsed fraction would go undetected.

\section{Radioactive K$\alpha$ Lines}\label{app:rka}

\begin{deluxetable}{ccccccccccccccc}
  \tablecaption{\xmm{} Observations Used for Limits on Radioactive K$\alpha$ Lines\label{tab:obs_xmm_rka}}
  \tablewidth{0pt}
  \tablenum{B.1}
  \tablehead{\colhead{Obs.\ ID} & \colhead{Start Date} & \colhead{Epoch} & \colhead{pn Exp.} & \colhead{MOS1 Exp.} & \colhead{MOS2 Exp.} \\
             \colhead{}       & \colhead{(YYYY-mm-dd)} & \colhead{(d)}   & \colhead{(ks)}    & \colhead{(ks)}      & \colhead{(ks)}    }
  \startdata
    0115690201 & 2000-01-19 & 4713 &            22 & \nodata{} & \nodata{} \\
    0115740201 & 2000-01-21 & 4715 &            33 & 43 & 32 \\
    0104660101 & 2000-09-17 & 4955 & \hphantom{0}4 & \nodata{} & \nodata{} \\
    0104660301 & 2000-11-25 & 5024 &     \nodata{} & 21 & 21 \\
    0083250101 & 2001-04-08 & 5158 &            14 & 13 & 13 \\ 
    0144530101 & 2003-05-10 & 5920 &            57 & 58 & 54 \\ 
    \enddata

    \tablecomments{All cameras were not operating simultaneously in
      science mode during some observations in the early \xmm{}
      cycles.}
    
\end{deluxetable}
We search early \xmm{} pn and MOS data for the \femn{} 5.9 and \tisc{}
4.1~keV lines. Only observations from 2000 to 2003
(Table~\ref{tab:obs_xmm_rka}) are used because the thermal emission
from the ER becomes overwhelming at later times. The data reduction is
similar to the reduction for the primary analysis
(Section~\ref{sec:xmm}). The differences are that we use both pn and
MOS data, perform no pile-up correction due to the lower fluxes, and
choose smaller source radii of 10--25\arcsec{} to improve
S/B. Furthermore, we bin to a minimum of 1 count per bin and use the
$C$~statistic \citep{cash79} because of the lower number of photons. A
background subtraction is not performed for these fits. Technically,
XSPEC employs the $W$~statistic for these fits, but it is activated
using the \texttt{statistic cstat}
command.\footnote{\url{https://heasarc.gsfc.nasa.gov/xanadu/xspec/manual/XSappendixStatistics.html}}

\subsection{The 5.9~keV $^{55}Mn$ Line}
Our methods follow \citet{leising06}, who perform an analogous
analysis using \chandra{} data. Focusing first on the \fe{} chain, we
limit the energy range to 4--8~keV and fit a model consisting of a PL
and a Gaussian line. The power-law normalization, power-law photon
index, and line flux are left free. The line energy is frozen to
5.9~keV and line width (represented by $\sigma$) is frozen to 0.06~keV
(3000\kmps{} Doppler width). We find a 3$\sigma$ upper limit on the
line flux of $4.2\times{}10^{-7}$~photons~s$^{-1}$~cm$^{-2}$. This is
similar to the approximate limit of
$<3\times{}10^{-7}$~photons~s$^{-1}$~cm$^{-2}$ found by
\citet{leising06}.

The \femn{} line is absorbed by the ejecta to a varying degree
sensitively depending on the spatial \femn{} distribution. The amount
of absorption is modeled by an average ``effective'' optical depth. At
the relevant epochs, only the outermost \femn{} on the near side is
observable \citep{alp18c, alp19}. Therefore, dedicated detailed
computations are necessary to estimate the amount of
absorption. \citet{leising06} estimated the amount of absorption of
the 5.9~keV line by the ejecta using the 10HMM SN model
\citep{pinto88b}. This model is one-dimensional and has additional
mixing introduced by hand. It was designed to successfully match many
observables \citep{mccray93}, in particular the \xray{} and gamma-ray
properties during the first few years \citep{alp19}. At the relevant
epochs, the effective optical depth is ${\sim}1.2$ at 5.9~keV for the
modeled \fe{} distribution. Combining this with the flux limit results
in a $3\sigma$ upper limit on the initial \fe{} mass of
$4.2\times{}10^{-4}$~\Msun{}.

However, we note that there is a hint of a line at 5.9~keV in the
data. Quantitatively, the estimated flux is
$1.2_{-0.9}^{+1.0}\times{}10^{-7}$~photons~s$^{-1}$~cm$^{-2}$ with a
1$\sigma$ confidence level. This indicates that there might be a weak
detection of the line, or possibly slight contamination from the
\femn{} pn energy calibration source \citep{struder01}. We inspect the
residuals but are unable to draw a firm conclusion. If astrophysical,
the flux would point toward an initial \fe{} mass of
$1.2_{-0.9}^{+1.0}\times{}10^{-4}$~\Msun{}. We stress that the
interval is a purely statistical 1$\sigma$ interval. Systematic
uncertainties are certainly also significant, particularly due to
uncertainties in the amount of absorption. For reference,
nucleosynthesis simulations predict \fe{} masses in the range
1--20$\times{}10^{4}$~\Msun{} \citep{leising06, popov14,
  sieverding20}.

\subsection{The 4.1~keV $^{44}Sc$ Line}
The analysis of the \tisc{} 4.1~keV line is analogous to that of the
\femn{} 5.9~keV line. For \tisc{}, we use the 3--5.5~keV range, shift
the line to 4.1~keV, and repeat the same procedure. The 3$\sigma$
upper limit on the line flux is
$3.0\times10^{-7}$~photons~s$^{-1}$~cm$^{-2}$. We note that
He\nobreakdash-like Ca could produce K\nobreakdash-shell lines near
4.1~keV, but we see no signs of any line around 4.1~keV in the
residuals. The thermal \xray{} emission rapidly rises toward lower
energies, which reduces the sensitivity. Since the initial \ti{} mass
is known to be ${\sim}1.7\times{}10^{-4}$~\Msun{}
(Section~\ref{sec:tle}), we compute the unattenuated flux to be
$7.9\times10^{-7}$~photons~s$^{-1}$~cm$^{-2}$. Thus, the 4.1~keV flux
limit does not constrain any parameters since it implies an effective
optical depth of $>1$, which clearly is consistent with the value of
${\sim}3$ obtained from the 10HMM model.

\section{Systematic Uncertainties}\label{app:sys}
The calibration of \xray{} instruments is known to be slightly
inaccurate with errors of approximately 5--10\,\% \citep{plucinsky08,
  plucinsky17, ishida11, madsen17}. In general, the calibration errors
are energy-dependent, implying that spectral shapes are affected in
addition to the overall normalization. We introduce free constants
between different instruments when fitting simultaneously, which
capture overall normalization errors. This approach does not address
the energy dependence, but is still conventionally used in \xray{}
analyses.

The calibrations are known to be variable and depend on epoch,
observing conditions, and underlying spectrum. Therefore, we
investigate the spectral difference between \nustar{} and \xmm{}/pn
using our data. We do this by fitting a simplified version of the
thermal shock model. This model is the same as the standard model but
with only one shock component. This is sufficient because we only
perform the comparison in a reduced energy range from 3--8~keV. To
quantify the calibration differences, the temperature of this single
component is untied between \nustar{} and pn (with $\tau$ and the EM
remaining tied).

We fit this model to all epochs and provide the resulting temperatures
and fit statistics in Table~\ref{tab:cal}. The average \nustar{}
temperature ($k_\mathrm{B}T_\mathrm{AB}$) is 2.79~keV compared with
the average pn temperature ($k_\mathrm{B}T_\mathrm{pn}$) of
2.39~keV. This difference is clearly systematic as all $T_\mathrm{AB}$
are significantly higher than $T_\mathrm{pn}$. 

When the temperatures are tied, the average cross-normalization
between \nustar{} ($C_\mathrm{AB}$) and pn ($C_\mathrm{pn}$) is
$0.88C_\mathrm{pn}=C_\mathrm{AB}$. The corresponding factor is 0.71
when the temperatures are untied. Consequently, the absolute magnitude
difference between \nustar{} and pn is approximately 0.88. The
remaining difference when the temperatures are untied is primarily due
to the connection between the temperature and the EM. This shows that
both the normalization and spectral shapes are different between
\nustar{} and pn.

Table~\ref{tab:cal} also shows the improvement in fit statistics
obtained by untying the \nustar{} and pn temperatures compared with
tying the temperatures. The average difference at each epoch of
$\Delta\chi^{2}\approx{}-16$ shows that the calibration differences
are significant. This value is only computed from the goodness within
the fitted 3--8~keV energy range and would be higher if a broader
energy range was used.

We note that the third \nustar{} and pn epochs are separated by only
one day. Any temporal evolution within this time interval is certainly
negligible. The fits from this epoch show similar results to other
epochs, which implies that the differences introduced by assuming
simultaneity within the epochs likely are small compared to the
systematic calibration differences.

Due to the calibration uncertainties, we omit the pn data from the
detailed spectral analysis to avoid biases. We motivate this further,
and explore the effects of including pn data and a number of other
choices for managing the calibration uncertainties in detail in
Appendix~\ref{app:pn} below. Finally, we also note that the pn data is
corrected for modest levels of pile-up (Section~\ref{sec:xmm}), which
could affect the accuracy.

In addition to the systematic differences between NuSTAR and pn, the
analysis is also affected by calibration uncertainties between pn and
RGS at low energies. The differences between these instruments have
previously been studied by dedicated calibration programs
\citep{plucinsky08, plucinsky17}. Small differences are clearly
visible in the observations of \sna{} (see Figures in
Appendix~\ref{app:pn}), but these cannot be easily modeled due to the
more complex spectrum at low energies. A simplified method of avoiding
part of the problem is to raise the lower limit of the pn energy range
slightly. We choose 0.8~keV as a trade-off between using as much data
as possible and introducing calibration tensions into the fits.

\begin{deluxetable}{cccccccc}
  \tablecaption{Comparison of Fitted Temperatures Between \nustar{} and \xmm{}/pn\label{tab:cal}}
  \tablewidth{0pt}
  \tablenum{C.1}
  \tablehead{\colhead{Epoch\tablenotemark{a}}  & \colhead{$\Delta{}t$\tablenotemark{b}} & \colhead{$k_\mathrm{B}T_\mathrm{AB}$}               & \colhead{$k_\mathrm{B}T_\mathrm{pn}$}               & \colhead{$T_\mathrm{pn}/T_\mathrm{AB}$} & \colhead{$\chi^2_\mathrm{AB}/\mathrm{DoF}$} & \colhead{$\chi^2_\mathrm{pn}/\mathrm{DoF}$} & \colhead{$\Delta \chi^{2}$\tablenotemark{c}} \\
             \colhead{(d)}                     & \colhead{(d)}                          & \colhead{(keV)}                                     & \colhead{(keV)}                                     & \colhead{}                              & \colhead{}                                  & \colhead{}                                  & \colhead{}}
  \startdata                                                                                                                                                                                       
     \hphantom{0}\hphantom{1,}9332\hphantom{0} & \hphantom{00} $-92$\hphantom{00}       & \hphantom{0}$2.78_{-0.15}^{+0.12}$\hphantom{0}      & \hphantom{0}$2.33_{-0.10}^{+0.11}$\hphantom{0}      & 0.84\hphantom{0}                        & 297/294=1.01 & 98/66=1.49 & $-21.40$ \\
     \hphantom{0}\hphantom{1,}9372\hphantom{0} & \hphantom{00} $-52$\hphantom{00}       & \hphantom{0}$2.79_{-0.16}^{+0.15}$\hphantom{0}      & \hphantom{0}$2.33_{-0.10}^{+0.11}$\hphantom{0}      & 0.84\hphantom{0}                        & 183/199=0.92 & 98/66=1.49 & $-19.25$ \\
     \hphantom{0}\hphantom{1,}9424\hphantom{0} & \hphantom{0000} $1$\hphantom{00}       & \hphantom{0}$2.65_{-0.15}^{+0.18}$\hphantom{0}      & \hphantom{0}$2.32_{-0.10}^{+0.11}$\hphantom{0}      & 0.88\tablenotemark{d}\hphantom{}        & 163/165=0.99 & 98/66=1.49 & $-10.09$ \\
     \hphantom{0}\hphantom{1,}9623\hphantom{0} & \hphantom{00} $199$\hphantom{00}       & \hphantom{0}$2.64_{-0.11}^{+0.12}$\hphantom{0}      & \hphantom{0}$2.35_{-0.10}^{+0.11}$\hphantom{0}      & 0.89\hphantom{0}                        & 221/241=0.92 & 98/66=1.49 & $-10.79$ \\
     \hphantom{0}\hphantom{1,}9920\hphantom{0} & \hphantom{0} $-221$\hphantom{00}       & \hphantom{0}$2.77_{-0.12}^{+0.10}$\hphantom{0}      & \hphantom{0}$2.40_{-0.10}^{+0.11}$\hphantom{0}      & 0.87\hphantom{0}                        & 316/309=1.02 & 64/67=0.96 & $-16.46$ \\
     \hphantom{0}\hphantom{1,}9978\hphantom{0} & \hphantom{0} $-163$\hphantom{00}       & \hphantom{0}$2.83_{-0.10}^{+0.11}$\hphantom{0}      & \hphantom{0}$2.38_{-0.10}^{+0.11}$\hphantom{0}      & 0.84\hphantom{0}                        & 331/338=0.98 & 64/67=0.96 & $-23.71$ \\
     \hphantom{0}           10,021\hphantom{0} & \hphantom{0} $-120$\hphantom{00}       & \hphantom{0}$2.78_{-0.12}^{+0.10}$\hphantom{0}      & \hphantom{0}$2.44_{-0.10}^{+0.12}$\hphantom{0}      & 0.88\hphantom{0}                        & 264/263=1.00 & 65/67=0.97 & $-13.02$ \\
     \hphantom{0}           12,147\hphantom{0} & \hphantom{00} $182$\hphantom{00}       & \hphantom{0}$3.10_{-0.13}^{+0.15}$\tablenotemark{e} & \hphantom{0}$2.56_{-0.18}^{+0.22}$\tablenotemark{e} & 0.83\hphantom{0}                        & 379/353=1.17 & 70/54=1.30 & $-13.06$ \\
    \enddata

    \tablecomments{Fitting a single shock component to the 3--8~keV
      range. The temperature $T_\mathrm{AB}$ is inferred from the
      \nustar{} data whereas $T_\mathrm{pn}$ is from pn data. The fits
      to the \xmm{} data show a rather variable goodness of fit at
      different epochs, likely primarily driven by random fluctuations
      and to a lesser extent by the spectral evolution. The data
      quality and reduction process are very similar for the different
      \xmm{} epochs.}

    \tablenotetext{a}{Epoch of the \nustar{} observation.}

    \tablenotetext{b}{Difference in time from the \xmm{} to the
      \nustar{} observation. Three different \xmm{} observations
      are used in these fits (see Table~\ref{tab:obs_xmm}).}

    \tablenotetext{c}{Improvement in total $\chi^{2}$
      ($\chi^2_\mathrm{AB}+\chi^2_\mathrm{pn}$) obtained by untying
      $T_\mathrm{AB}$ from $T_\mathrm{pn}$.}

    \tablenotetext{d}{Used for connecting the \nustar{} and \xmm{}
      temperatures in Appendix~\ref{app:pn}.}

    \tablenotetext{e}{The fitted temperatures rise in the last epoch,
      in contrast to the results of the main spectral analysis
      (Figure~\ref{fig:com}). This is because the soft part of the
      spectrum fades whereas the hard part brightens
      (Figure~\ref{fig:lc}). The combined effect causes the 3--8~keV
      range to harden.}

\end{deluxetable}

In addition to the instrumental calibration errors, there are
systematics introduced by the data reduction and methods. For example,
\citet{frank16} also analyzed the \xmm{} observations up to
10,141~d. We find better agreement within a few percent at 0.5--2~keV
and at earlier epochs (around 8000~d), but up to 10\,\% higher
3--8~keV fluxes at later times. There is a large number of factors
that could contribute to these differences: source and background
regions; data calibration version; choice of underlying model; free
parameters; ISM and model abundances; amount of absorption; cross
sections; and pile-up corrections.

For these reasons, we add a systematic uncertainty to our \xray{}
fluxes. This is added in quadrature to the systematic error in
Figure~\ref{fig:lc}, but we report both separately in tables. We adopt
a fiducial systematic uncertainty of 8\,\% (of the total flux), which
we believe is a reasonable estimate. This particular value is the
typical difference we find in our comparisons between independent
analyses. It also agrees with cross-calibration programs between
different instruments (e.g.\ \citealt{plucinsky17, madsen17}). We do
not rely on this value for any quantitative analysis. The purpose is
solely to serve as a reminder of the systematics, which are often left
implicit.

Finally, we note that these systematics only apply to comparisons
between parameters from different instruments, data reductions, or
methods. The systematic errors are likely significantly lower for
analyses of homogeneously estimated parameter values. Importantly,
this implies that the fluxes derived from the same instrument (and
analysis) are comparable.

\section{Alternative Set-ups for the Spectral Analysis}\label{app:pn}
\begin{deluxetable}{lcccccccccccccc}
  \tablecaption{Fit Results for Different Data Sets and Models\label{tab:pn}}
  \tablewidth{0pt}               
  \tablenum{D.1}
  \tablehead{\colhead{Model\tablenotemark{a}} & \colhead{RGS} &  \colhead{pn} & \colhead{NuS.} &  \colhead{$\chi^2_{3}/\mathrm{DoF}_{3}$} &  \colhead{$\chi^2 _\Sigma{}/\mathrm{DoF}_\Sigma{}$} & \colhead{$k_\mathrm{B}T_1$} & \colhead{$k_\mathrm{B}T_2$} & \colhead{$k_\mathrm{B}T_3$} \\
                  \colhead{}                  &    \colhead{} &    \colhead{} &     \colhead{} &                               \colhead{} &                                          \colhead{} &             \colhead{(keV)} &             \colhead{(keV)} &             \colhead{(keV)}}\startdata
  2 shocks$^\text{I}$                         &      \cmark{} &      \cmark{} &       \cmark{} &                         $2559/1749=1.46$ &                                  $20741/14520=1.43$ &                      $0.71$ &                   \nodata{} &                      $3.06$ \\
  2 shocks                                    &      \cmark{} &      \xmark{} &       \cmark{} &                         $2093/1623=1.29$ &                                  $17223/13515=1.27$ &                      $0.66$ &                   \nodata{} &                      $3.25$ \\
  2 shocks                                    &      \xmark{} &      \cmark{} &       \cmark{} &                           $573/328=1.75$ &                                    $5815/3776=1.54$ &                      $0.76$ &                   \nodata{} &                      $3.14$ \\
  2 shocks                                    &      \cmark{} &      \cmark{} &       \xmark{} &                                \nodata{} &                                  $17126/11709=1.46$ &                      $0.70$ &                   \nodata{} &                      $2.66$ \\
  3 shocks$^\text{II}$                        &      \cmark{} &      \cmark{} &       \cmark{} &                         $2059/1746=1.18$ &                                  $17111/14496=1.18$ &                      $0.54$ &                      $0.96$ &                      $3.39$ \\
  3 shocks$^\text{III}$                       &      \cmark{} &      \xmark{} &       \cmark{} &                         $1771/1620=1.09$ &                                  $14810/13491=1.10$ &                      $0.53$ &                      $0.97$ &                      $3.84$ \\
  3 shocks                                    &      \xmark{} &      \cmark{} &       \cmark{} &                           $451/325=1.39$ &                                    $4676/3752=1.25$ &                      $0.63$ &                      $0.98$ &                      $3.58$ \\
  3 shocks                                    &      \cmark{} &      \cmark{} &       \xmark{} &                                \nodata{} &                                  $13694/11685=1.17$ &                      $0.52$ &                      $0.93$ &                      $2.82$ \\
  $T_\mathrm{X}=0.88T_3$$^\text{IV}$          &      \cmark{} &      \cmark{} &       \cmark{} &                         $2019/1746=1.16$ &                                  $16756/14496=1.16$ &                      $0.52$ &                      $0.94$ &                      $3.43$ \\
  \enddata

  \tablecomments{The data sets used are indicated by the checkmarks
    and crosses. The fit statistics for the third epoch when the
    \nustar{} and \xmm{} observations are separated by one day are
    given by
    $\chi^2_{3}/\mathrm{DoF}_{3}$. The summed fit statistics across
    all epochs are given by $\chi^2
    _\Sigma{}/\mathrm{DoF}_\Sigma{}$. The temperatures are the
    averages across epochs for the different shock components in order
    of increasing temperature, identified by the subscripts. The last
    model has different \nustar{} and \xmm{} temperatures for the
    hottest component. The \nustar{} temperature is
    $T_\mathrm{3}$, whereas the \xmm{} temperature is
    $T_\mathrm{X}$. The statistical errors (not shown) for the
    temperatures vary by approximately a factor of 2 for the different
    set-ups. More data generally reduce the uncertainties and the
    different data sets naturally primarily affect the components
    within the energy ranges of the instruments.}

  \tablenotetext{a}{The Roman numeral superscripts identify the models
    among the subfigures in Figure~\ref{fig:cal2}. Model III is the
    standard model that is used for the spectral analysis in the main
    text.}
  
\end{deluxetable}
In this section, we explore different possibilities for handling the
cross-calibration uncertainties in the spectral analysis. We try
different spectral models, treatments of instrument differences, and
data sets. A summary is provided in Table~\ref{tab:pn} and the fits
are described below. Importantly, even though the best-fit values are
slightly different, our main conclusions remain unchanged regardless
of how the data are analyzed.

\subsection{Models Used for Comparisons}
The models we present here are the standard model with three shocks
and a simplified version with only two shock components. We fit this
model to RGS, pn, and \nustar{} data simultaneously, as well as
separately to data from only two of the instruments. This leads to
four possible set-ups for each of the models. All parameters apart
from constant cross-normalization factors are tied between the
instruments in these fits. The fitted energy ranges are the same as
presented in Section~\ref{sec:met}.

In addition to the fits with two and three shock components, we
investigate a variation of the three-shock model. For this model, the
temperatures of the hottest component are coupled as
$T_\mathrm{X}=0.88T_3$, where $T_3$ is the \nustar{} temperature and
$T_\mathrm{X}$ is the \xmm{} temperature. The factor 0.88 is obtained
from the fit to the overlapping 3--8~keV region of the third epoch at
9424~d (Table~\ref{tab:cal}). This version of the model with different
temperatures for the hottest components is only fitted to RGS, pn, and
\nustar{} data simultaneously.

There are a number of other possible variations for connecting the
\nustar{} and pn data. In addition to the fits presented below, we
have explored constraining the cross-normalizations, completely
decoupling the temperatures, and additionally untying the EMs of the
hottest components. The trend is naturally that gradually increasing
the amount of freedom improves the fit quality, but the results are
similar for all set-ups.

\subsection{Results}
\begin{figure}
  \centering
  \figurenum{D.1}
  \includegraphics[width=\textwidth]{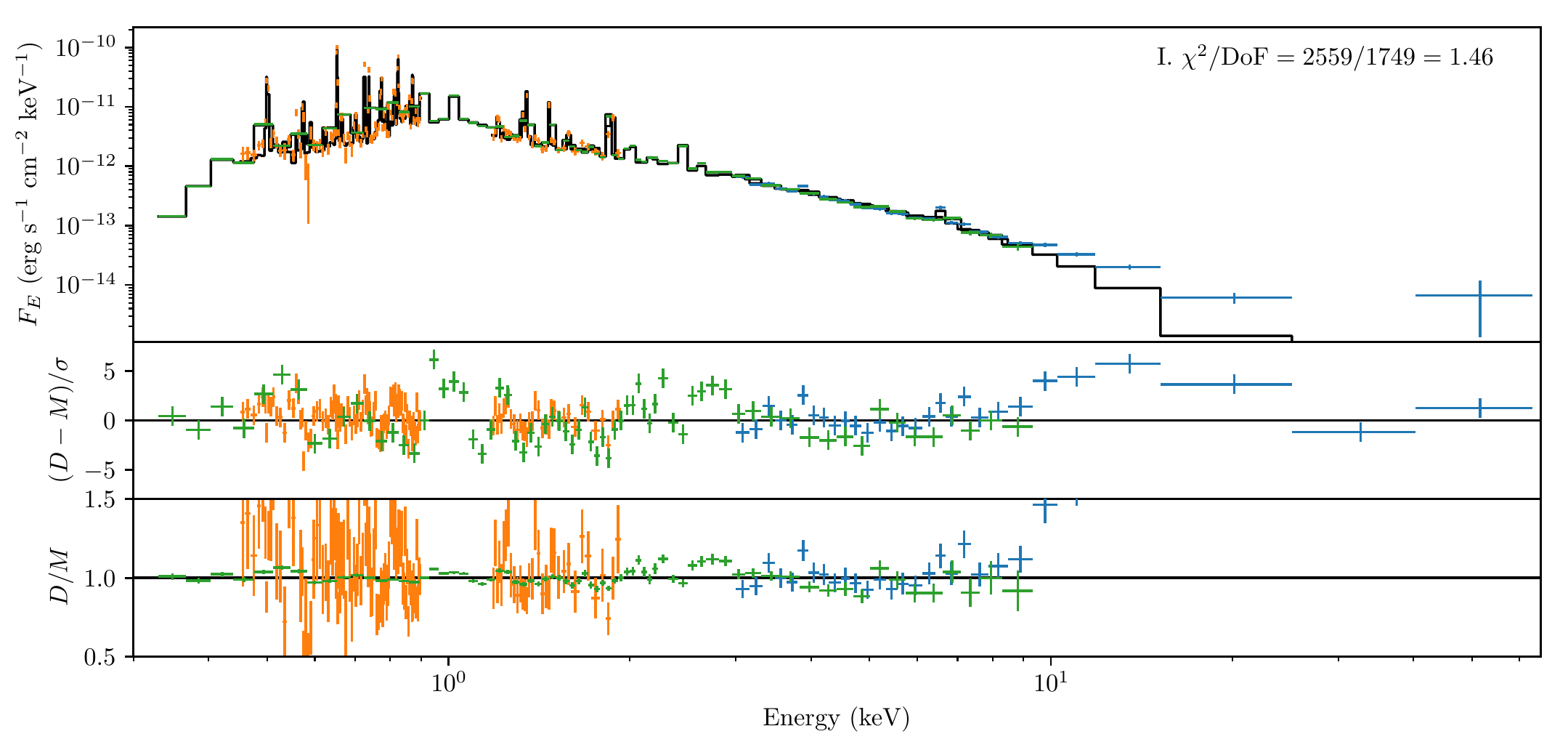}
  \includegraphics[width=\textwidth]{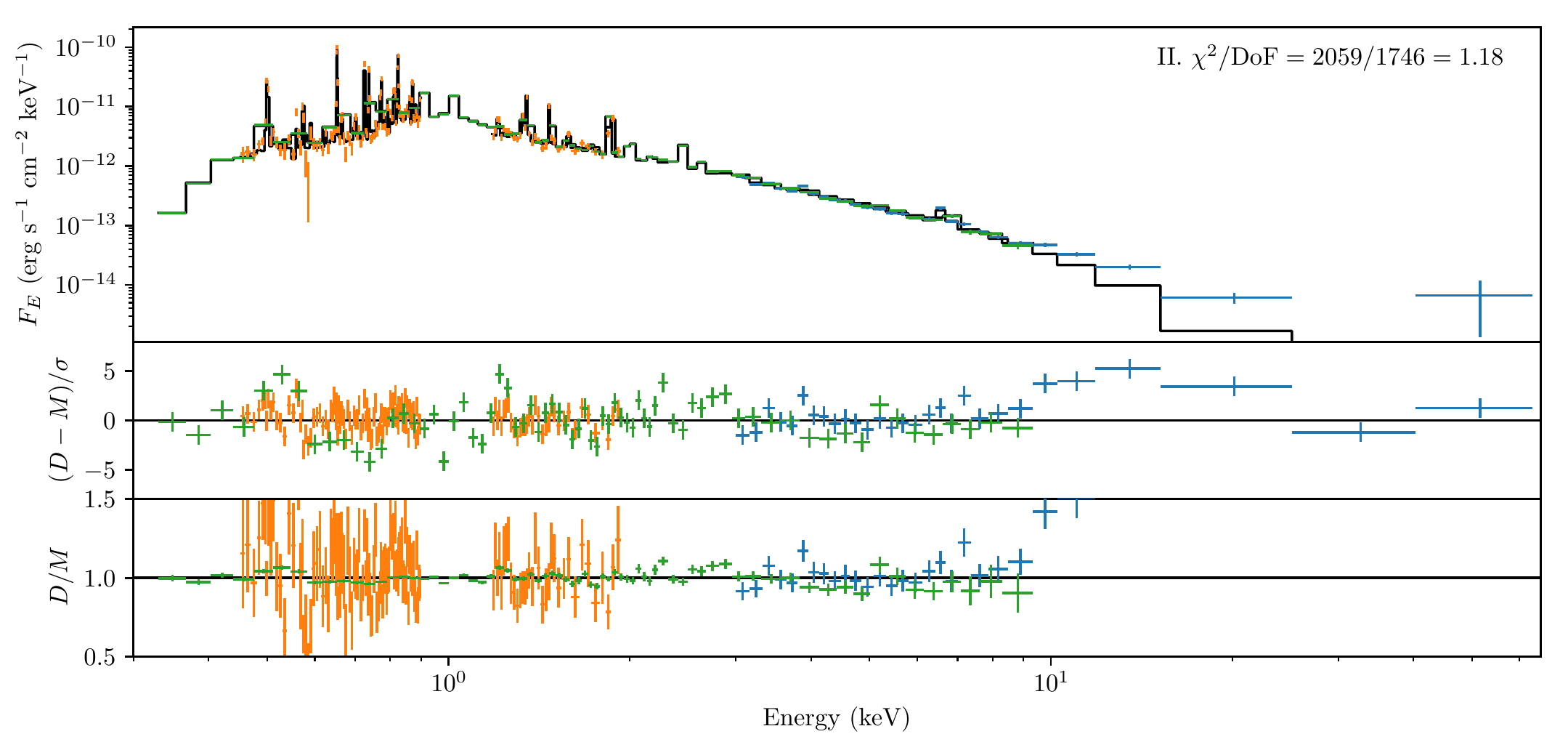}
  \caption{\nustar{} (blue), RGS (orange), and pn (green) spectra
    together with the best-fit model (black) from the third epoch at
    9424~d. The four subfigures (continues on the next page) show
    different set-ups (Table~\ref{tab:pn}): Two-shock model fitted to
    all data (I). Three-shock model fitted to all data
    (II). Three-shock model fitted to \nustar{} and RGS with pn
    overplotted (III). This is the standard model used throughout the
    main text. Three-shock model fitted to all data with the
    temperatures of the hottest component rescaled as
    $T_\mathrm{X}=0.88T_3$ between \xmm{} and \nustar{} (IV;
    Appendix~\ref{app:sys}). The top panels in the individual
    subfigures show the spectra, the middle panels show the normalized
    residuals, and the bottom panels show the data-to-model ratio. The
    two upper panels of each figure are analogous to
    Figure~\ref{fig:con}. The pn data are shown from 0.3~keV, but only
    0.8--10~keV data are used for the analysis throughout the
    paper. \label{fig:cal2}}
\end{figure}
\begin{figure}
  \centering
  \figurenum{D.1}
  \includegraphics[width=\textwidth]{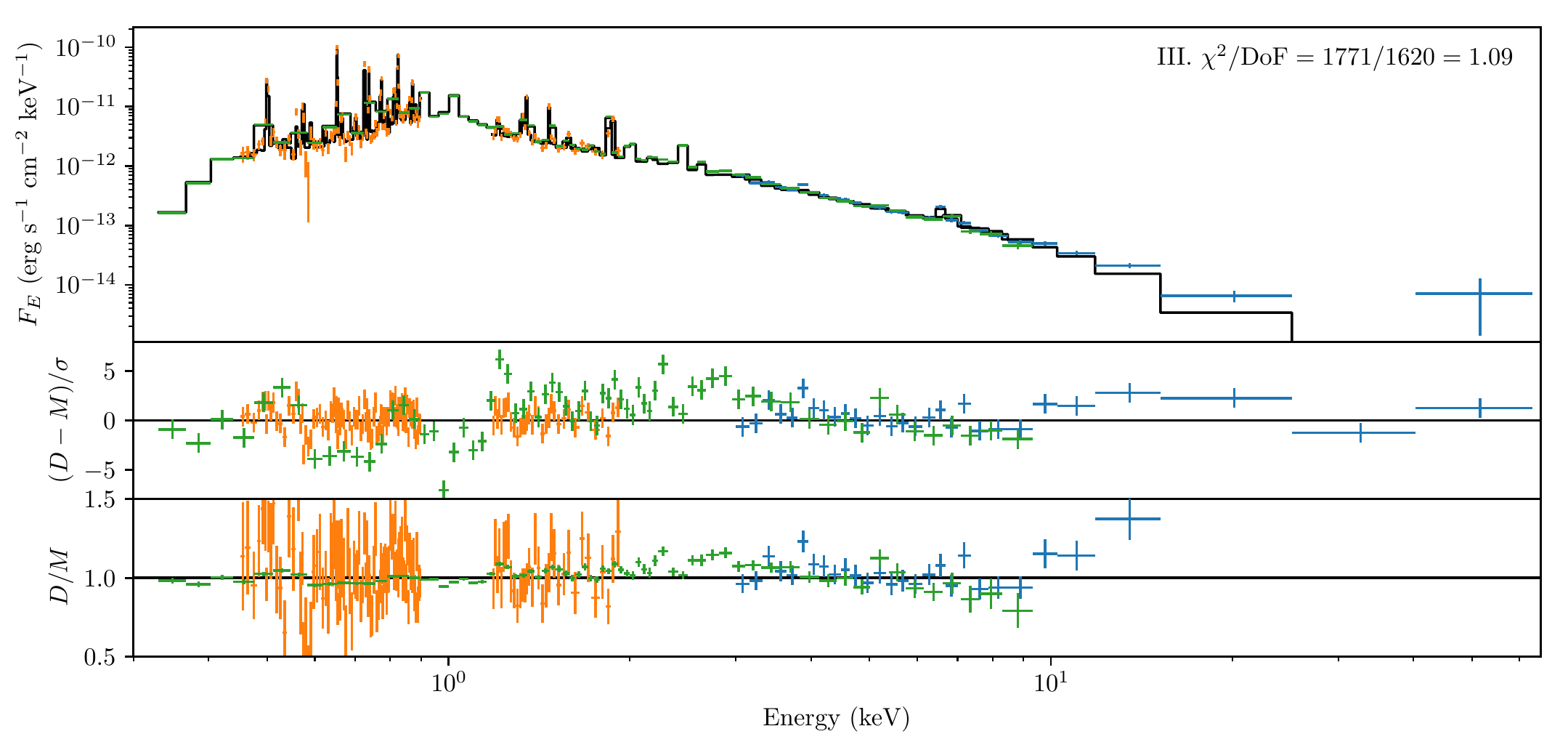}
  \includegraphics[width=\textwidth]{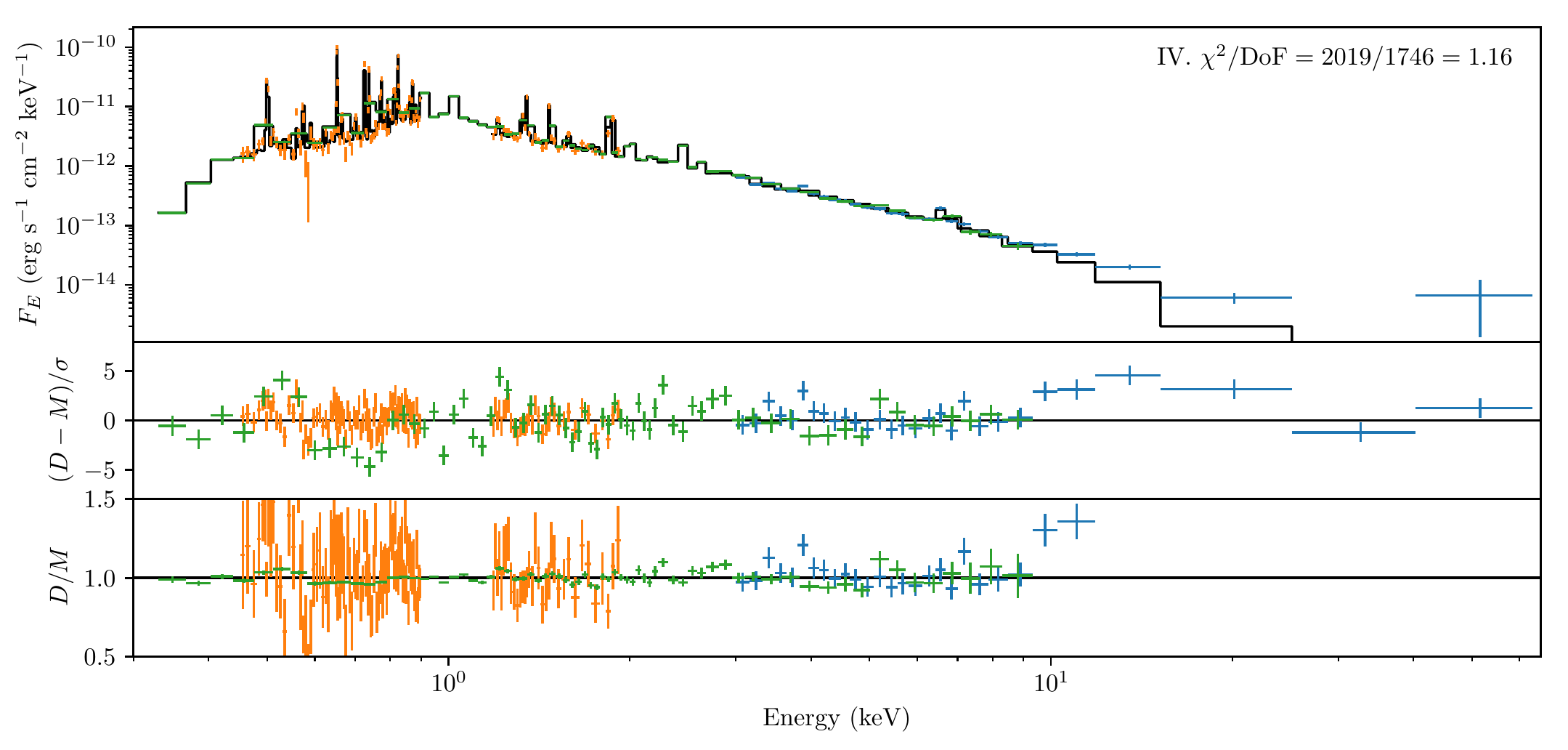}
  \caption{(Continued.)}
\end{figure}
The fit statistics and temperatures of the above set-ups are provided
in Table~\ref{tab:pn}. We present the fit statistics to all data
combined, the average temperatures, and the goodness of fits to the
third epoch (9424~d; \nustar{} and \xmm{} are separated by one
day). We also show fits of four of the models in Figure~\ref{fig:cal2}
where the models are identified by Roman numerals (see
Table~\ref{tab:pn}).

First, we compare Models~I and~II. These are the models with two and
three shock components, respectively, that are fitted to data from
\nustar{}, RGS, and pn. It is clear that two components (Model~I) are
insufficient to model the combined spectrum and that including a third
component (Model~II) results in a substantial improvement in the
goodness of fit. Specifically, the improvement for the third epoch
alone is $\Delta \chi^{2}=-500$ for an additional 3~DoF, and
$\Delta \chi^{2} = 3630$ for an additional 24~DoF across all epochs
combined. The reason for the inferior fit using two components is
primarily driven by the $<2$~keV range, as can be seen in
Figure~\ref{fig:cal2}. This is also clear from the temperatures of the
two cooler components of the three-shock model. Both cool components
have temperatures below ${\sim}$1~keV and primarily influence the RGS
energy range. The first two subfigures of Figure~\ref{fig:cal2} also
show that the addition of a third component does not significantly
change the broadband model continuum shape. This implies that
conclusions regarding the continuum are relatively independent of the
third shock component. However, a third component is clearly motivated
by the statistical improvement. We note that improved fits can be
achieved by allowing for more freedom by freeing the constant
parameters (Section~\ref{sec:abu}) or more conservative data filtering
criteria (which reduces the number of photons; e.g.\ \citealt{sun21}),
but exploring all these possibilities are not within the current
scope.

Figure~\ref{fig:cal2} subfigure~II can also be compared with
subfigure~III. Both use the same standard three-shock model. The
difference between these models is that Model~II is fitted to all
data, whereas Model~III is fitted to \nustar{} and RGS data. We note
that Model~III is the primary standard model used for the main
analysis. The pn data in subfigure~III have been overplotted and only
rescaled by a factor of 0.95 to minimize the residuals. It is clear
from the residuals that there are systematic offsets between the
\nustar{} and pn data (as quantified in Appendix~\ref{app:sys}
above). This is especially clear in the overlapping
region. Furthermore, the pn data are in excess of the model in the
3~keV region, which is a consequence of the differences in spectral
slope. The consequences of these difference can be seen in the
residuals of subfigure~II and the worse fit statistic.

Given the known differences between \nustar{} and pn, we also show the
model variant with the \nustar{} and \xmm{} temperatures rescaled as
$T_\mathrm{X}=0.88T_3$ (Model~IV). Model~IV can be compared to
Models~II and~III in Figure~\ref{fig:cal2}. The rescaling of
temperatures alleviates a significant part of the tension but still
drastically reduces the overall quality of the fit when compared to
fits without pn data. Comparing Models~III and~IV, introducing the pn
data to the fits introduces 1005 DoF in total across all epochs
(Table~\ref{tab:pn}). The change in total $\chi^{2}$ is 1946, implying
that the fit is significantly worse despite the $T_\mathrm{X}=0.88T_3$
scaling. This is not surprising since the calibration errors between
all three instruments are more complicated than what can be captured
by our simplified treatment.

In addition to the decrease in overall fit quality for Model~IV, the
average cross-normalization constant between the \nustar{} and \xmm{}
data is 0.82. These fits use the three-shock model, but \nustar{} is
quite insensitive to the two cooler components. This means that the EM
of the hottest component is degenerate with the cross-normalization
constant, which implies that the connection between the \nustar{} and
pn data is relatively weak. Consequently, the hottest shock components
are largely decoupled between \nustar{} and pn, which partly defeats
the purpose of including the pn data. In light of this, we conclude
that adding the pn data worsens the fit quality, introduces biases,
and adds little information since the components are required to be
relatively uncoupled.

Despite the above arguments, including pn data remains a possible
alternative. I principle, it is possible that the pn data are more
accurate than \nustar{} data. However, the inconsistencies in the
overlapping region between \nustar{} and pn show that there must be
significant calibration errors. The excess above 10~keV in \nustar{},
which is more pronounced when pn data are included, cannot be a result
of instrumental calibration since the ratio between the data and model
is much higher than $10$\,\%. It is possible that the fit attempts to
capture calibration inconsistencies at the expense of producing an
artificial excess above 10~keV. This could happen if both the
instrumental inconsistencies and artificial excess are of comparable
statistical significance.

Importantly, including pn data do not affect the scientific
conclusions qualitatively. The pn data lowers the temperature of the
hottest component from ${\sim}$4 to ${\sim}$3--3.4~keV. Naturally,
this leads to an increased excess in the \nustar{} data above
10~keV. This does not imply that the origin of such an excess is
physically different. Even in this case, we argue that it most likely
is the hottest part of the thermal emission. Simply adding a fourth
thermal component to the model does not provide stable fits as the
temperatures vary irregularly across epochs. We note that the
continuous temperature model has sufficient freedom to allow for a
broader temperature distribution. This model fits the data well and is
discussed in Section~\ref{sec:cal_con} below.

Instead, to characterize the shapes of the \nustar{} spectra, we fit a
two-shock model to all \nustar{} epochs with no other data sets. The
fitted temperatures are $1.8\pm 0.3$ and $5.9\pm0.7$~keV (these
intervals are standard deviations among the eight epochs) with a total
fit statistic of $\chi^{2}=2599$ for 2771 DoF. We also inspect the
residuals and find that the fits are very good. This shows that the
\nustar{} spectra can be well-fitted with a thermal model with
temperatures up to ${\sim}$6~keV. Temperatures around 6~keV are not
unreasonably high since a faint, hot component with temperatures of up
to 20--49~keV is expected from the reverse shock
(Section~\ref{sec:cp}).

Moreover, in Section~\ref{sec:lim}, we show that the \nustar{} data
alone do not favor the inclusion of an additional PL component,
implying that thermal models are fully sufficient to model the
\nustar{} data. This mean that there is no support for two physically
distinct components based on the \nustar{} data alone. Therefore, it
seems more likely that an excess, which is more significant when pn
data are included, is a result of calibration uncertainties or a
high-energy tail of the thermal distribution. Based on the above
discussion, we favor excluding the pn data from the spectral analysis,
while noting that including the pn data would not affect the
conclusions qualitatively.

\subsection{Implication of Adding pn Data for the Continuous Model}\label{sec:cal_con}
\begin{figure}
  \centering
  \figurenum{D.2}
  \includegraphics[width=0.39\textwidth]{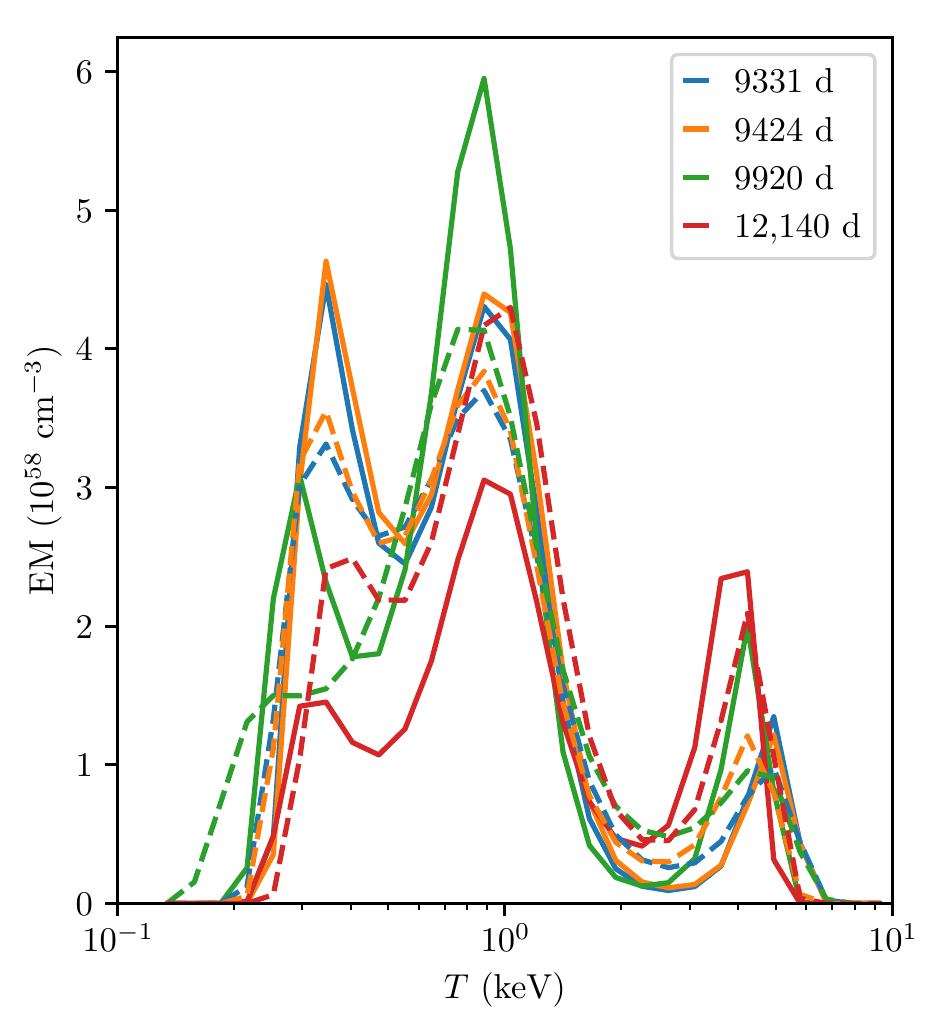}
  \includegraphics[width=0.59\textwidth]{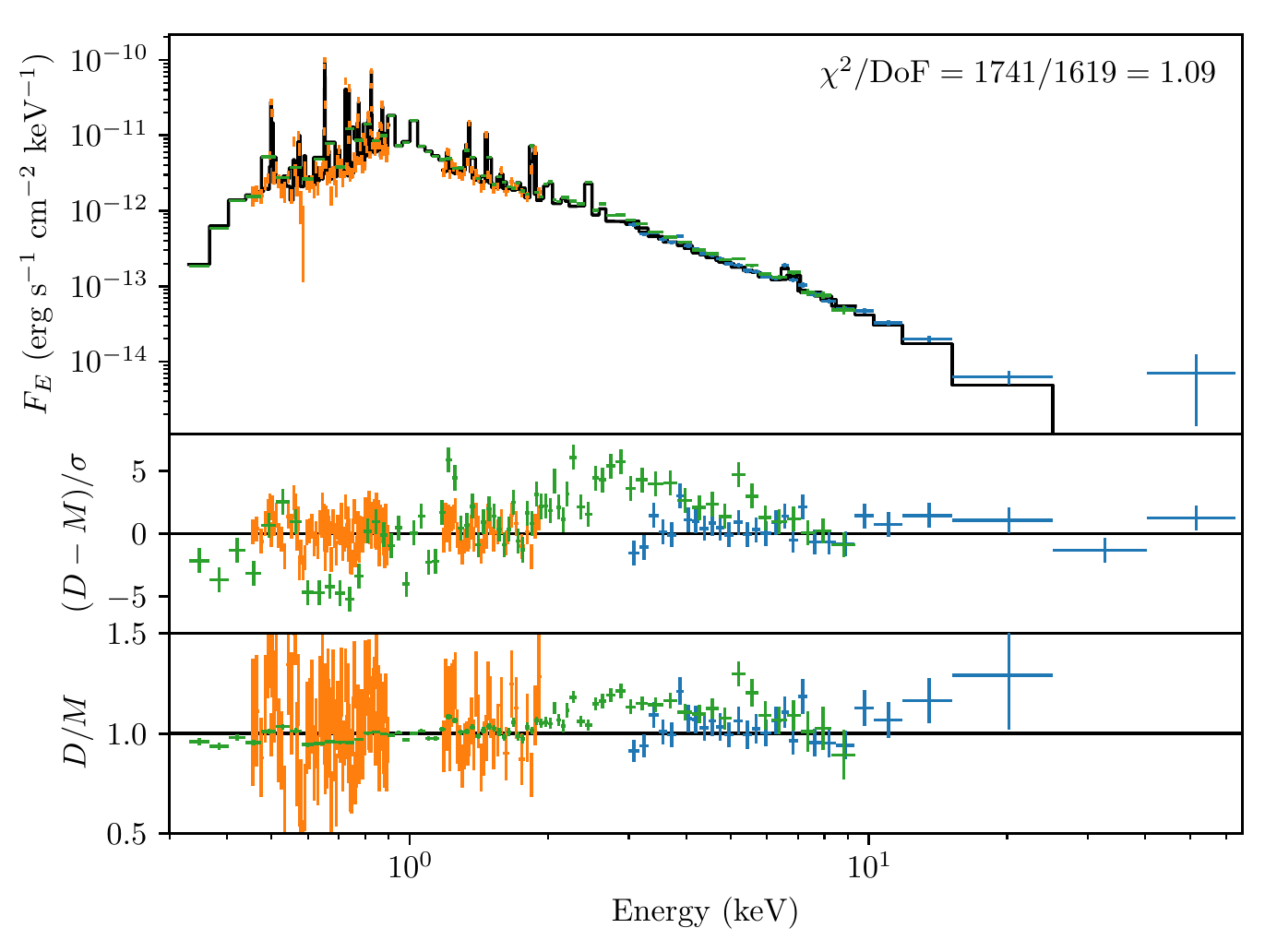}
  \includegraphics[width=\textwidth]{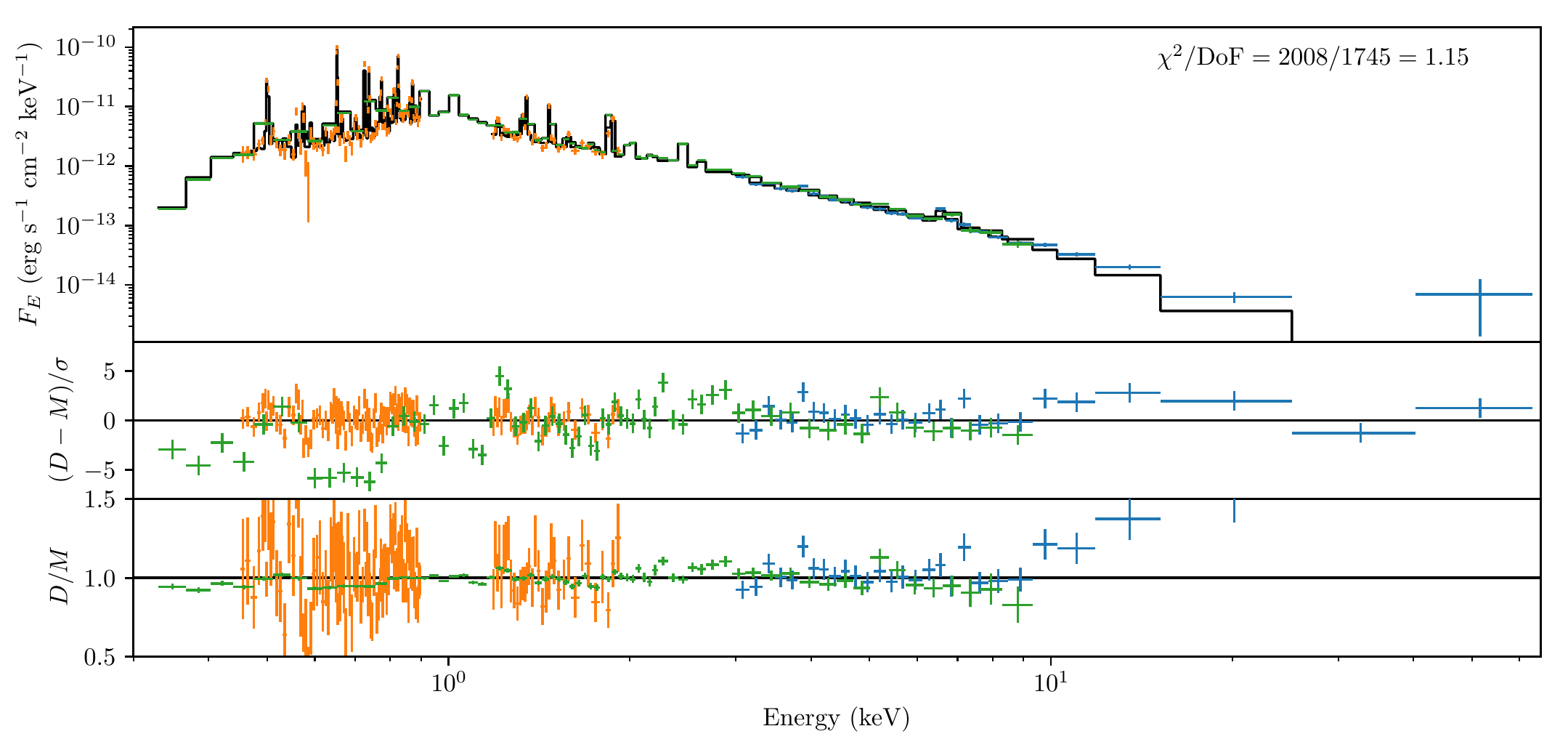}
  \caption{Upper left panel: Fitted EM distributions
    (Eq.~\ref{eq:phi}) for the continuous shock temperature
    model. Solid lines are for the fits to only RGS and \nustar{} data
    (same as in Figure~\ref{fig:dos}). Dashed lines are the
    corresponding fits to all data, including pn. Only a
    cross-normalization is allowed to vary between the
    instruments. Only four epochs are shown for visual clarity but the
    remaining epochs show similar behaviors. Upper right panel:
    \nustar{} (blue), RGS (orange), pn (green) and the model (black)
    spectra from the third epoch at 9424~d. The fit is only performed
    to \nustar{} and RGS data. The pn data are only overplotted and
    rescaled by 1.06 to minimize the residuals. Lower panel: Same as
    upper right but also fitted to the pn data. The upper-right fit
    corresponds to the solid orange line in the upper left plot,
    whereas the lower-panel fit corresponds to the dashed orange
    line. As in Figure~\ref{fig:cal2}, the pn data are shown from
    0.3~keV, but only 0.8--10~keV data are used for the analysis
    throughout the paper.\label{fig:dos_cal}}
\end{figure}
It is also illuminating to explore fits of the continuous temperature
model in addition to the discrete temperature models above. We show
the results of fits of the continuous temperature model to RGS, pn,
and \nustar{} data in Figure~\ref{fig:dos_cal}. Only a
cross-normalization constant is allowed to vary between the
instruments (standard set-up). The results are qualitatively similar
to fits without the pn data, and primarily show a well-separated
high-temperature peak and a broader, possibly bimodal, peak below
2~keV. The differences are expected due to the large amount of freedom
in the fits and the amount of data added by pn. This further
strengthens the argument that the pn data do not qualitatively affect
the conclusions.

Figure~\ref{fig:dos_cal} also shows spectra and residuals of the
continuous model for the third epoch at 9424~d. We show fits both with
and without the pn data. The temperature distributions for these
models both show contributions up to approximately 5~keV, above which
the distributions quickly decline. It is clear that an excellent fit
can be achieved using the continuous model when pn data are not
included. In particular, we note that the slight excess visible above
10~keV is well-captured when the model includes a faint temperature
tail to ${\sim}$6~keV.

The fit including the pn data is also relatively good. It is evident
that the main issues for this fit are the tensions between the
instruments, as can be seen in the overlapping regions. This can also
be seen in the fit statistics. When pn data are included, the fit
statistic for the third epoch increases by 267 for 126 DoF, for a
total increase of 1954 for 1005 DoF across all epochs. Despite the
tensions, the fit is reasonable and the resulting temperature
distribution is very similar to when pn data are excluded.

This further corroborates the conclusion that a purely thermal model
is sufficient. Of course, the investigations above do not prove that
other components are excluded. They only show that additional
components are not necessary.

\section{\xmm{} Fluxes}\label{app:xmm_flx}
\begin{deluxetable}{ccccccccccccccc}
  \tablecaption{\xmm{} Fluxes\label{tab:xmm_flx}}
  \tablewidth{0pt}
  \tablenum{E.1}
  \tablehead{\colhead{Obs.\ ID} & \colhead{Epoch} &                \colhead{$F_\text{0.45--0.7}$} &                   \colhead{$F_\text{0.5--2}$} & \colhead{$F_\text{3--8}$}                     & \colhead{$F_\text{0.5--8}$} \\
             \colhead{}      & \colhead{(d)}   & \colhead{($10^{-13}$~erg~s$^{-1}$~cm$^{-2}$)} & \colhead{($10^{-13}$~erg~s$^{-1}$~cm$^{-2}$)} & \colhead{($10^{-13}$~erg~s$^{-1}$~cm$^{-2}$)} & \colhead{($10^{-13}$~erg~s$^{-1}$~cm$^{-2}$)}}
  \startdata
  0144530101 &   5920 & $1.78_{-0.09}^{+0.07}\pm0.14$ & $8.0_{-0.1}^{+0.1}\pm0.6$ & $1.66_{-0.02}^{+0.02}\pm0.13$ & $10.5_{-0.2}^{+0.2}\pm0.8$ \\
  0406840301 &   7268 & $5.75_{-0.12}^{+0.10}\pm0.46$ & $34.1_{-0.2}^{+0.2}\pm2.7$ & $3.93_{-0.02}^{+0.02}\pm0.31$ & $40.7_{-0.3}^{+0.3}\pm3.3$ \\
  0506220101 &   7627 & $7.01_{-0.09}^{+0.10}\pm0.56$ & $44.4_{-0.2}^{+0.2}\pm3.6$ & $5.32_{-0.02}^{+0.02}\pm0.43$ & $52.9_{-0.3}^{+0.3}\pm4.2$ \\
  0556350101 &   8012 & $8.39_{-0.14}^{+0.09}\pm0.67$ & $54.9_{-0.2}^{+0.2}\pm4.4$ & $6.35_{-0.03}^{+0.03}\pm0.51$ & $66.2_{-0.3}^{+0.3}\pm5.3$ \\
  0601200101 &   8327 & $8.71_{-0.10}^{+0.09}\pm0.70$ & $62.6_{-0.2}^{+0.2}\pm5.0$ & $7.73_{-0.03}^{+0.03}\pm0.62$ & $77.6_{-0.2}^{+0.3}\pm6.2$ \\
  0650420101 &   8693 & $9.44_{-0.13}^{+0.12}\pm0.76$ & $70.7_{-0.3}^{+0.3}\pm5.7$ & $9.45_{-0.04}^{+0.04}\pm0.76$ & $86.9_{-0.4}^{+0.4}\pm7.0$ \\
  0671080101 &   9048 & $9.78_{-0.13}^{+0.11}\pm0.78$ & $76.4_{-0.2}^{+0.3}\pm6.1$ & $10.91_{-0.04}^{+0.04}\pm0.87$ & $95.0_{-0.3}^{+0.4}\pm7.6$ \\
  0690510101 &   9423 & $9.87_{-0.10}^{+0.10}\pm0.79$ & $80.1_{-0.3}^{+0.2}\pm6.4$ & $12.34_{-0.04}^{+0.05}\pm0.99$ & $103.2_{-0.3}^{+0.2}\pm8.3$ \\
  0743790101 & 10,141 & $9.13_{-0.09}^{+0.09}\pm0.73$ & $79.9_{-0.2}^{+0.3}\pm6.4$ & $14.29_{-0.05}^{+0.05}\pm1.14$ & $103.5_{-0.4}^{+0.4}\pm8.3$ \\
  0763620101 & 10,492 & $8.59_{-0.08}^{+0.09}\pm0.69$ & $78.9_{-0.3}^{+0.2}\pm6.3$ & $15.36_{-0.06}^{+0.06}\pm1.23$ & $104.7_{-0.4}^{+0.4}\pm8.4$ \\
  0783250201 & 10,845 & $8.56_{-0.12}^{+0.08}\pm0.68$ & $76.3_{-0.3}^{+0.3}\pm6.1$ & $16.12_{-0.06}^{+0.06}\pm1.29$ & $102.1_{-0.4}^{+0.4}\pm8.2$ \\
  0804980201 & 11,192 & $7.67_{-0.08}^{+0.07}\pm0.61$ & $73.6_{-0.3}^{+0.4}\pm5.9$ & $16.66_{-0.08}^{+0.08}\pm1.33$ & $98.5_{-0.5}^{+0.5}\pm7.9$ \\
  0831810101 & 11,965 & $6.64_{-0.13}^{+0.13}\pm0.53$ & $65.2_{-0.4}^{+0.4}\pm5.2$ & $17.27_{-0.11}^{+0.11}\pm1.38$ & $90.7_{-0.6}^{+0.6}\pm7.3$ \\
  \enddata

  \tablecomments{The flux $F_{a\text{--}b}$ denotes the flux from $a$
    to $b$~keV. Asymmetric error bars are statistical and the
    symmetric uncertainties are systematic (Appendix~\ref{app:sys}).}

\end{deluxetable}

We provide fluxes in four energy bands for all \xmm{} observations in
Table~\ref{tab:xmm_flx}. These fluxes, except for the full 0.5-8~keV
flux, are plotted in Figure~\ref{fig:lc}. The 0.45--0.7~keV range is
dominated by the strong N\textsc{\,vii} Ly$\alpha$ (0.5003~keV) and
O\textsc{\,viii} Ly$\alpha$ (0.6537~keV) lines. In the literature,
0.5--2~keV is commonly referred to as the ``soft'' band, while
3--8~keV is the ``hard'' band.

\section{Best-Fit Parameters}\label{app:fit}
Table~\ref{tab:par} provides all best-fit parameters for fits using
the standard three-shock model (Sections~\ref{sec:mod}
and~\ref{sec:sa}). The temperatures are shown in
Figure~\ref{fig:com}. We note that all $\tau_2$ and the last epoch
$\tau_1$ only have lower limits. There also appears to be slight
degeneracies between $\tau$ and EM between different shock components,
which can be seen by the difference at 12,140~d relative to the other
epochs. However, the decrease in $T_3$ in the last epoch appears
robust. This is supported by the light curves, hardness ratios, and
continuous model fits.

In Table~\ref{tab:com}, we provide the observed (absorbed) fluxes of
the individual components and goodness-of-fit measures. These fluxes
are the bolometric fluxes of the components. Consequently, they are
much more uncertain than the fluxes in energy bands
(Table~\ref{tab:nus_flx}) because the component fluxes are more
sensitive to the underlying model.

\begin{deluxetable}{cccccccccccccc}
  \tablecaption{Best-Fit Parameters of the Three-Shock Model to \nustar{} and RGS Data\label{tab:par}}
  \tablewidth{0pt}
  \tablenum{F.1}
  \tablehead{
    \colhead{Group} & \colhead{Epoch} & \colhead{$k_\mathrm{B}T_1$} & \colhead{$k_\mathrm{B}T_2$} & \colhead{$k_\mathrm{B}T_3$} &                \colhead{$\tau_1$} &                \colhead{$\tau_2$} &                \colhead{$\tau_3$} & \colhead{EM$_1$}                & \colhead{EM$_2$}                & \colhead{EM$_3$}                \\
    \colhead{}      & \colhead{(d)}   &                  \colhead{} &             \colhead{(keV)} &                  \colhead{} &                        \colhead{} & \colhead{($10^{12}$~s~cm$^{-3}$)} &                        \colhead{} &                      \colhead{} & \colhead{($10^{58}$~cm$^{-3}$)} &                      \colhead{} }
  \startdata
1 &   9331 & $0.52_{-0.03}^{+0.02}$ & $0.98_{-0.05}^{+0.05}$ & $4.1_{-0.2}^{+0.4}$ & $0.92_{-0.11}^{+0.17}$  & $>2$ & $0.33_{-0.08}^{+0.29}$ & $14.9_{-0.5}^{+0.8}$ & $11.8_{-0.9}^{+1.1}$ & $3.5_{-0.8}^{+0.6}$ \\
2 &   9372 & $0.50_{-0.02}^{+0.02}$ & $0.96_{-0.06}^{+0.02}$ & $3.7_{-0.2}^{+0.2}$ & $0.97_{-0.13}^{+0.58}$  & $>2$ & $0.36_{-0.08}^{+0.21}$ & $14.5_{-1.0}^{+0.7}$ & $11.4_{-0.8}^{+1.1}$ & $4.3_{-0.7}^{+0.7}$ \\
3 &   9424 & $0.51_{-0.04}^{+0.03}$ & $0.97_{-0.05}^{+0.06}$ & $3.8_{-0.2}^{+0.4}$ & $0.93_{-0.13}^{+0.33}$  & $>9$ & $0.34_{-0.08}^{+0.39}$ & $14.7_{-1.0}^{+0.8}$ & $11.7_{-0.9}^{+1.0}$ & $3.8_{-1.0}^{+0.6}$ \\
4 &   9623 & $0.52_{-0.02}^{+0.02}$ & $0.99_{-0.02}^{+0.03}$ & $4.0_{-0.2}^{+0.4}$ & $0.95_{-0.05}^{+0.20}$  & $>2$ & $0.27_{-0.06}^{+0.10}$ & $15.0_{-0.8}^{+0.5}$ & $11.9_{-0.8}^{+1.0}$ & $3.2_{-0.5}^{+0.3}$ \\
5 &   9920 & $0.55_{-0.01}^{+0.02}$ & $0.98_{-0.01}^{+0.04}$ & $3.9_{-0.1}^{+0.1}$ & $0.86_{-0.09}^{+0.13}$  & $>4$ & $0.36_{-0.09}^{+0.20}$ & $12.4_{-0.9}^{+0.7}$ & $12.9_{-1.0}^{+0.9}$ & $3.8_{-0.3}^{+0.5}$ \\
6 &   9976 & $0.55_{-0.03}^{+0.03}$ & $0.98_{-0.02}^{+0.04}$ & $4.1_{-0.2}^{+0.3}$ & $0.85_{-0.12}^{+0.11}$  & $>4$ & $0.36_{-0.10}^{+0.32}$ & $12.4_{-0.9}^{+1.4}$ & $12.9_{-1.1}^{+1.0}$ & $3.8_{-0.7}^{+0.5}$ \\
7 & 10,021 & $0.55_{-0.02}^{+0.02}$ & $0.98_{-0.05}^{+0.02}$ & $4.0_{-0.2}^{+0.2}$ & $0.93_{-0.12}^{+0.22}$  & $>5$ & $0.26_{-0.06}^{+0.08}$ & $12.4_{-1.2}^{+0.8}$ & $12.8_{-1.0}^{+1.1}$ & $3.5_{-0.4}^{+0.7}$ \\
8 & 12,140 & $0.49_{-0.09}^{+0.06}$ & $0.98_{-0.07}^{+0.05}$ & $3.5_{-0.1}^{+0.1}$ & $>0.8$                  & $>2$ & $0.51_{-0.14}^{+0.23}$ & $6.9_{-0.8}^{+0.8}$ & $10.2_{-1.1}^{+1.2}$ & $7.6_{-1.0}^{+0.9}$ \\
  \enddata

  \tablecomments{The subscripts on the parameters in the header refer
    to the three individual shock components in terms of increasing
    temperature. The goodness-of-fit measures of these fits are
    provided in Table~\ref{tab:com}.}
\end{deluxetable}

\begin{deluxetable}{ccccccccccccccc}
  \tablecaption{Observed Bolometric Fluxes of the Shock Components in Table~\ref{tab:par}\label{tab:com}}
  \tablewidth{0pt}
  \tablenum{F.2}
  \tablehead{
    \colhead{Group} & \colhead{Epoch} &                               \colhead{$F_1$} &                               \colhead{$F_2$} &                        \colhead{$F_\text{3}$} & $\chi^2/\mathrm{DoF}$ \\
    \colhead{}      & \colhead{(d)}   & \colhead{($10^{-13}$~erg~s$^{-1}$~cm$^{-2}$)} & \colhead{($10^{-13}$~erg~s$^{-1}$~cm$^{-2}$)} & \colhead{($10^{-13}$~erg~s$^{-1}$~cm$^{-2}$)} &            \colhead{}}
  
  \startdata
1 &   9331 & $30.8_{-4.4}^{+2.3}$ & $30.8_{-2.5}^{+3.4}$ & $23.8_{-4.6}^{+5.1}$ & $1952/1785=1.09$ \\
2 &   9372 & $29.2_{-6.5}^{+1.7}$ & $29.7_{-2.5}^{+4.1}$ & $27.5_{-4.0}^{+6.7}$ & $1833/1664=1.10$ \\
3 &   9424 & $30.2_{-5.6}^{+2.6}$ & $30.3_{-1.4}^{+3.9}$ & $25.1_{-5.5}^{+4.6}$ & $1771/1620=1.09$ \\
4 &   9623 & $30.9_{-3.7}^{+2.1}$ & $31.2_{-2.2}^{+2.7}$ & $22.0_{-3.3}^{+3.8}$ & $1894/1748=1.08$ \\
5 &   9920 & $27.4_{-3.1}^{+3.1}$ & $33.7_{-3.1}^{+2.1}$ & $24.4_{-2.2}^{+4.8}$ & $2097/1872=1.12$ \\
6 &   9976 & $27.5_{-3.1}^{+3.5}$ & $33.8_{-3.2}^{+3.4}$ & $25.0_{-4.2}^{+3.9}$ & $2102/1895=1.11$ \\
7 & 10,021 & $26.7_{-3.6}^{+2.9}$ & $33.1_{-2.2}^{+2.1}$ & $24.8_{-3.0}^{+4.0}$ & $2030/1814=1.12$ \\
8 & 12,140 & $12.7_{-4.3}^{+3.5}$ & $26.6_{-3.1}^{+3.0}$ & $44.1_{-5.7}^{+7.7}$ & $1241/1093=1.14$ \\
  \enddata
  
  \tablecomments{The subscripts refer to the three individual shock
    components in terms of increasing temperature.}
  
\end{deluxetable}

\bibliography{ref}{}
\bibliographystyle{aasjournal}

\end{document}